# Why Emission Spectral Profile of Diverse Fluorophores Look Alike with Thick Polymer Overlayers?

*Vinayak Kumar Tiwari,* [†,§] *Srishti Batra,*[#] *Aranyak Sarkar,* [†] *Vinu Namboodiri,* [†] *Ravindra Venkatramani,* [#] *and Manoj Kumbhakar*[†,§,]*

[†]Radiation & Photochemistry Division, Bhabha Atomic Research Centre, Mumbai 400085, India

[§]Homi Bhabha National Institute, Training School Complex, Anushaktinagar, Mumbai 400094, India

[#] Department of Chemical Sciences, Tata Institute of Fundamental Research, Homi Bhabha Road, Colaba, Mumbai 400005, India

**Corresponding Author**

*Corresponding author: manojk@barc.gov.in

**ABSTRACT:** Fluorescence from fluorophore-polymer overlayers reflects an interplay between molecular photophysics and thickness dependent optical propagation. Combining spectrally resolved single molecule measurements, transition dipole moment orientation analysis and ensemble fluorescence, we show that structurally and photophysically distinct fluorophores can exhibit similar red shifted emission profiles under thick polymer overlayers even without additional reflecting surface. This similarity indicates a common optical transfer function imposed by the overlayer rather than a purely fluorophore specific molecular response. Although local molecular interactions, excited state dynamics and vibronic coupling govern intrinsic emission, the overlayer modifies detected spectra through refractive index contrast, dielectric boundary reflections, cavity effects and collection geometry. Vibronic energy spacing distributions remain largely preserved as thickness changes, whereas spectral positions and relative intensities vary systematically. Thus, polymer thickness primarily reshapes detected spectrum through optical propagation and outcoupling while retaining vibrational modes. Time dependent density functional theory identifies candidate Franck-Condon and Herzberg-Teller active dye modes contributing to spectral fluctuations and dipole reorientation.

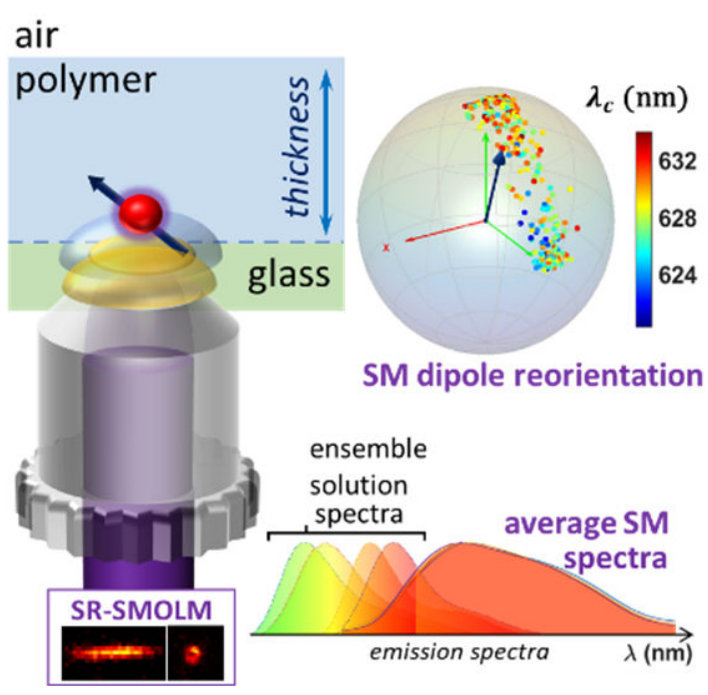

**TOC GRAPHICS**

**TEXT:** Fluorescent molecules embedded within solid polymer matrices serve as versatile model systems for probing how molecular photophysics responds to heterogeneous condensed phase environments.[1, 2] Polymer hosts inherently restrict large amplitude translational and rotational motion of embedded chromophores while retaining localized variations in free volume, molecular packing, microviscosity and non-covalent intermolecular interactions. These structural heterogeneities directly influence fundamental electronic processes, leading to spatial and temporal variations in fluorescence lifetimes, excited state relaxation pathways and spectral line shapes. Because bulk ensemble measurements average across these microscopic environments, critical sub-ensemble distributions and dynamic fluctuations are obscured. Single molecule (SM) fluorescence spectroscopy overcomes this ensemble averaging by resolving individual molecular trajectories, thereby unmasking the discrete photophysical variability and temporal dynamics of single emitters.[1-6]

A key manifestation of matrix heterogeneity is SM fluorescence spectral diffusion, in which the emission spectrum of an individual fluorophore undergoes continuous or abrupt temporal fluctuations.[7-9] These spectral jumps reflect local conformational shifts, matrix dipolar realignments and evolving non-radiative decay channels. Recent developments in multidimensional SM tracking,[10] combining emission spectroscopy with three dimensional transition dipole moment (TDM) orientation analysis, have linked spectral fluctuations to dipolar reorientations associated with changes in dominant vibronic emission contributions.[11] These findings highlight molecular orientation as a potentially coupled observable in the interpretation of SM spectral dynamics.

Despite these conceptual advancements, two primary gaps persist in the literature. First, how molecular level spectral orientational relationships manifest within spatially constrained polymer overlayers remains insufficiently understood. Second, fluorescence detected with thick polymer coating is intrinsically shaped by the photonic properties of the host layer.[6, 12-16] Finite film thicknesses introduce optical interference, polarization dependent out coupling, and substrate guided optical modes that modulate emission spectra.[17,

[18] Consequently, thickness dependent spectral variations cannot be unequivocally attributed to intrinsic molecular dynamics without assessing host layer contribution.

Here, we employ spectrally resolved SM orientation localization microscopy (SR-SMOLM[10, 11]) (Figure 1; Section S3, Supporting Information) to investigate a diverse set of fluorophores (Rh6G, RhB, Atto 520, MC540, Cy3, Cy3b, and NR; Figure 2) sandwiched between glass coverslip (Schott D 263 M borosilicate glass) and relatively thick polymer (Figure 2) layer without additional reflecting interfaces under ambient laboratory conditions. SM spectroscopy[19-25] and TDM orientation analysis[26] are combined with ensemble fluorescence measurements to examine spectral reshaping and its relationship to molecular reorientation. The vibrational origin of spectral modulation provides a complementary microscopic perspective. In the Franck-Condon (FC) framework, vibronic intensities are determined by overlap between vibrational wavefunctions on the two electronic potential energy surfaces, whereas Herzberg-Teller (HT) coupling accounts for coordinate dependent electronic transition moments. Both contributions can influence vibronic structure and emission intensities, particularly when the transition moment varies substantially along relevant normal modes. [27-29] Therefore, we carried out complementary time dependent density functional theory (TDDFT) calculations of the $S_1 \rightarrow S_0$ transition in the set of eight dyes (including different ionic forms) to identify active vibrational modes and evaluate FC contributions and HT transition moment derivatives, followed by direct calculations of electronic gap, transition strength, and transition axis responses. The thickness dependent optical response of the glass-polymer-air structure is considered alongside these molecular observables. By integrating molecular spectroscopy, orientation analysis, ensemble measurements and vibrationally informed calculations, this work examines how intrinsic fluorophore dynamics and finite thickness optical effects jointly contribute to fluorescence spectra in polymer overlayers. The novelty of this work resides in establishing a unified framework that treats host film thickness as an explicit coupling variable linking intrinsic fluorophore dynamics, vibronic coupling and overlayer optical boundary conditions.

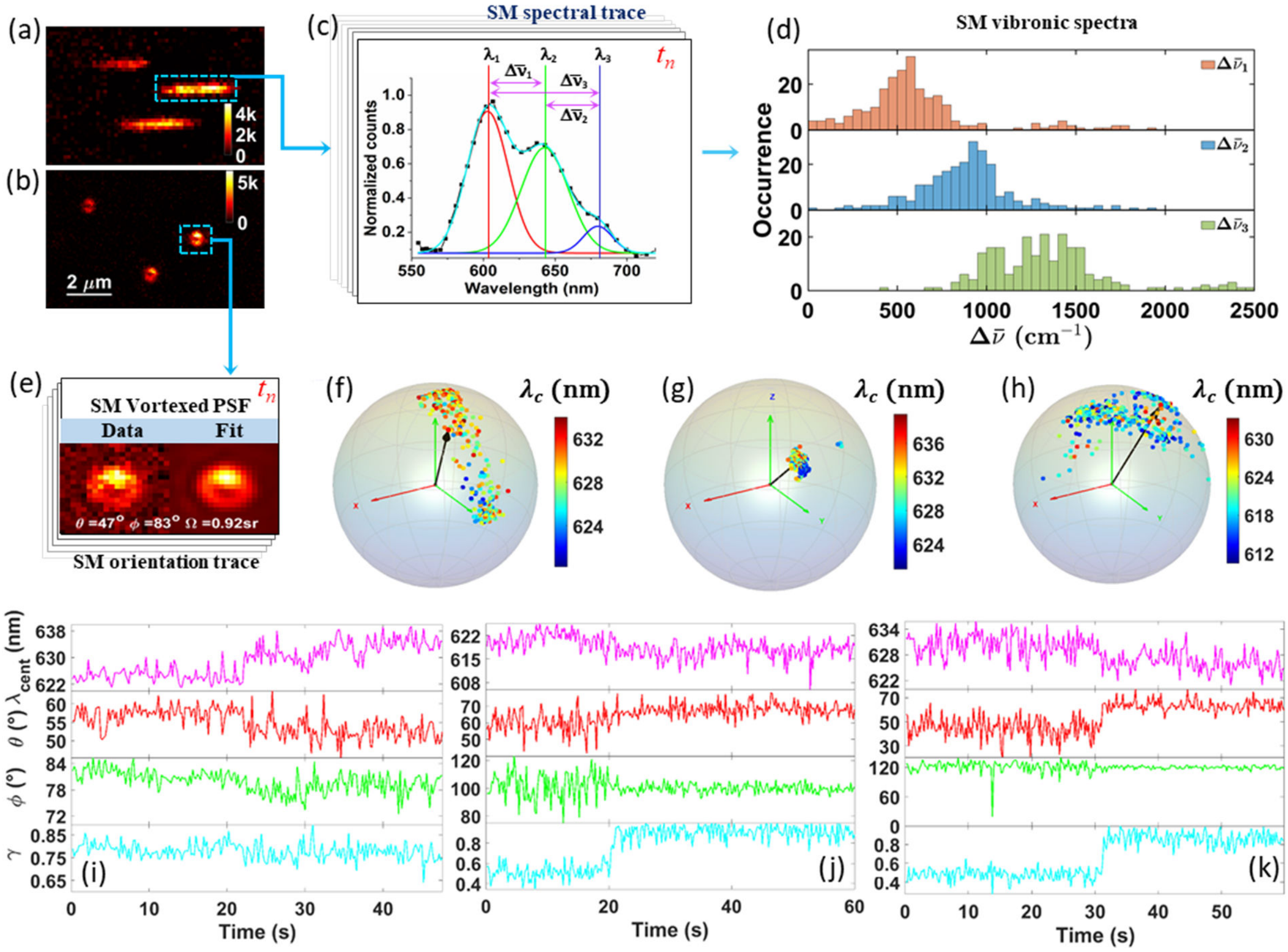

**Figure 1.** Simultaneous spectral and orientational tracking via SR-SMOLM. (a) Prism-dispersed spectral PSF and (b) vortex DSF images simultaneously capture SM spectra and 3D TDM orientation. Given photon budget constraints in the dual detection channel, with single NA 1.46 objective collection setup (Section S3, Supporting Information), acquisitions were performed under ~100 W/cm² circularly polarized 532 nm CW excitation in highly inclined and laminated optical sheet (HILO) geometry. Data were recorded at 4.5 nm/pixel dispersion with 200 ms exposure times over ~300–500 frames per stack. Extracted SM traces for (c) spectra and (e) orientation were generated after filtering out dim emitters (<800 photons). The spectral precision of our setup is ~3 nm.[10, 11] (c) Fitting SM emission bands yields vibronic maxima ($\lambda_i$) and band separations ($\Delta\bar{\nu}_i$) to identify active vibronic modes. (d) Histogram of mode occurrences highlighting the most frequently accessed vibronic states. (f–h) 3D projection maps correlating intensity weighted spectral centroids ($\lambda_c$) with TDM orientation (black arrows) for Cy3b, Atto 520, and Rh6G, respectively, revealing behaviors ranging from purely spectral fluctuations without reorientation (g) to extensive configurational dynamics (f). (i–k) SM traces comparing $\lambda_c$ and rotational constraint $\gamma$ for Cy3b, MC540, and RhB, respectively. Inset coordinate system defines polar ($\theta$) and azimuthal ($\phi$) angles alongside the rotational constraint $\gamma$ (an inverse function of the wobbling area Ω).[10, 11, 30]

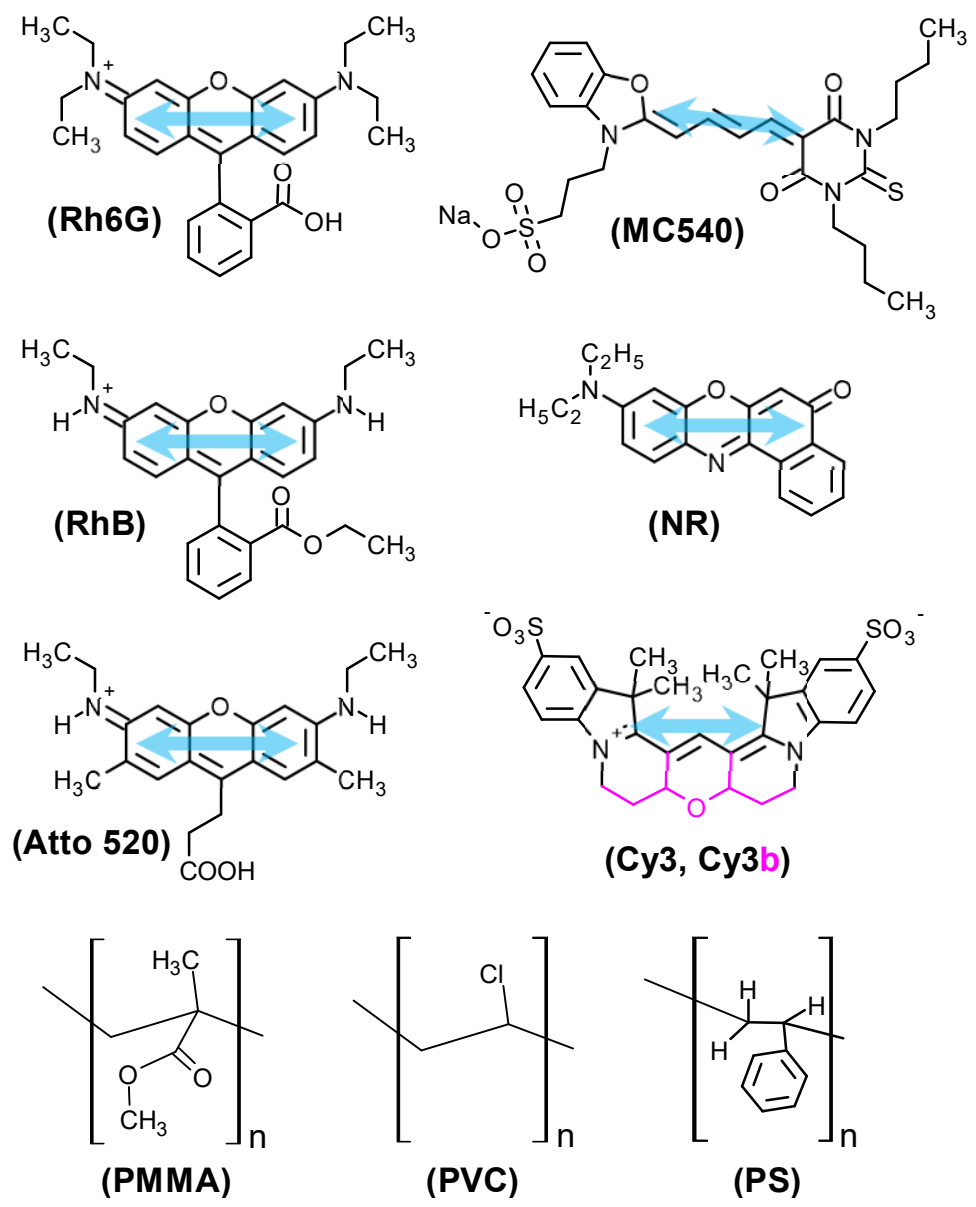

**Figure 2.** Structures of Rhodamine 6G (Rh6G), Rhodamine B (RhB), Atto 520, Merocyanine 540 (MC540), Cyanine 3 (Cy3), rigidized Cy3 (Cy3b), and Nile Red (NR) overlaid with theoretically calculated $S_1$-$S_0$ TDM vectors (blue double headed arrows) (Section S7, Supporting Information). The series features diverse molecular architectures, rigidities, conformational freedom, charge distributions and local environment sensitivities. Typical solution fluorescence emission maxima: Rh6G (~560 nm), RhB (~568-575 nm), Atto 520 (~538 nm), MC540 (~573 nm in aqueous solution), Cy3 (~570 nm) and strongly solvatochromic NR (polarity dependent shifts spanning few tens of nanometers).[31] Polymers used are poly(methyl methacrylate) (PMMA), poly(vinyl chloride) (PVC), polystyrene (PS).

The structurally and photophysically diverse fluorophores provide a systematic framework for separating molecular and optical contributions to fluorescence in thick polymer coatings (by systematically probing polymer films beyond ~50 nm, Figure S1). Environment sensitive NR probes local polarity and polymer rigidity, while highly emissive rhodamine derivatives enable assessment of waveguiding and optical amplification. Comparisons among Cy3, MC540, and conformationally restricted Cy3b further reveal the roles of molecular flexibility, while Atto 520-Rhodamine comparisons probe the influence of central phenyl ring substitution. Their emission maxima span approximately 540-590 nm under typical conditions, while NR exhibits most pronounced solvatochromism in the range of ~540-640 nm. Their complementary photophysical properties, together with the solvatochromism of NR, enable ensemble and SM measurements to distinguish molecular heterogeneity from coating scale optical effects.

Our SR-SMOLM measurements reveal broadly similar orientational behavior and spectral diffusion across chemically distinct fluorophores (Figure 1f-k & Figures S10-S14). More strikingly, average SM spectra obtained with relatively thick PMMA (MW 350,000, $n$=1.49-1.50[32, 33]) overlayers converge toward remarkably similar red-shifted spectral profiles ($\lambda_c$~615 nm), including closely related vibronic energy spacing distribution (Figure 3a,c; the emission bands exhibit a vibronic progression with a dominant characteristic spacing of ~1050 cm$^{-1}$), despite substantial differences in molecular structure and solution phase emission wavelengths (Figure 3b). The effect is particularly pronounced for Atto 520, whose emission is red-shifted by more than 80 nm. Such convergence is difficult to reconcile solely with dye specific solvatochromism, which should retain distinct spectral responses according to the electronic structure, excited state dipole, and environmental sensitivity of each fluorophore. The similar response of weakly solvatochromic rhodamines and carbocyanines from the strongly solvatochromic NR further supports this distinction.[34-37] The observed broad spectral profile reflects local heterogeneity of polymer matrix experienced by SMs.[2]

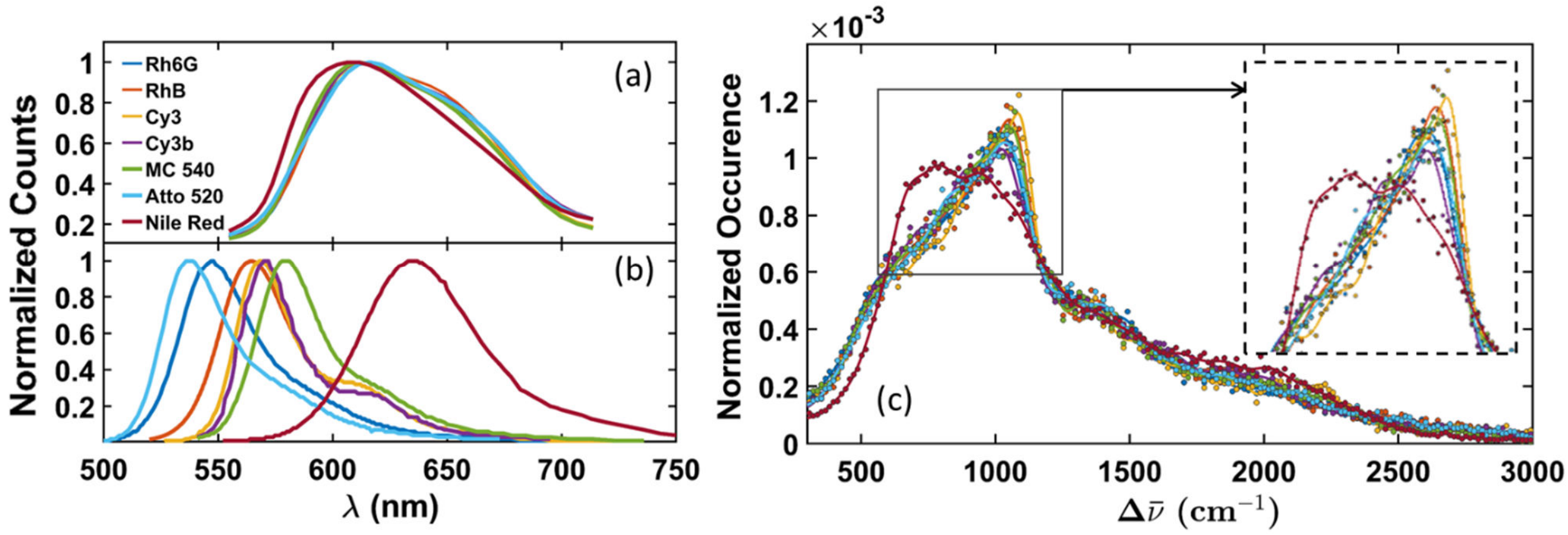

**Figure 3.** Average SM spectra (a) with thick PMMA overlayer is nearly similar for all the dyes irrespective of their chemical identity though their ensemble emission spectra (b) are distinctly different in ethanol. All the dyes also exhibit similar distribution of vibronic mode spacings (c) constructed from SM spectra. The solid lines represent multi-Gaussian fits. Only the solvatochromic probe NR shows slight exception from the general behavior.

The experimental conditions and optical configuration were identical for all the fluorophores, apart from minor excitation power adjustments. The comparable red shift observed for chemically diverse dyes therefore suggests that, in addition to molecular photophysics, the polymer overlayer may contribute to the detected spectrum. Possible contributions include dielectric interference, wavelength- and angle-dependent out-coupling, reabsorption, waveguiding and modification of the local electromagnetic environment.[38-40] Intrinsic fluorescence generated by a molecule in the glass-dye-polymer structure propagates through the finite thickness dielectric layer before collection; consequently, refractive index, film thickness, dipole orientation and polarization can influence its angular distribution and collection efficiency. Fluorophores near dielectric interfaces can also couple to substrate or guided modes, with the associated optical response depending sensitively on polymer film thickness.[38-40]

The detected spectrum can therefore be considered phenomenologically as

$$I_{det}(\lambda, d) = I_{mol}(\lambda, E_{loc})\, T_{optical}(\lambda, d, \theta, p)$$

where $I_{det}$ is the detected spectrum, $I_{mol}$ represents the molecular emission determined by the local environment, $d$ is the polymer thickness, and $T_{optical}$ describes the thickness-, wavelength-, angle- and polarization-dependent optical transfer function of the glass-dye-polymer structure. The measured spectrum thus contains both molecular and coating mediated contributions.

To distinguish these effects, we varied the PMMA thickness using ten times dilute polymer solutions (≤0.1% PMMA in $CHCl_3$; Figure 4). Reducing the overlayer thickness (~30 nm) produced a systematic blue shift of the average SM spectra toward the bare fluorophore spectra for all the dyes, accompanied by increasingly dye specific spectral profiles (Figure 4a). This continuous evolution is significant because the interfacial fluorophore environment remains essentially unchanged while polymer thickness is the principal experimental variable. The red-shifted spectrum in thick PMMA is therefore not an intrinsic, thickness independent property of the fluorophore, but evolves with the optical thickness of the surrounding dielectric layer. Comparison of NR with the other fluorophores further suggests a relatively larger molecular contribution for NR, consistent with its greater environmental sensitivity[8] (Figure S15). The bare fluorophore measurements provide an internal reference (Figure 4c & Figure S16). In the absence of PMMA, selected fluorophores recover solution like emission positions and characteristic spectral profiles.

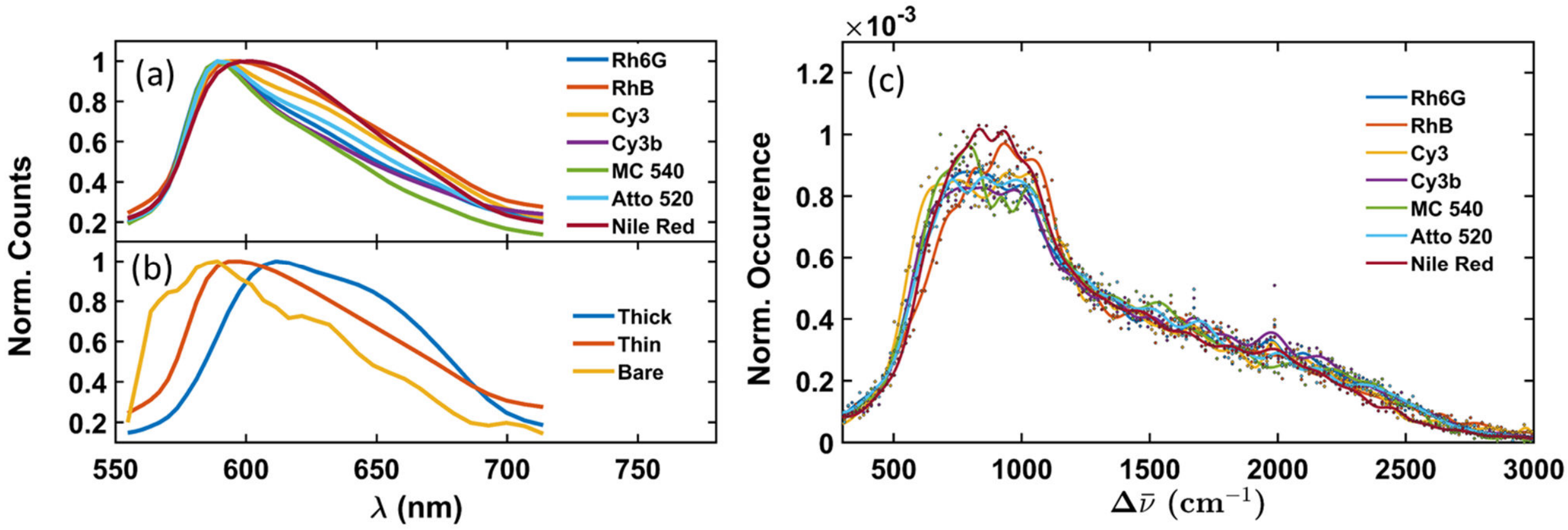

**Figure 4.** Average SM spectra (**a**) with thin PMMA overlayer is degenerate and blue shifted compared to thick overlayer case. The apparently broad vibrational mode distribution with thin PMMA overlayer also shows essentially same specific feature (b) (Figure S15). Comparison of average SM spectra for RhB with thick & thin coating and with bare condition (c) highlights the interplay between local molecular photophysics and optical coating effects. We have not considered SMs showing truncated spectra at blue side (Figure S7) in any of our analysis (arising due to 561 long pass filter).

The histogram of $\lambda_c$ distribution for all the dyes (Figures S20, S21 for thin and thick configurations, respectively) reflect SM spectral heterogeneity, but apparently with similar kind of distribution positions in most of the cases. A bootstrap Kolmogorov-Smirnov (KS) analysis of SM spectral snapshots was performed to check whether the thick and thin spectral data sets come from the same underlying distribution profile (i.e. optical response) or not, as exemplified with RhB and Cy3 in Figure 5 (and Figure S22) to assess for closed and open frame chromophores', respectively. Exceedingly low p-value ($p << 0.0001$) even in thick overlayer samples indicates that the $\lambda_c$ distribution profiles of RhB and Cy3 are statistically distinct and independent in both regimes. This further highlight the presence of independent molecular response from the dyes alongside optical response due to coating thickness. A measured spectral shift of mean position with overlayer thickness, evident from Figure 5a clearly indicates wavelength dependent spectral attenuation with overlayer thickness. The sequence of bare fluorophore → thin PMMA → thick PMMA therefore corresponds qualitatively to a transition from predominantly molecular emission to a mixed molecular/optical response and finally to a strongly coating influenced spectrum. The systematic thickness dependence supports the presence of a wavelength dependent optical transfer contribution superimposed on the molecular emission.

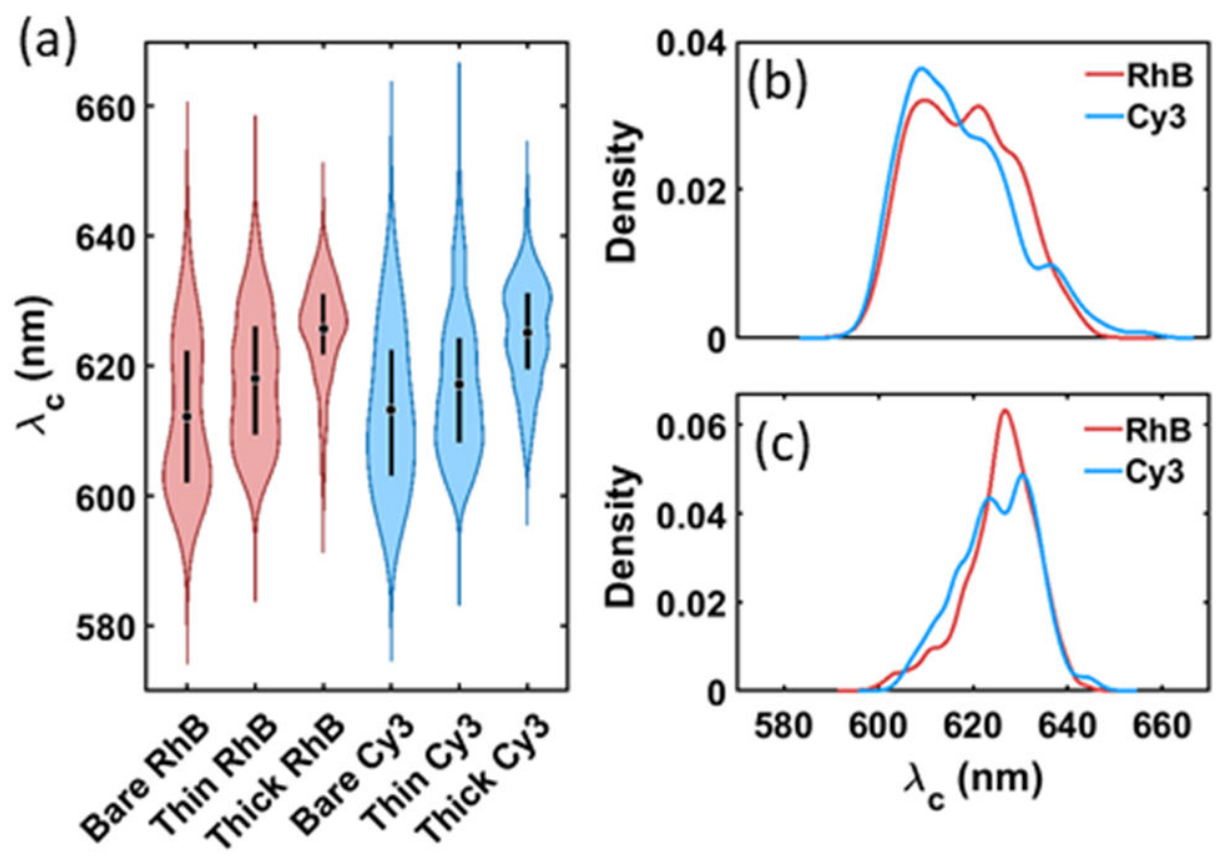

**Figure 5.** Probability density distributions of $\lambda_c$ across bare, thin and thick PMMA configurations for RhB (crimson) and Cy3 (cyan). Solid black vertical bars represent the interquartile range ($25^{th}$–$75^{th}$ percentiles), while white outlined black markers denote the means. A clear shift toward higher mean $\lambda_c$ value is observed for transitioning from bare to thick overlayer for both fluorophores. Two sample KS density comparison plots for RhB and Cy3 with thin (b) and thick (c) overlayer configurations. The calculated test statistics (D) and p-values ($D < 0.05$, $p \ll 0.0001$) confirm that the spectral distributions of RhB and Cy3 are statistically distinct and independent in both regimes.

Finally, fixed instrumental effects are unlikely to explain this evolution. HILO illumination primarily controls the excitation geometry and does not intrinsically impose a wavelength selective emission response. Likewise, supercritical angle fluorescence (SAF) depends on emitter interface separation and emission angle and would require angular discrimination for selective detection.[41-46] The 561 nm long-pass emission filter does introduce sampling bias against shorter wavelength fluorescence ( and also rejection of truncated spectra from further analysis, Figure S7), but its spectral response remains essentially unchanged during the thickness series. PMMA has extremely high optical clarity and very low absorption in the 500-700 nm region, therefore attenuation of molecular emission due to re-absorption by the polymer overlayer is improbable. These static contributions therefore cannot readily account for the systematic recovery of solution like spectra as the PMMA thickness decreases. Interface mediated emission is more appropriately regarded as part of $T_{optical}$, whose thickness dependence is directly probed by the polymer overlayer series.

The glass-PMMA-air structure constitutes a multilayer optical environment in which weak reflections (Section S2, Supporting Information) at the interfaces may produce interference effects. Previous studies have demonstrated thickness dependent spectral and angular modulation of fluorescence through self-interference in dye doped polymer films in front of a mirror.[16, 47-49] Therefore, interference and out-coupling at the PMMA interfaces may contribute to the observed fluorescence spectral variations in the present system.

Contrary to present spectral red shift with polymer overlayer, more than 20 nm spectral blue shift is reported with Rh6G for underlayer PMMA thickness of ~44 nm.[50] Interestingly this was not concurrently followed by change in fluorescence lifetime of Rh6G, prompting lifetime and spectral response were not related events. Distinction of these two geometries, i.e. underlayer and overlayer thickness, is important because the fluorophore's distance from each interface determines the phase and amplitude of the reflected dipole fields. Since phase shift depends on $d/\lambda$ (equation S1), observed spectral shift is a result of relative attenuation of $T_{optical}(blue, d)/T_{optical}(red, d)$. However, the sign of the spectral shift cannot be predicted from thickness alone. Depending on the fluorophore's position, which interface is involved, dipole orientation, detection geometry and increasing thickness can preferentially enhance either the blue or red portion of the spectrum. Reweighting of relative vibronic band intensities are known to display spectral fluctuations. A quick simulation of spectral line shape of Rh6G, Cy3, NR and Atto 520 (Figure 6) for an asymmetric thin film dielectric microcavity without additional reflecting surface (described in Section S5, Supporting Information) clearly establishes spectral alteration as a direct consequence of Purcell effect.[51-53] This is further supported by the observation of shift to lower values for the full distribution of SM fluorescence lifetimes with increase in polymer thickness (Figure S23).[16] Another possibility is that increasing polymer thickness changes the local physical environment experienced by the fluorophore, e.g. polymer chain packing around the fluorophore, local free volume, segmental mobility and conformational relaxation, etc. Distribution of rotational constraint indicate relative relaxation in thick overlayer compared to thin case (Figure S19), though influence of polymer molecular weight is negligible (Figure S27).

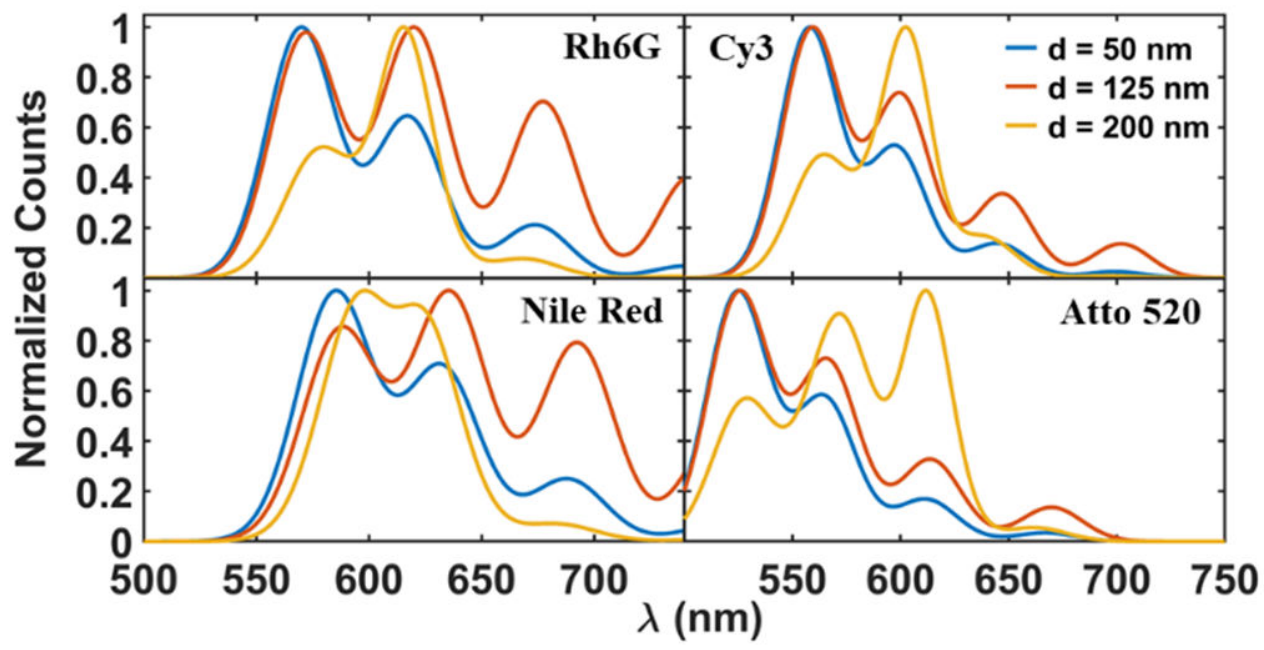

**Figure 6.** Simulation of spectral line shape for a SM in optical microcavity based on the Fabry-Pérot thin film interference as function of film thickness (*d*) clearly indicates reweighting of relative vibronic band intensities in favor of longer emission wavelengths. This is irrespective of fluorophore identity. Simulations were performed for $\theta = 60^0, \gamma = 1, NA = 1.46, n_{glass} = 1.53, n_{PMMA} = 1.49, n_{air} = 1,$ HILO excitation with 532 nm laser excitation. The cavity pulling effect along with shorter wavelength spectral filtering by the 561 nm long pass filter clearly explains the observed extreme red-shift for Atto 520.

The vibronic energy spacing distributions reconstructed from SM emission bands exhibit broadly similar profiles for thin and thick PMMA overlayers, dominated by 700-1200 $cm^{-1}$ spacings with weaker high energy contributions (Figure S15). Although relative feature intensities vary with overlayer thickness, the overall distribution and characteristic energy scales remain nearly conserved. This preservation alongside thickness dependent emission shifts indicates that the overlayer primarily modifies fluorescence spectral position and relative weighting via a thickness dependent optical propagation and detection response, leaving dominant molecular vibrational scales stable. Polymer supported optical modes depend sensitively on polymer film thickness and redistribute fluorescence according to wavelength and emission angle.[39, 48, 49, 54] Furthermore, the SM brightness distribution shifts toward higher values with increasing overlayer thickness (Figure S17) without profile narrowing, conceivably due to local environmental heterogeneity, including PMMA thickness variations.

SR-SMOLM further reveals a correlation between spectral diffusion and molecular orientation fluctuations in the glass-dye-polymer architecture, see Figure 1(f-h), Figures S10-S14, and Movies S1-S3. Principal component analysis (PCA) of few selected traces (Figures S24-S26) helps in highlighting the SM response into several clusters. Most molecules undergo small angular reorientations (≤10°, e.g. Figure 1g,i ), consistent with transitions among vibronically active states,[11] whereas occasional larger TDM reorientations coincide with pronounced changes in rotational constraint (e.g. Figure 1f,h). These later excursions last for a very short period, mostly single frame of 200 ms (Figures S12, S14). These rare events may reflect access to otherwise hindered conformational or physical states enabled by the local polymer interactions. Sterically hindered planarization of aromatic rings by torsional motion at the inter-ring C–C linkage in Rhodamine's is not expected to alter TDM direction (Figure S31) and also has very high activation barrier. Prospective dye polymer interaction dynamics in case of NR showing large scale spectral shift in tune of nearly the entire spectral range may trigger TDM reorientation. Photo/matrix induced twisting in polyene bridged molecules remain another viable possibility. Spectral diffusion of single fluorophores embedded in polymer matrices can also arise from conformational changes and dynamic polymer cage rearrangements of the surrounding polymer environment.[9, 11, 55] Although conformational transitions or chemical state changes of the dyes[56, 57] cannot be entirely excluded, their contribution is expected to be limited for the present fluorophores (Section S6, Supporting Information). Spectral diffusion may therefore arise predominantly from dynamic rearrangements of the local polymer environment, coupled to fluctuations in molecular orientation.

Interestingly, transitions in the averaged SM emission spectra can be classified according to their spectral and TDM signatures. In some cases, transitions produce a modest spectral shift ($\Delta\lambda \approx 8$ nm) with little change in spectral shape or relative vibronic intensities, despite measurable TDM reorientation ($\Delta\theta \approx 5°$; Figure S10). This behavior is consistent with a change in the $S_0$–$S_1$ energy gap without substantial alteration of the underlying potential energy surfaces. In other cases, the transition is accompanied by redistribution of the dominant vibronic features and a shift in the spectral centroid ($\Delta\lambda \approx 10$ nm, $\Delta\theta \approx 4°$, $\Delta\bar{\nu} \approx 170$ cm$^{-1}$; Figure S11), indicating a concomitant change in the ground state potential energy surface and the associated vibrational structure. The same molecule subsequently undergoes a transition with pronounced TDM reorientation ($\Delta\theta \approx 15°$, $\Delta\varphi \approx 20°$) and a smaller spectral shift ($\Delta\lambda \approx 5$ nm), while the vibronic spacing remains essentially unchanged (state I→II in Figure S11). This decoupling of spectral and orientational responses further supports a change in the electronic energy gap without substantial modification of the vibrational landscape. Transitions involving changes in the relative vibronic peak intensities are also observed (Figure S12), pointing to variations in FC factors and consequently in the relative equilibrium geometries sampled by the ground and excited states.

In order to separate intrinsic dye responses from contributions not captured without an explicit polymer environment, we used electronic structure calculations with continuum solvation (Section S7, Supporting Information) to identify whether, and which intramolecular degrees of freedom of the dyes can contribute to the experimentally observed SM spectral diffusion and transition dipole reorientation. Essentially, FC and HT active modes define the vibrational pathways coupled to the $S_1 \rightarrow S_0$ transition and can potentially modulate both the excited state geometry and emission response. Modes were therefore screened according to their normalized Huang-Rhys factors ($S_k$) and reorganization energies ($\lambda_k$) together with molecular-frame HT activity to identify those most relevant to spectral modulation and TDM reorientation. Across the set of eight dyes (including different ionic forms) the largest TDM orientation changes are driven by soft, low-frequency collective deformations (butterfly, flapping, seesaw and framework wagging modes) with characteristic frequencies in the ~15–85 $cm^{-1}$ window (examples: Cy3 collective modes at 15–30 $cm^{-1}$ produce axis deviations up to ~5.4°; Cy3b flapping near 20 $cm^{-1}$ gives ~5.6°; RhB zwitterion rocking/flapping ~30–70 $cm^{-1}$ yields short-axis deviations approaching ~1.9° and relaxed torsional spans up to ~4.4°). By contrast, high-frequency core stretching/compression modes in the ~1400–1730 $cm^{-1}$ region strongly modulate vertical gap and oscillator strength but generally induce smaller TDM rotations (commonly <~1.8°, often ~0.2–1.7°; e.g., Cy3 stretches near 1500–1690 $cm^{-1}$ give axis deviations ≤~1.6°; Nile Red core stretches up to ~1.7°). Notably, the dominant contributions to TDM reorientation arise from very low-frequency modes (< 100 $cm^{-1}$) that are not directly resolved within the recorded SM vibronic progressions, highlighting a vibrational pathway for orientational fluctuations that can occur with only weak spectral signatures. Consequently, observed large scale spectral and reorientation dynamics possibly arises from the dye-polymer interactions within the local nanocavity at the interface of glass and polymer.

The molecular weight of PMMA has no additional influence on the spectral behavior of SMs, except slight alteration in their parameter distributions (Figure S27). To determine whether the pronounced red shift and spectral convergence observed with thick PMMA overlayers are specific to PMMA or represent a more general effect, we performed analogous measurements using PVC (*n*~1.54[58, 59]) and PS (*n*~1.58–1.59[60]) with Rh6G and RhB under comparable conditions (Figure S28-S30). As shown in Figure 7, all three polymers exhibit substantial wavelength shifts in the glass-dye-polymer architecture, indicating that the effect is not specific to PMMA. The modest spectral differences likely reflect polymer specific dye-matrix interactions arising from their distinct chemical functionalities, namely the polar ester group in PMMA, ionic C–Cl groups in PVC, and hydrophobic phenyl groups in PS. These results further support the combined roles of local dye-polymer interactions and the optical transfer function of the finite thickness polymer overlayer.

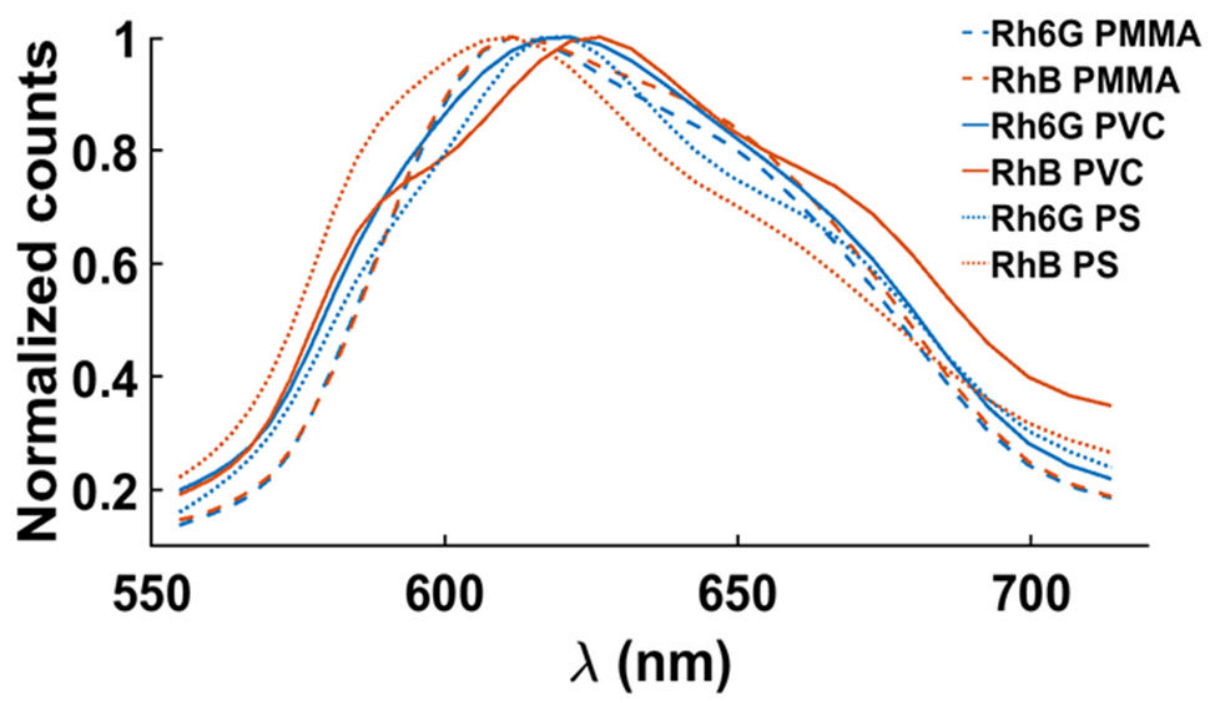

**Figure 7.** Average SM spectra for Rh6G (blue) and RhB (red) with thick overlayers of PMMA (dashed line), PVC (sloid line) and PS (dotted line).

In short, the fluorescence spectrum of thick polymer coated samples is therefore a coupled molecular-optical observable: local fluorophore-polymer interactions govern the intrinsic emission and molecular dynamics, while the finite thickness overlayer modifies the detected spectrum through wavelength dependent propagation, interference, mode coupling, and out-coupling. Thus, film thickness is not merely a geometrical parameter, even in absence of an additional reflecting surface, but a key design variable for controlling and tailoring the optical response of fluorescent polymer coatings for spectral filtering, wavelength conversion, amplification, and controlled emission. Consequently, rigorous spectral interpretation requires holistic molecular-optical modelling.

**ASSOCIATED CONTENT**

**Supporting Information**

Additional figures for spectral & orientation distributions, analysis methodology, PCA analysis of SM traces, computational methods and intrinsic dye response and sample preparation (PDF).

**AUTHOR INFORMATION**

**Notes**

The authors declare no competing financial interests.

This work is a part of PhD Thesis of Vinayak Kumar Tiwari.

**Author contributions:** M.K. conceived the study, supervised the research and provided scientific guidance and the necessary infrastructure. V.K.T. performed the experiments, data acquisition and processing, formal analysis and preparation of figures and graphical representations. A.S. formulated the image analysis methodology. A.S. and V.N. contributed to the optimization of the experimental setup and microscopy methodology. V.N. performed simulation for cavity effect. S.B. performed the DFT calculations and their analysis. R.V. supervised the DFT calculations and provided scientific guidance on their analysis. M.K. interpreted the results and prepared the original manuscript. All authors contributed to reviewing and editing the manuscript and approved the final version.

**ACKNOWLEDGMENT**

This work is funded and supported by BARC (DAE), India.

**Supporting Information**

# Why Emission Spectral Profile of Diverse Fluorophores Look Alike with Thick Polymer Overlayers?

**Vinayak Kumar Tiwari,** [†,§] **Srishti Batra,**[#] **Aranyak Sarkar,** [†] **Vinu Namboodiri,** [†] **Ravindra Venkatramani,** [#] **and Manoj Kumbhakar**[†,§,*]

[†]Radiation & Photochemistry Division, Bhabha Atomic Research Centre, Mumbai 400085, India

[§]Homi Bhabha National Institute, Training School Complex, Anushaktinagar, Mumbai 400094, India

[#] Department of Chemical Sciences, Tata Institute of Fundamental Research, Homi Bhabha Road, Colaba, Mumbai 400005, India

* manojk@barc.gov.in

## Section S1

**Materials**: All the dyes used (i.e. Rh6G, RhB, MC540, Atto 520, and NR) were purchased from Sigma-Aldrich. The dyes Cy3 and Cy3b were a gift from Prof. J. Enderlein (George August University, Göttingen). Poly(methyl methacrylate) (PMMA; MW ≈ 350,000 and 100,000 g/mol), polyvinyl chloride (PVC; MW ≈200,000 g/mol), and polystyrene (PS; MW ≈150,000 g/mol) were also obtained from Sigma-Aldrich. Spectroscopic grade chloroform ($CHCl_3$), HPLC grade water, tetrahydrofuran (THF), and toluene were purchased from Merck Life Science Pvt. Ltd. All chemicals were used as received without further purification. Cover slips (Ted Pella Inc., No. 1 thick) were cleaned as described earlier.[1, 2] Ensemble solution spectra were recorded using a JASCO model V530 spectrophotometer (Tokyo, Japan) for ground-state absorption spectra and HITACHI model F-4010 spectrofluorimeter (Tokyo, Japan) for steady-state (SS) fluorescence spectra at a dye concentration of ~ μM.

## Section S2

**Sample Preparation:** Polymers such as PMMA, PVC, PS, and related glassy materials are particularly relevant because of their optical transparency, mechanical stability, and compatibility with fluorescent molecules.[3-9] PMMA is widely used as a protective matrix in single molecule (SM) studies, reducing photobleaching and background fluorescence.[1, 10]

Beyond molecular interactions in ultrathin layers, sufficiently thick polymer films can function as active optical components, with thickness, refractive index, interfaces, and optical confinement governing fluorescence through interference, waveguiding, reabsorption, and cavity effects. Dye-doped polymer films and waveguides have consequently exhibited thickness dependent emission,[11, 12] amplified spontaneous emission (ASE) wavelength tuning,[5, 6] amplification,[13, 14] spectral narrowing and directional emission.[15, 16] These findings highlight film thickness as a key optical parameter rather than merely a geometrical property.

In multilayer structures and microcavities, dielectric confinement can significantly modify emitter spectra and radiative rates, as demonstrated in dye-doped polymer microcavities and related systems exhibiting spectral narrowing and Purcell-type effects.[7, 17, 18] Thus, fluorescence from sufficiently thick polymer coatings may reflect both intrinsic molecular emission and optical filtering by the surrounding film. Beyond concentration and aggregation effects,[4, 19] polymer matrices exhibit spatial and temporal heterogeneity that can alter local polarity, rigidity, and confinement, leading to spectral fluctuations and changes in molecular rotational dynamics.[8, 20, 21]

In order to elucidate how molecular level fluorophore-polymer interactions translate into measurable optical responses within polymer coatings, we systematically probed polymer films beyond ~50 nm (**Figure S1**), we examine the transition from a predominantly molecular regime in bare fluorophore case to one where film thickness alters optical response.

Dilute dye solutions (~10 pM) were prepared in $CHCl_3$ by successive dilution from stock. In every step the solution was vortexed for at least 15 minutes to ensure complete dissolution. The solution was subsequently spin-coated onto pre-cleaned glass coverslips. The required polymer layer was then spin-coated directly onto the dye-coated coverslip. For polymer coating, requisite amount of PMMA or PVC or PS was dissolved in $CHCl_3$, THF or toluene, respectively. The overlayer fabrication was carried out using a two-step spin coating procedure. First, the solution was transferred at 2,500 rpm for 15 seconds, after which the speed was increased to 6,000 rpm for 100 seconds. Finally, the polymer coated fluorophore samples were either placed under dark condition in a vacuum desiccator or air dried for at least an hour prior to measurements.

**Thick Polymer Coating:** A stock solution of ~1 wt% PMMA in chloroform was prepared and homogenized via vortex mixing for ~ 30 minutes prior to spin coating. AFM topographic analysis revealed a non-uniform film thickness ranging from 50 nm to 200 nm. Because these dimensions significantly exceed typical molecular boundaries, this regime was classified as a thick polymer coating. Similar preparation steps were followed for PVC and PS coating.

**Thin Polymer Coating:** To achieve a thinner overlayer, an additional around ten times dilution of the PMMA polymer solution with chloroform was performed. Following spin coating, AFM imaging showed a reduced, non-uniform thickness ~30 nm. Since this spatial profile approaches the physical dimensions of individual polymer chains, it was categorized as a thin PMMA coating.

**Reflectivity Calculation:** Using the refractive indices of $n_{\text{glass}} \simeq 1.523$ and $n_{\text{PMMA}} \simeq 1.4935$, the normal incidence reflectivity at the glass-PMMA interface is $R_{glass/PMMA} = \frac{n_{\text{glass}} - n_{\text{PMMA}}}{n_{\text{glass}} + n_{\text{PMMA}}} \simeq 9.6\text{x}10^{-5}$; i.e. only about 0.01%. At the PMMA–air interface, $R_{glass/air} = \frac{n_{\text{glass}} - 1}{n_{\text{glass}} + 1} \simeq 3.9\%$. So, one interface is essentially index matched and the other is only weakly reflecting. Consequently, it is a natural candidate for contributing to weak interference effects involving fields reflected within or radiated through the PMMA layer.

**Thickness-dependent phase: 50–200 nm**

For a one-way propagation path in PMMA,

$$\varphi(\lambda, \theta_p, d) = \frac{2\pi n_p d \cos\theta_p}{\lambda} \qquad \text{(eq. S1)}$$

The round-trip propagation phase is twice this value, before including reflection phase shifts.

At normal incidence and λ~600 nm, the calculated one-way phase ranges from approximately 0.78 rad for a 50 nm layer to 3.13 rad for a 200 nm layer. The phase therefore changes substantially as the layer thickness varies.

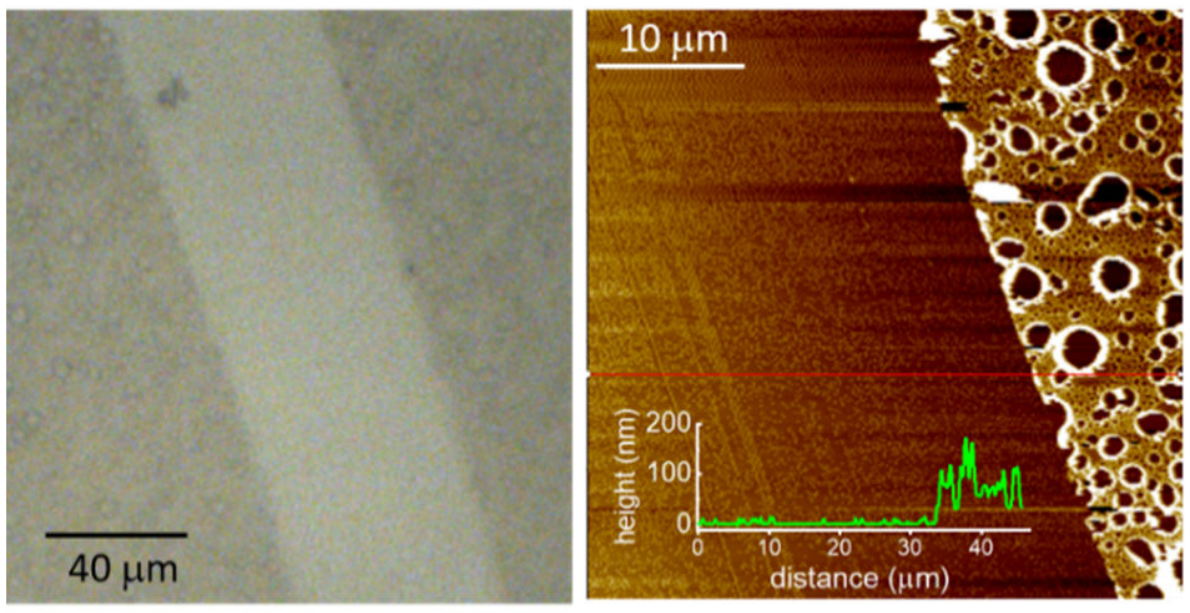

**Figure S1.** Polymer coating thickness was characterized using optical profilometer (MFT2000, RTec Instruments) and AFM (NX100, Park Systems). Representative images for thick PMMA (350K) are shown in the left and right side, respectively. Height profile after a scratch over the coverslip, as shown in the inset, indicated a thickness range of ~50-200 nm for PMMA thick overlayer. The same measurements with thick PVC and PS show thickness variations in the range of ~40-120 nm and ~35-150 nm, respectively. The regional thickness variations (non-uniformity) observed across both preparation regimes are attributed to the solvent evaporation rate, fluid dynamics and disc rotation speeds.

## Section S3

**SR-SMOLM Setup**: SM fluorescence experiments were carried out on a custom built wide-field epifluorescence microscope (Axio Observer D1, Carl Zeiss GmbH) configured for circularly polarized excitation in highly inclined and laminated optical sheet (HILO) illumination geometry (just before reaching near total internal reflection).[1, 2] The excitation light (532 nm continuous wave; Shanghai Dream Lasers Technology Co., Ltd.) was focused onto the back focal plane of a high numerical aperture oil-immersion objective lens (100x, NA 1.46; Carl Zeiss GmbH). The emitted fluorescence signal was collected through the same objective and filtered using a dichroic mirror (FF538-FDi01, Semrock) along with a filter combination comprising a 532 nm notch filter, a 561 nm long-pass filter (Semrock). As schematically illustrated in **Figure 1**, the optical detection path split into two distinct channels: a spectral detection channel utilizing prism-based dispersion, and an orientation-localization channel incorporating a vortex phase plate (VPP; V-593-10-1, Vortex Photonics) placed at the Fourier plane of a 4*f*-imaging system.[22] This optical layout provided an additional 2x magnification, yielding an effective pixel size of 80 nm at the camera plane. After passing through the dichroic mirror, a beam splitter divided the fluorescence signal, which was then directed via relay mirrors and lenses onto an electron-multiplying charge-coupled device (EMCCD) camera (DU971N-UVB, Andor) across both detection paths. Optical alignment, calibration, and image processing procedures were executed as outlined in our earlier publications.[1, 2] Image acquisition was controlled using Andor Solis software. Data were collected continuously in kinetic mode with an exposure time of 200 ms per frame and an electron-multiplying (EM) gain setting of 200. For each measurement, image stacks comprising 50 sequential frames for bare fluorophore case and ~300-1000 sequential frames for fluorophores with overlayer were recorded.

**Orientation imaging of single molecule:** For the extraction of orientation parameters, we have adopted the methodology elucidated by Hulleman et al.[22] and discussed with respect to our SR-SMOLM setup.[1, 2] We have considered only the isotropic rotationally diffusing for molecular dipoles. Single-molecule fitting was performed using a full vectorial point spread function (PSF) model. The observed intensity pattern of a single-molecule dipole spread function (DSF) on the detector is jointly determined by eight physical parameters: the 3D spatial coordinate vector, r = (x, y, z), the 3D orientation vector, Θ = (θ, φ, γ), the total signal photon count, *N*, and the background noise per pixel, *b*.

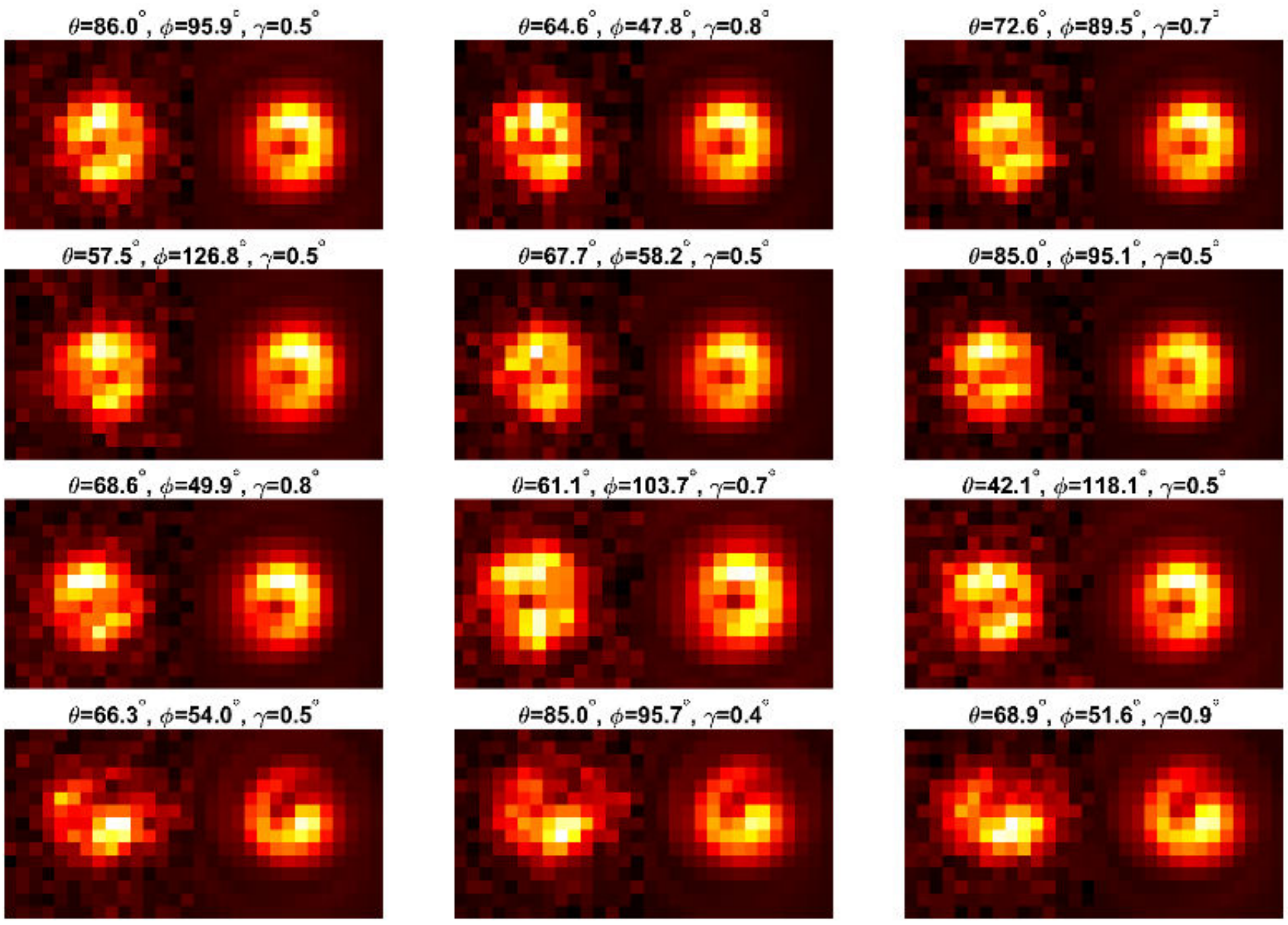

**Figure S2.** Twelve fitted DSFs are shown in the right panels of each subfigure, while their corresponding recorded DSFs for demonstration purposes are displayed in the left panels. The orientation parameters associated with each subfigure are specified above them.

**Precision of parameters with respect to numbers of detected photons:**

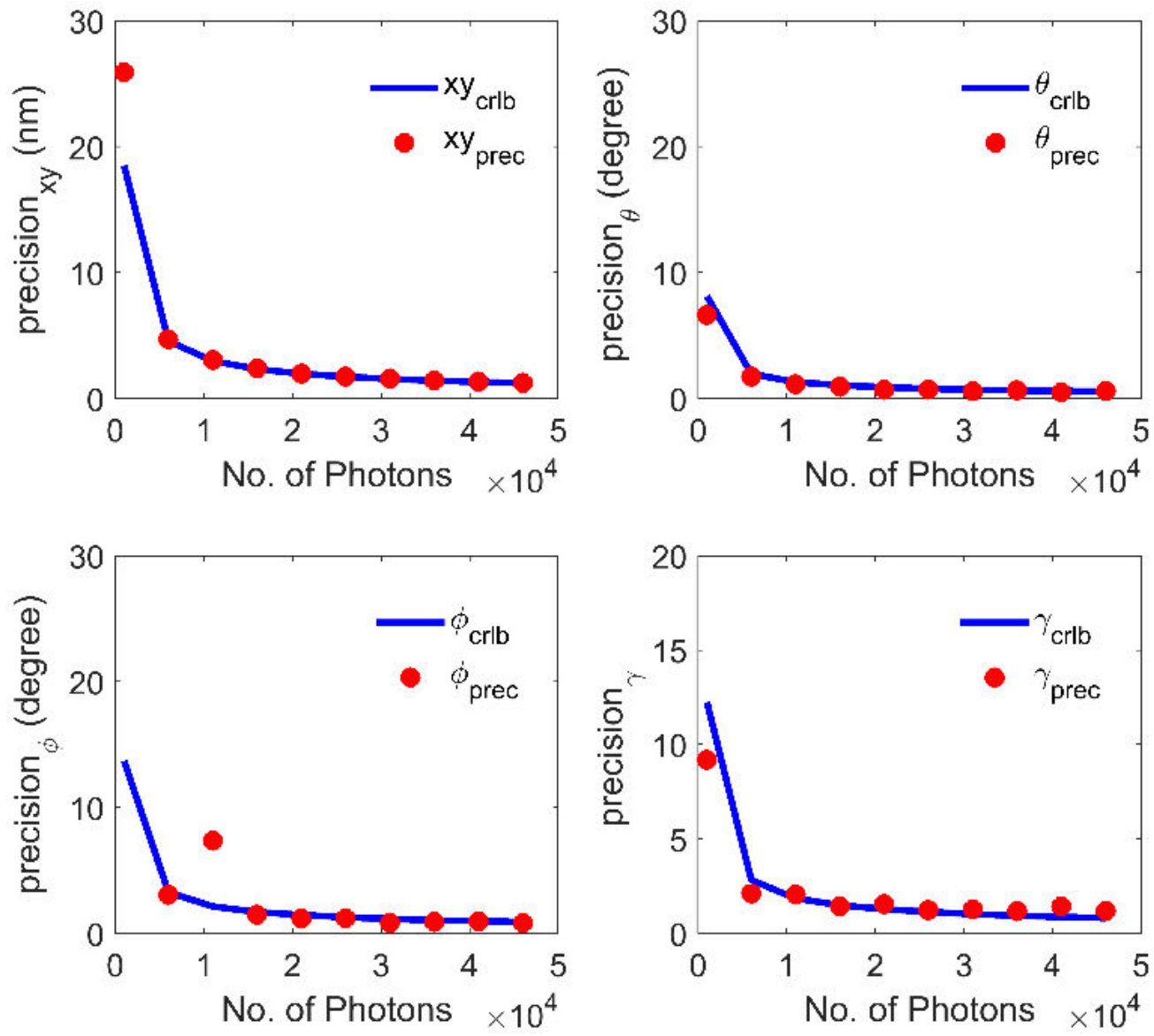

**Figure S3.** To systematically evaluate the impact of photon count on spatial and orientation determination under microscopic conditions, we performed a comprehensive simulation study. We evaluated four key parameters as a function of the molecule's orientation ($\boldsymbol{\theta}$, $\boldsymbol{\Phi}$) and constraint ($\boldsymbol{\gamma}$): (a) average lateral localization precision, (b) average axial localization precision, (c) orientation precision, and (d) wobble angle ($\boldsymbol{\gamma}$) precision. The simulation parameters were varied across the following ranges: photon count ($N$) - 1000 to 50,000 in increments of 5,000; polar angle ($\boldsymbol{\theta}$) - 0° to 180 ° in steps of 20°; azimuthal angle ($\boldsymbol{\Phi}$) - 0° to 180° in steps of 20°; wobble constraint ($\boldsymbol{\gamma}$) - 0 to 1 in increments of 0.2; replicates - 10 independent iterations per condition state.

**Spectral Imaging of Single Molecule:** To resolve the spectrum, the signal is passed through a prism, which disperses the light into its constituent wavelengths. The primary challenge in this setup is accurately quantifying the wavelength-dependent pixel shift caused by the prism's dispersive power. Knowing this exact shift allows for the precise mapping of specific wavelengths to individual pixels on the detector. Setup calibration and image analysis for spectrum reconstruction are described in Ref. [1, 2].

**Spectral Image Alignment and Coordinate Transformation:** Once the pixel shifts are quantified, a spatial transformation matrix is calculated to achieve an exact, one-to-one mapping between the single-molecule orientation channel and the spectral channel. Because each dispersed spectrum must be precisely matched to its corresponding molecule's spatial position, accurate alignment is critical. For this we are using tetra speck bead and inserting 590/10 nm bandpass filter.

**The calibration and mapping workflow consists of three main steps:**

i. **Subpixel Localization**: First, the precise coordinates of the molecules in both the orientation and spectral channels are determined using subpixel localization algorithms.

ii. **Control Point Selection**: To find matching pairs across the both the channels, candidate molecules are identified using MATLAB's "*cpselect*" function. This tool allows for the manual or semi-automated selection of corresponding control points, providing their precise coordinates.

iii. **Nearest-Neighbour Matching**: To match these selected candidates with the entire pool of subpixel-localized coordinates, the "*dsearchn*" function is employed. This function executes a nearest-neighbour search to pair each selected control point with its closest subpixel-localized counterpart.

iv. Finally, these matched coordinate pairs are used to compute the transformation matrix, enabling seamless spatial registration between the orientation and spectral data.

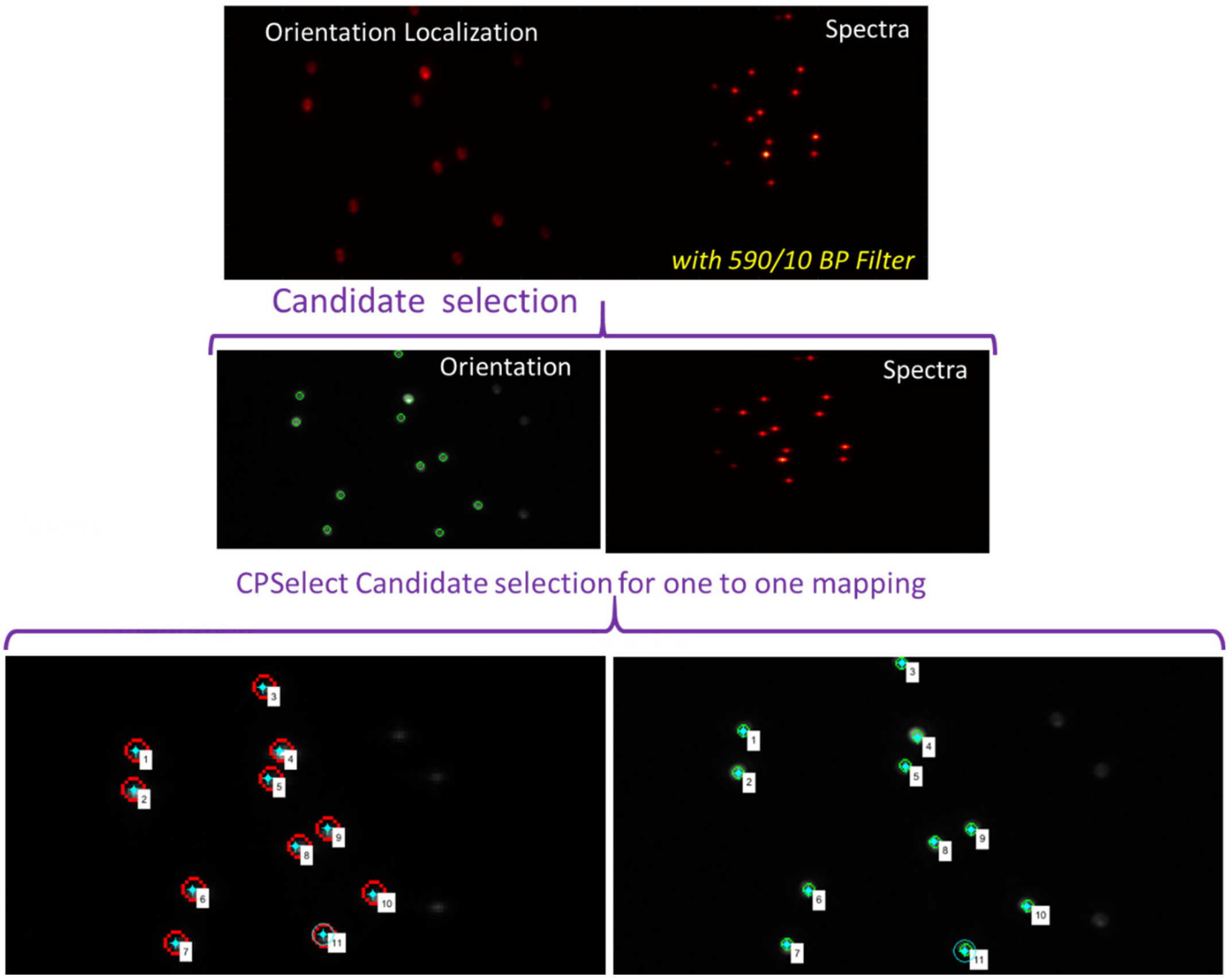

**Figure S4.** Image registration and sub-pixel alignment of TetraSpeck fiducial markers using control point selection. (Top) Representative fluorescence image of multi-spectral TetraSpeck beads used as spatial reference markers. (Middle) Sub-pixel localization of bead centre coordinates extracted to achieve high-precision spatial tracking. (Bottom) One-to-one coordinate mapping between optical channels established via fine control point selection (cpselect) to compute the affine transformation matrix for image registration.

**Reconstruction of Spectra:** Once the spatial transformation matrix and pixel-shift calibration function are established, they are applied to reconstruct individual single-molecule spectra. Because the transformation matrix was generated using a 590 nm band-pass filter as the reference, the workflow for extracting the spectrum is as follows:

i. **Reference Coordinate Mapping**: The subpixel coordinates of a molecule localized in the spatial channel are multiplied by the transformation matrix to calculate its baseline position (the 590 nm reference pixel coordinate) in the spectral channel.

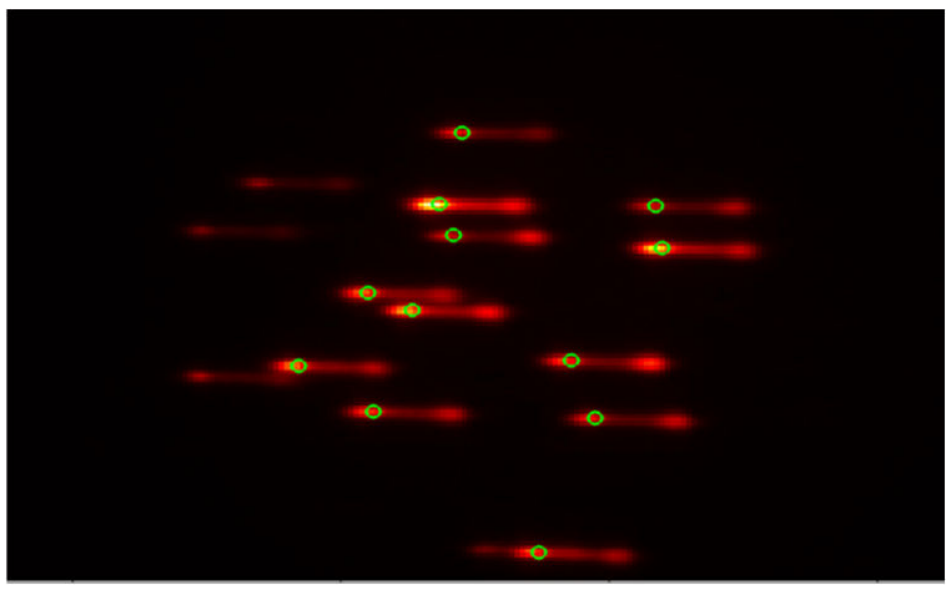

**Figure S5.** Mapped spatial coordinates of 590 nm spectral position following matrix transformation.

ii. **Spectral Profiling along the Dispersion Axis**: The intensity values are integrated along the y-axis. From the baseline 590 nm reference position, pixel shifts are tracked in both the positive and negative directions. Finally, the previously calibrated polynomial fit is applied to these pixel coordinates to calculate their corresponding wavelength values, yielding the absolute spectrum.

**Fitting of SM Vibronic Emission Spectra:** To determine the vibronic emission maxima and intensity weighted spectral centroid for further analysis we employ multi-Gaussian fitting function (with 3 or more vibronic bands) as given below

All the single molecule normalized spectra are fitted with double Gaussian

$$y = y_0 + \frac{1}{\sqrt{\pi/2}} \sum_{i=1}^{n} \frac{A_i}{w_i} e^{-\frac{2(\lambda-\lambda_i)^2}{w_i^2}} \quad \text{(eq. S2)}$$

Here $y_0$ is a constant offset, $A_i$ are the area enclosed by the Gaussian profiles centred at ($\lambda_i$, $y_{ci}$ ) and $w_i$ are the respective widths where

$$y_{c_i} = y_0 + \frac{A_i}{w_i\sqrt{\pi/2}} \quad \text{(eq. S3)}$$

The intensity-weighted centroid of the spectral intensity distribution across the camera pixels were calculated following $\lambda_c = \frac{\sum_i \lambda_i S_i}{\sum_i S_i}$, where $\lambda_i$ and $S_i$ are the wavelengths and signal of $i^{th}$ pixel.

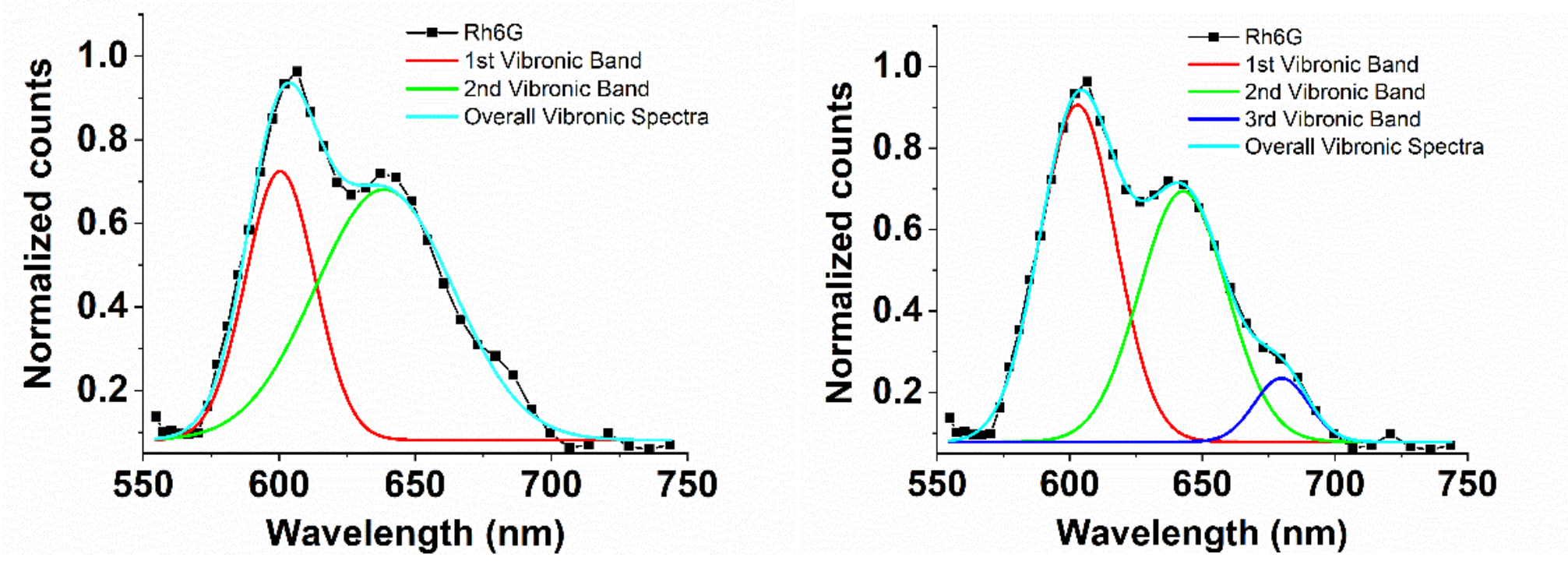

**Figure S6.** Comparison of two vs. three Gaussian fitting for Rh6G. In the present case we used 3 or more vibronic bands for SM spectral fittings.

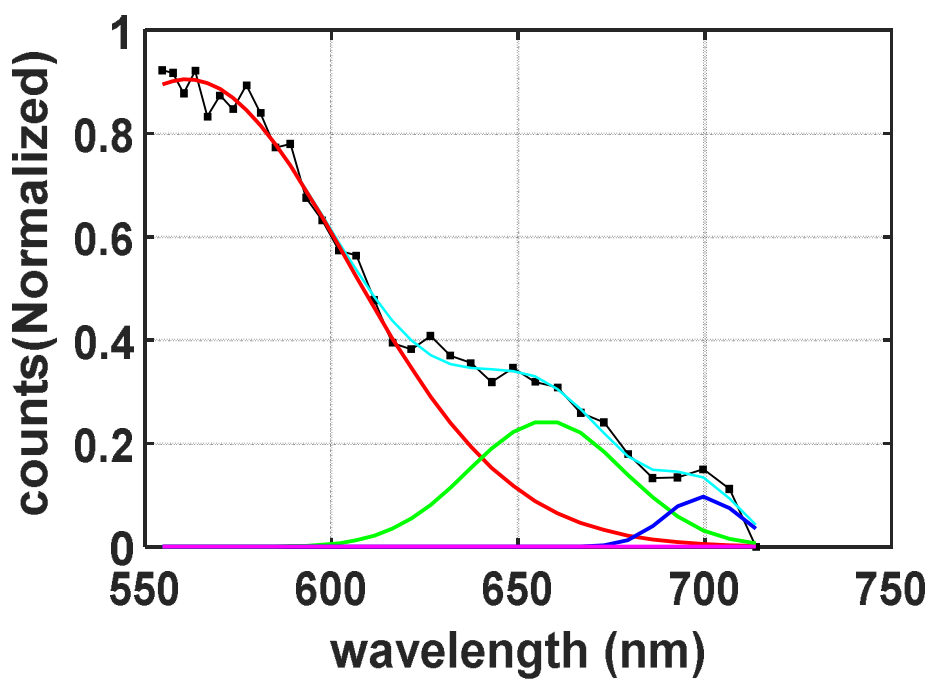

**Figure S7.** Such SM spectra were not considered for any further analysis and statistics. Reflects selective filtering of spectra for the 561 nm long pass filter.

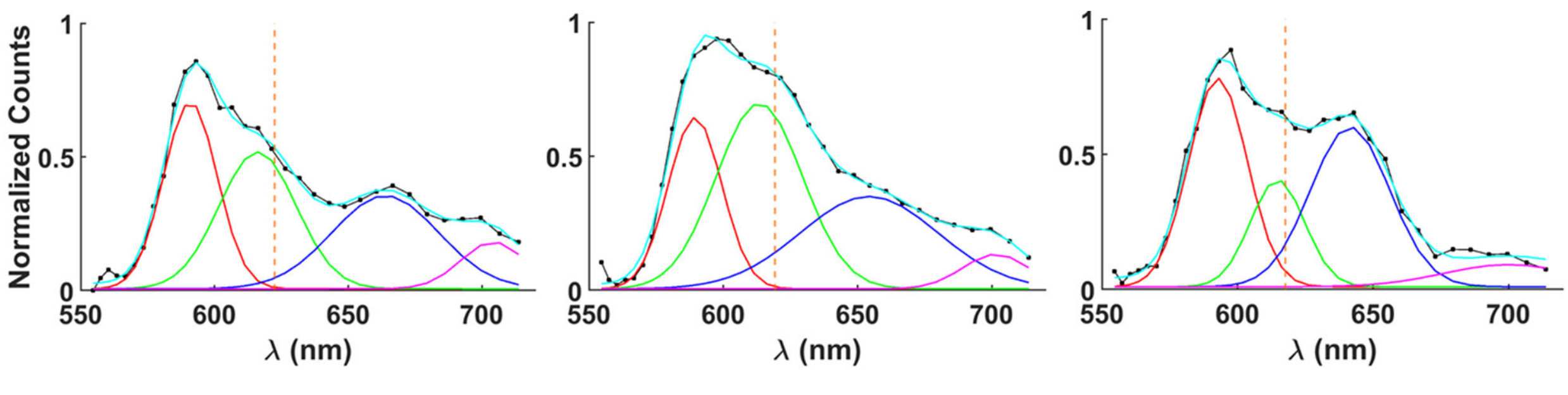

**Figure S8.** Few representative fittings of SM spectra with three or more Gaussian vibronic bands. We mostly used the intense bands for further calculations and analysis. The dashed line represents the intensity weighted spectral centroid.

**Table S1: Summary of number of analysed SMs and SM spectral snapshots under each specific polymer overlayer condition.**

| Fluorophore | Number of analyzed SM spectral snapshots (Number of SMs analyzed) | | | | | |
|---|---|---|---|---|---|---|
| | ***Thin, PMMA Coating (MW 300K)*** | ***Thick, PMMA coating (MW 300K)*** | ***Bare dye on coverslip*** | ***Thick PMMA coating (MW 100K)*** | ***Thick PVC coating (MW 200K)*** | ***Thick PS coating (150K)*** |
| Rh6G | 16123 (159) | 31353 (97) | – | 29901 (99) | 18923 (81) | 18813 (86) |
| RhB | 21538(163) | 18836 (73) | 268 (268) | 33142 (104) | 11729 (53) | 19585 (100) |
| Cy3 | 7000 (108) | 14031 (65) | 249 (249) | 74187 (248) | – | – |
| Cy3b | 15334 (218) | 28770 (113) | 311 (311) | 33471 (104) | – | – |
| MC540 | 6768 (67) | 23494 (79) | – | 29017 (104) | – | – |
| Atto 520 | 13242 (143) | 26983 (98) | – | 44450 (130) | – | – |
| Nile Red | 23156 (146) | 41647 (189) | 187 (187) | 30247 (103) | – | – |

To ensure statistical reliability across all experimental conditions, a total of tens of thousands of individual fluorescence spectra were analyzed for each fluorophore–polymer combination. However due to poor photo-stability of bare fluorophores (**Figure S9**), we could record only few hundred spectral snapshots cases where the probe shows emission after discarding initial 10 frames.

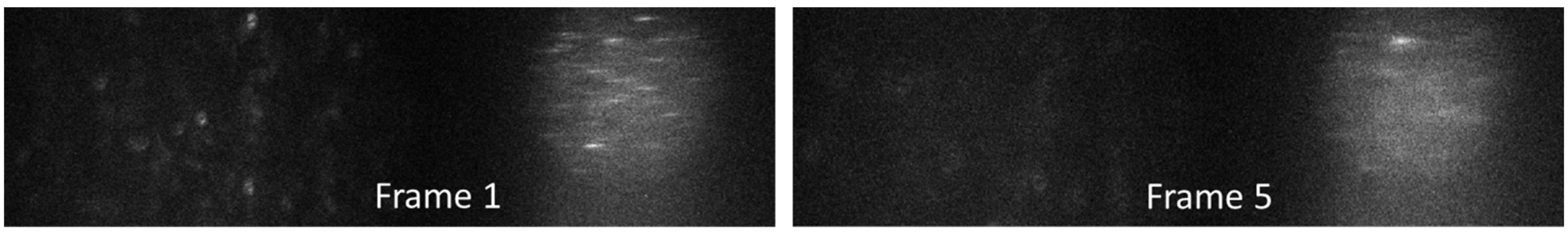

**Figure S9.** Fast photobleaching of dyes in bare fluorophore condition.

SM studies reveal that even nominally homogeneous polymer films contain heterogeneous environments, causing variations in emission wavelength and spectral dynamics.[4, 8] SM studies show that local probe–polymer interactions and slow polymer rearrangements cause spectral fluctuations.[20] Studies with Nile Red (NR) in PMMA show that local dielectric environments cause inhomogeneous fluorescence broadening and excitation dependent spectral shifts.[23, 24]

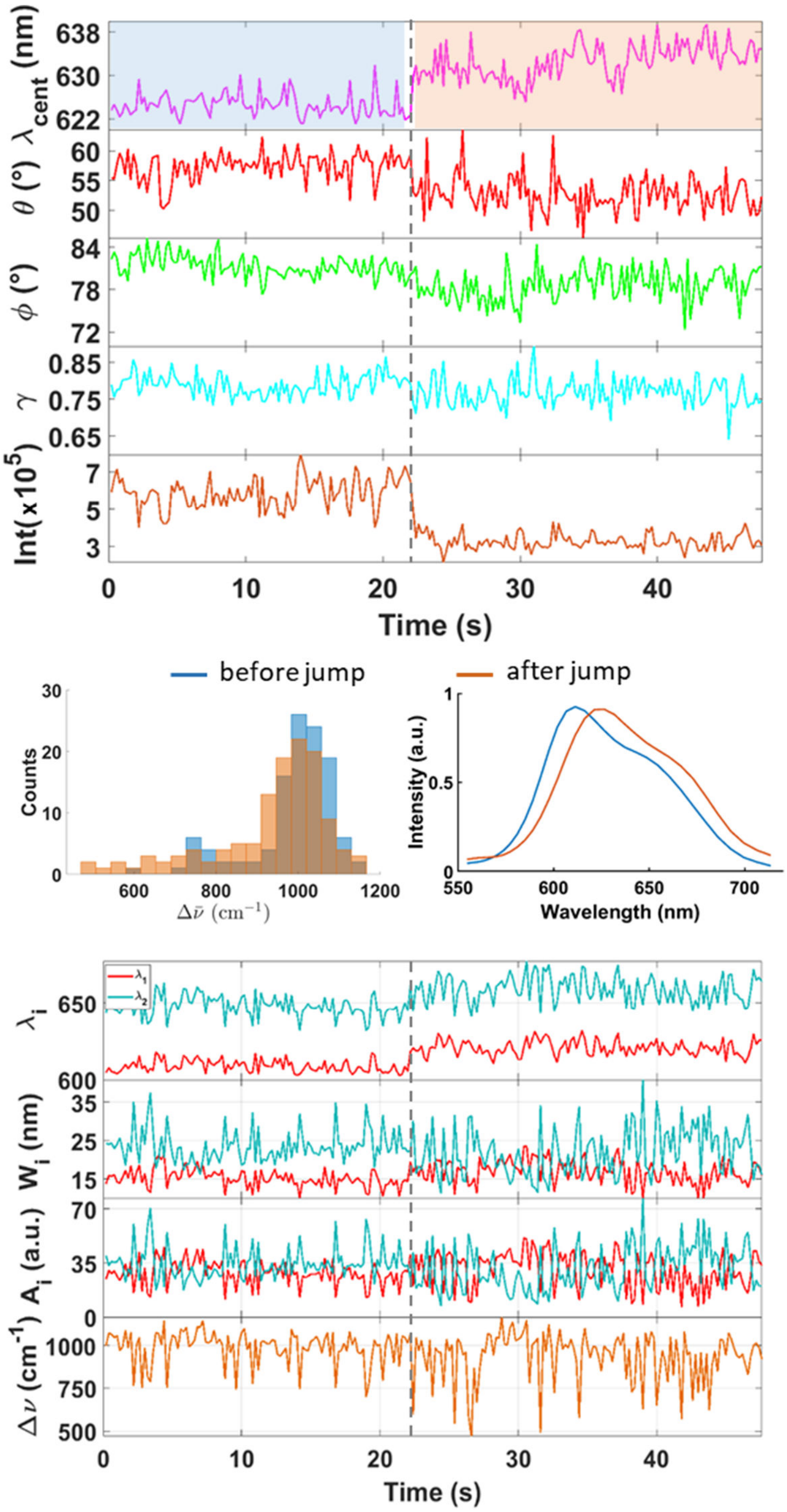

**Figure S10.** A close look at a SM trajectory of Cy3b with thick PMMA (MW 350K) overlayer (same as **Figure 1i**). It clearly depicts SM spectral ($\Delta\lambda \sim 8$ nm) and orientation jump ($\Delta\theta \sim 5°$) after 22 second with only subtle change is rotational constraint ($\gamma$) unlike drastic alterations observed other SM traces shown in **Figure 1j,k**. Interestingly the average SM spectra before and after jump indicates spectral shift without much change is spectral shape or relative vibronic peak intensities, i.e. change in energy separation between ground and excited potential wells without appreciable change in its hypersurfaces. The distribution of energy spacing between main two vibronic bands are marginally shifted to lower energies after the jump. A closer look at the two prominent vibronic emission band features obtained from Gaussian fitting, i.e. maxima ($\lambda_i$), width ($w_i$) and area ($A_i$) clearly highlights the photophysical dynamics at the local environment. Interestingly, the vibronic progression after the jump shows increased wondering to low energy modes.

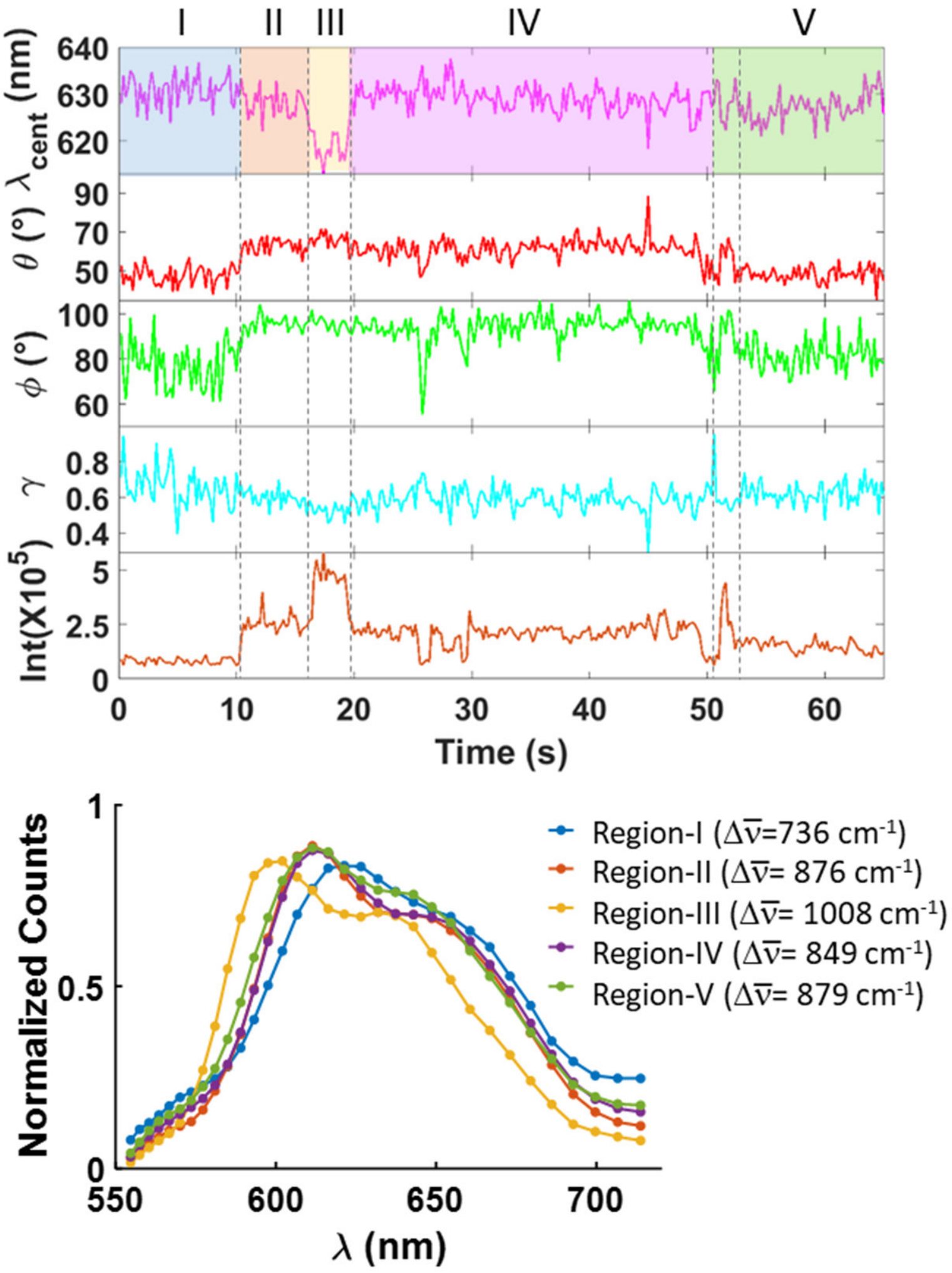

**Figure S11.** A SM trajectory of MC540 with thick PMMA (MW 350K) overlayer. It clearly depicts multiple spectral (I→II→III→IV) and orientation jumps (I→II, IV→V) without any distinct variations in rotational constraint ($\gamma$). The average SM emission spectra for the transitions II→III & III→IV shows substantial change in dominant vibronic modes and spectral centroid but without any significant TDM reorientation. This indicates change in ground state potential hypersurface along with variation in energetic distance between the ground and excited state potential wells which alters the energetic separation of the vibronic progression. While during I→II state transition there is large scale TDM reorientation with dominant spectral shift but nominal change in vibronic spacings – signifies only variation in energetic distance between the ground and excited state potential wells. Interestingly, this molecule shows different types of correlative transitions.

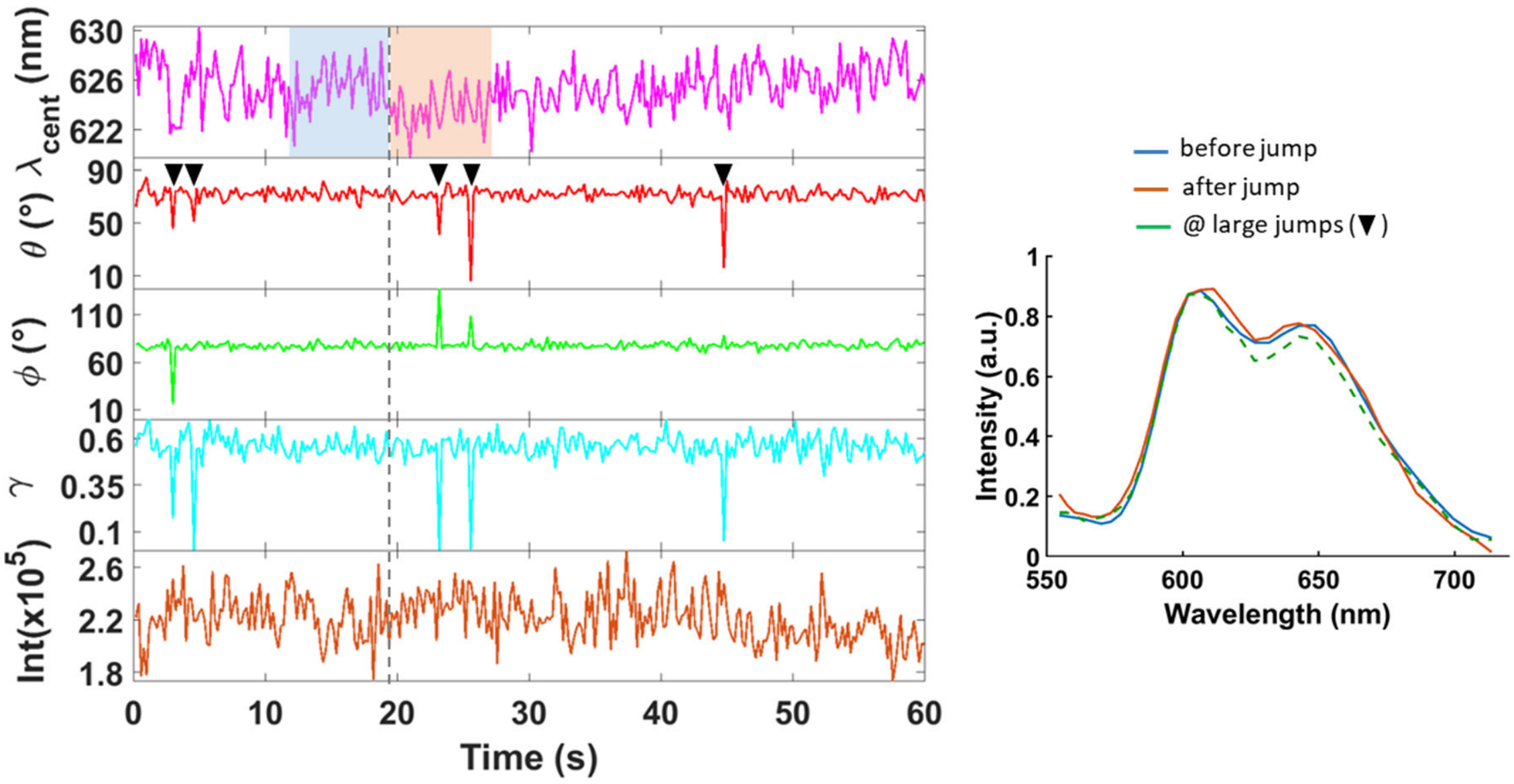

**Figure S12.** Another trajectory for MC540 with thick PMMA (MW 350K) overlayer. The black triangles highlight the occasional single frame excursions with large scale TDM reorientations along with large rotational mobility. Comparison of average spectra for these regions possibly indicate variations of the intensity ratios within the vibronic progression, i.e. variations of Franck-Condon (FC) factors instigated by relative isoenergetic displacement of the ground and excited state potential wells.

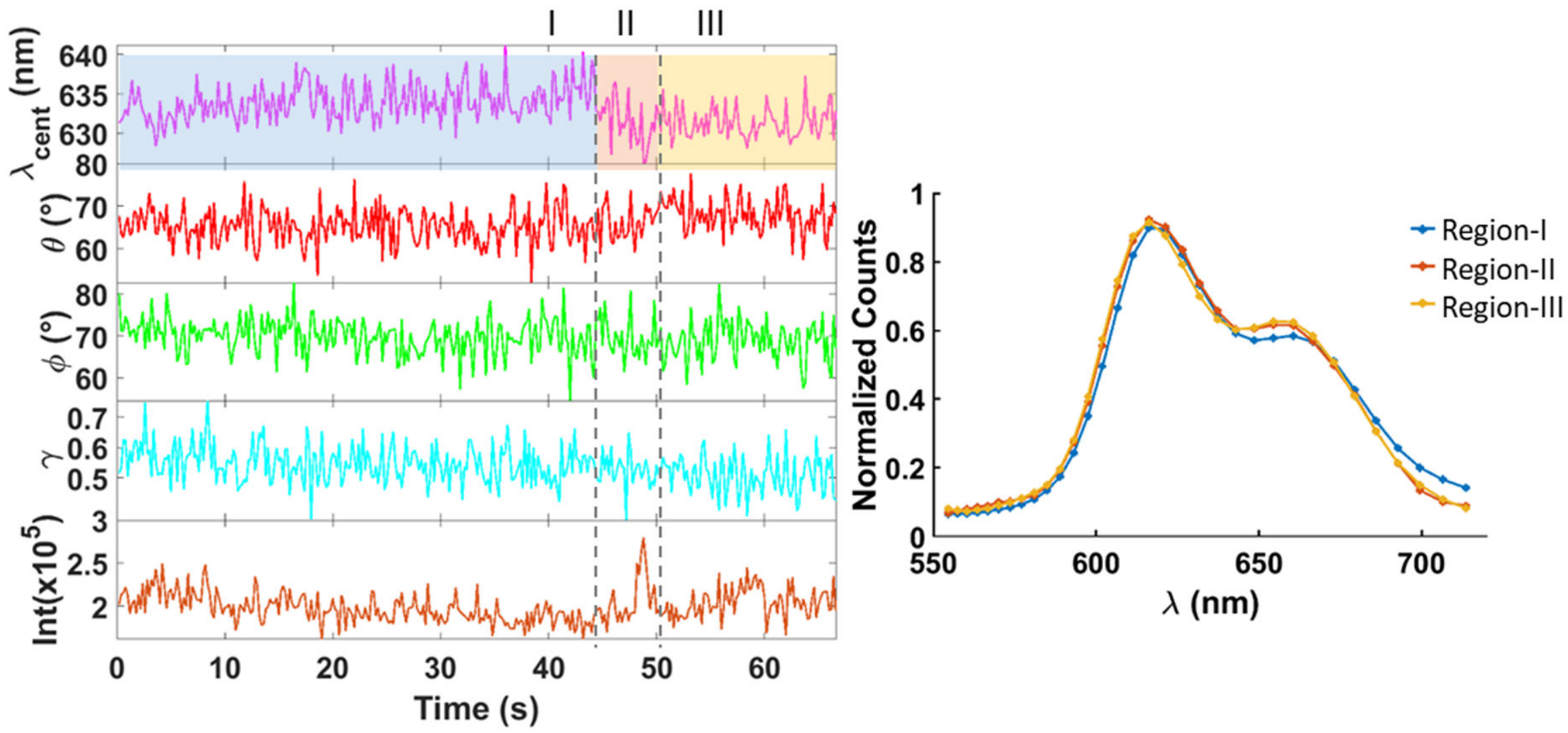

**Figure S13.** Representative trajectory for Cy3 with thick PMMA (MW 350K) overlayer. There are subtle changes in polar orientation in the II→III state transition without much change in average SM spectra or vibronic energy progression. Whereas I→II state transition shows spectral shift and slight alteration in vibronic energy spacing, but without any TDM reorientation.

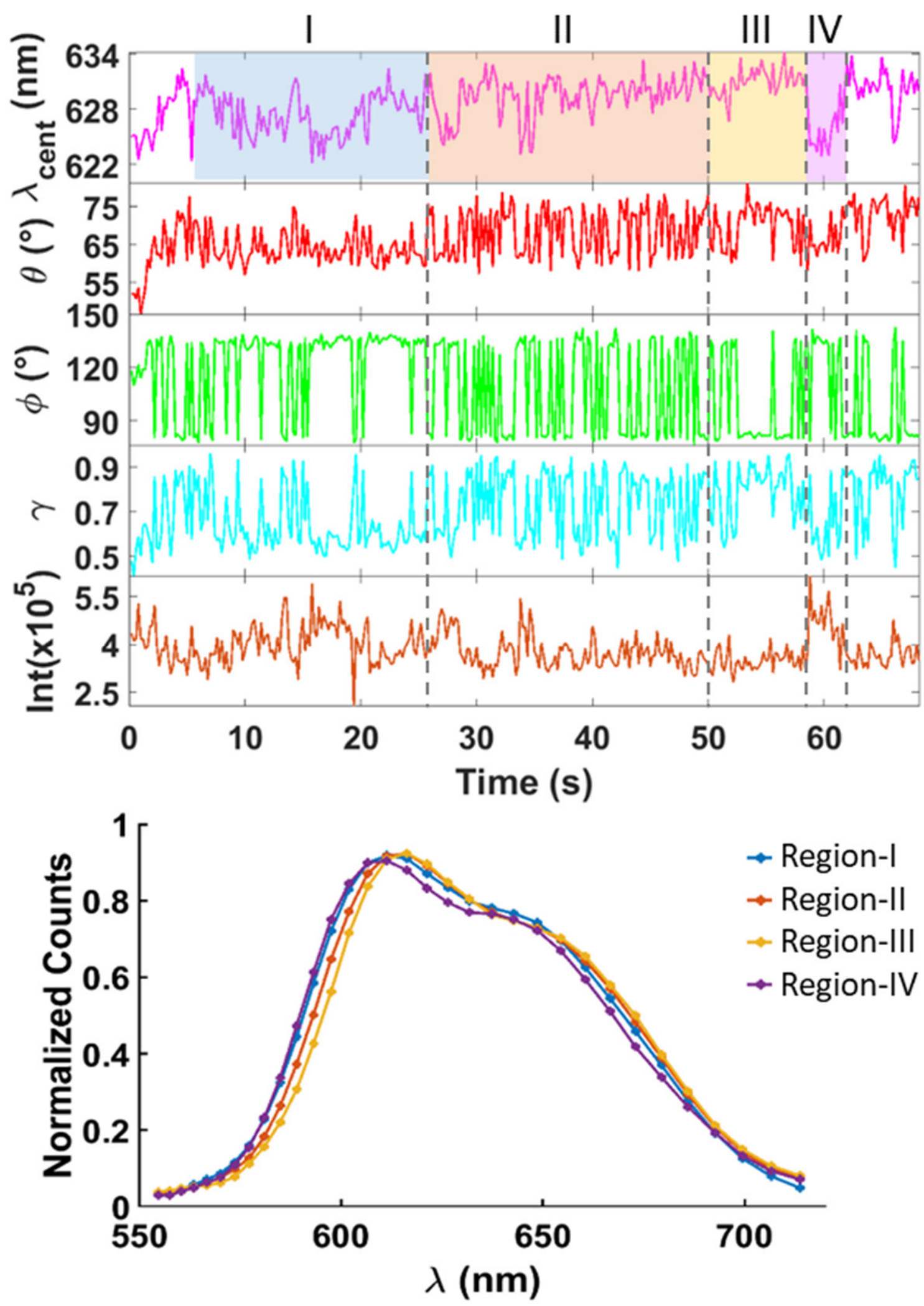

**Figure S14.** Representative trajectory for NR with thick PMMA (MW 350K) overlayer shows repeated large scale TDM reorientation along with spectral fluctuation. Such molecular response perhaps is linked to the behaviour of dye-polymer coupled systems where sensitivity of the probe in its local environment play crucial role in determining the momentary TDM. In region II, the molecule exhibits increased wondering of TDM. The SM average spectra for all these regions indicates variations of FC factors made by isoenergetic displacement of the ground state potential wells relative to that of excited state.

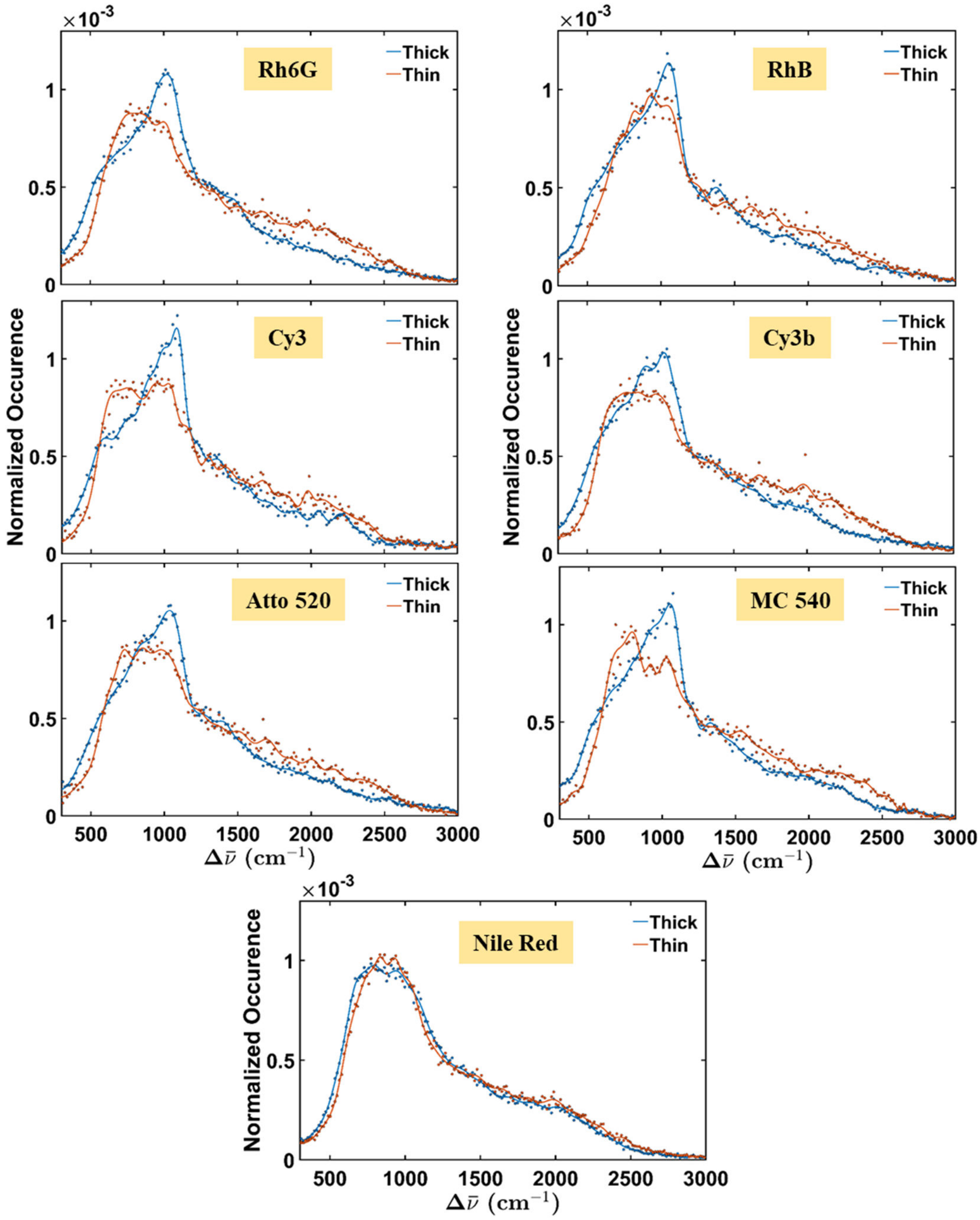

**Figure S15.** Distribution of vibronic band separations for thick and thin PMMA (MW 350K) overlayers. The thick-film spectrum shows a somewhat stronger feature near ~1050–1100 cm⁻¹, whereas the thin-film distribution is relatively broader around ~900–1100 cm⁻¹. The persistence of the dominant 700–1200 cm⁻¹ spacing range across PMMA thicknesses suggests that the principal vibrational energy scales contributing to the reconstructed emission bands remain conserved. TDDFT calculations are therefore used to identify candidate vibrational modes within or near these energy ranges and to examine their potential contributions to transition dipole modulation and vibronic emission. The thickness-dependent changes in feature intensity are interpreted as changes in vibronic weighting that may reflect both molecular coupling and the optical response of the overlayer.

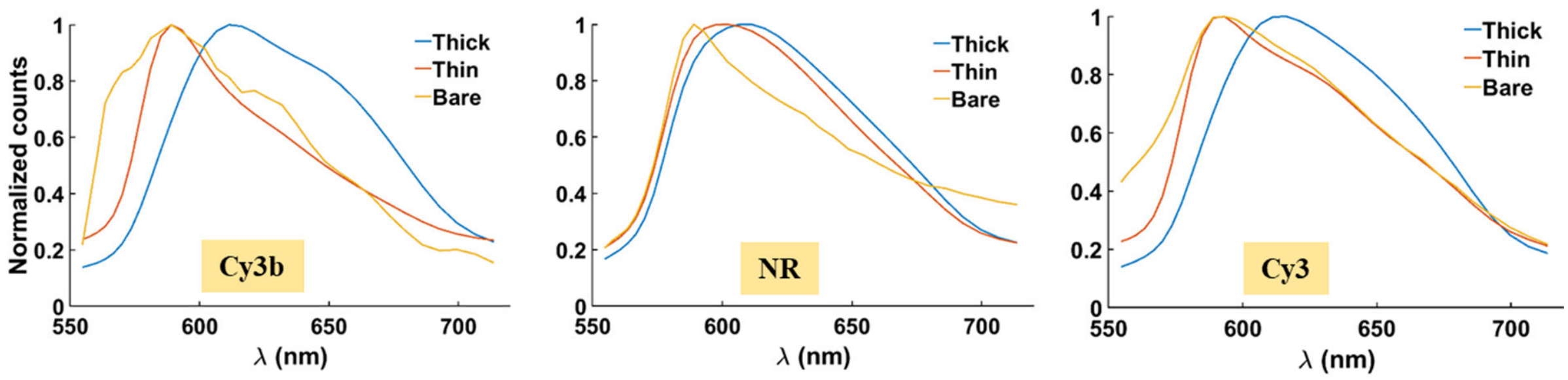

**Figure S16.** Comparison of average SM spectra with thick & thin coating and with bare condition (c) highlights the interplay between local molecular photophysics and optical coating effects.

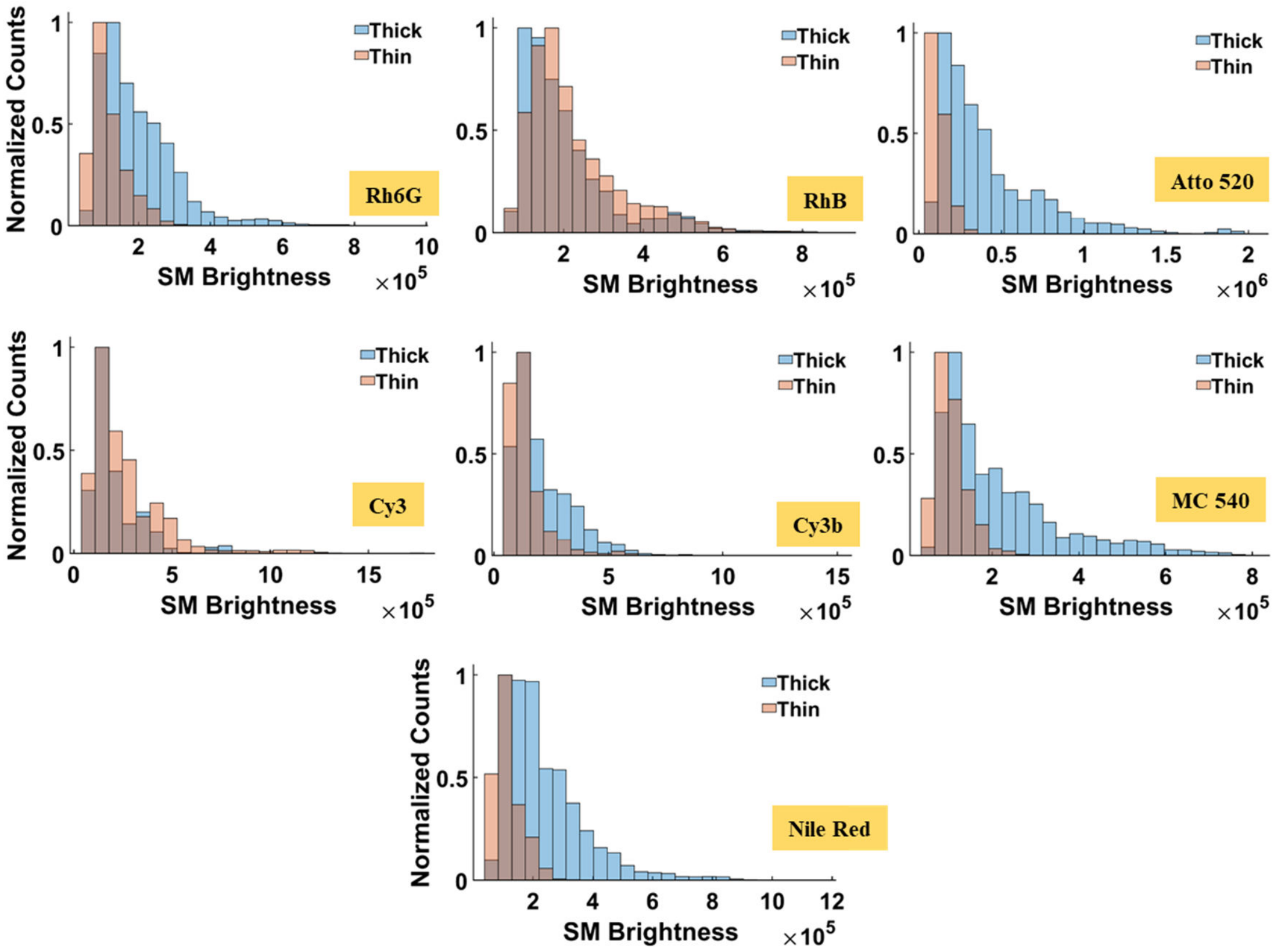

**Figure S17.** Distribution of SM brightness with PMMA (MW 350K) overlayer indicates increase in brightness with coating thickness.

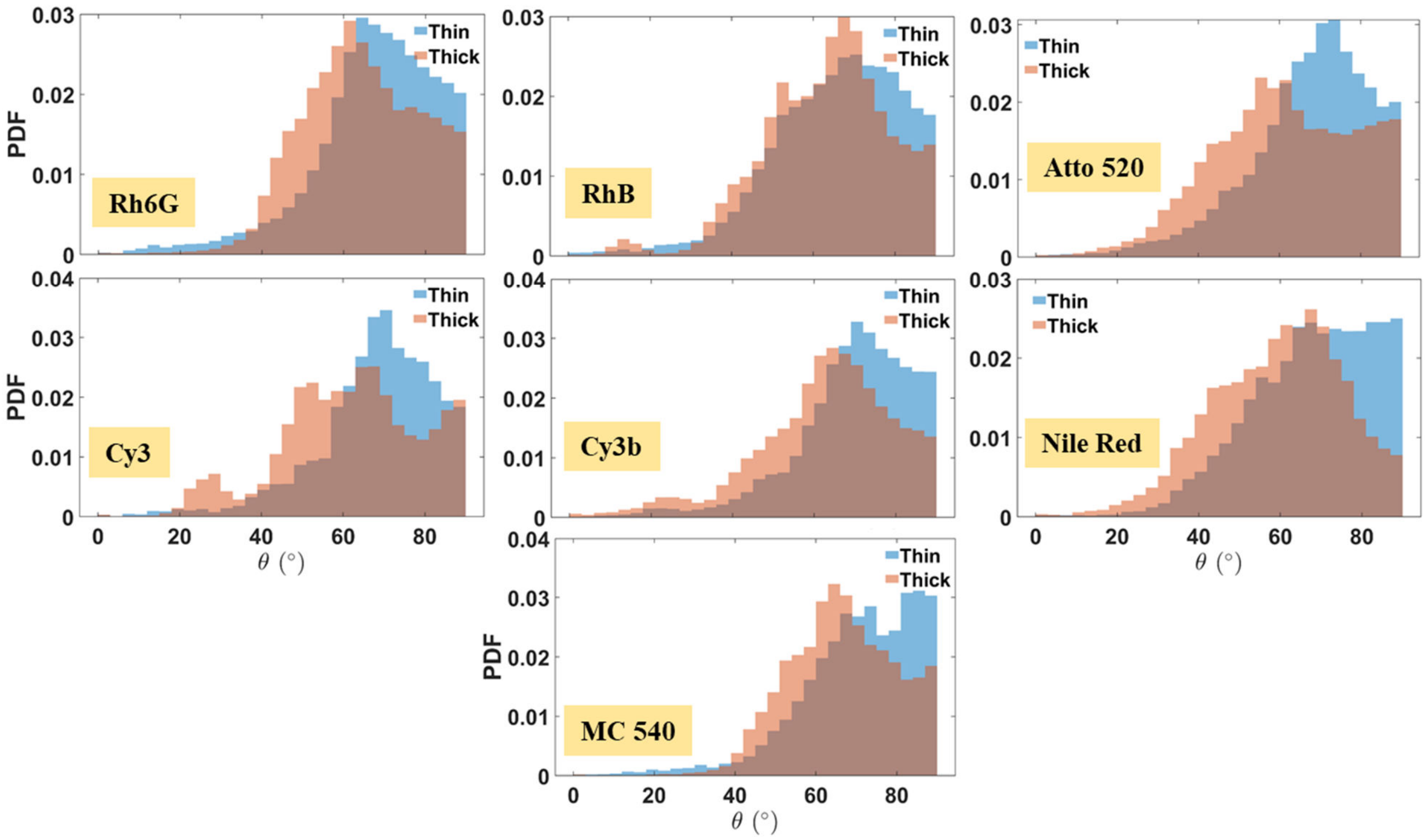

**Figure S18.** Comparison of polar angle distribution in thick and thin overlayers of PMMA (MW 350K).

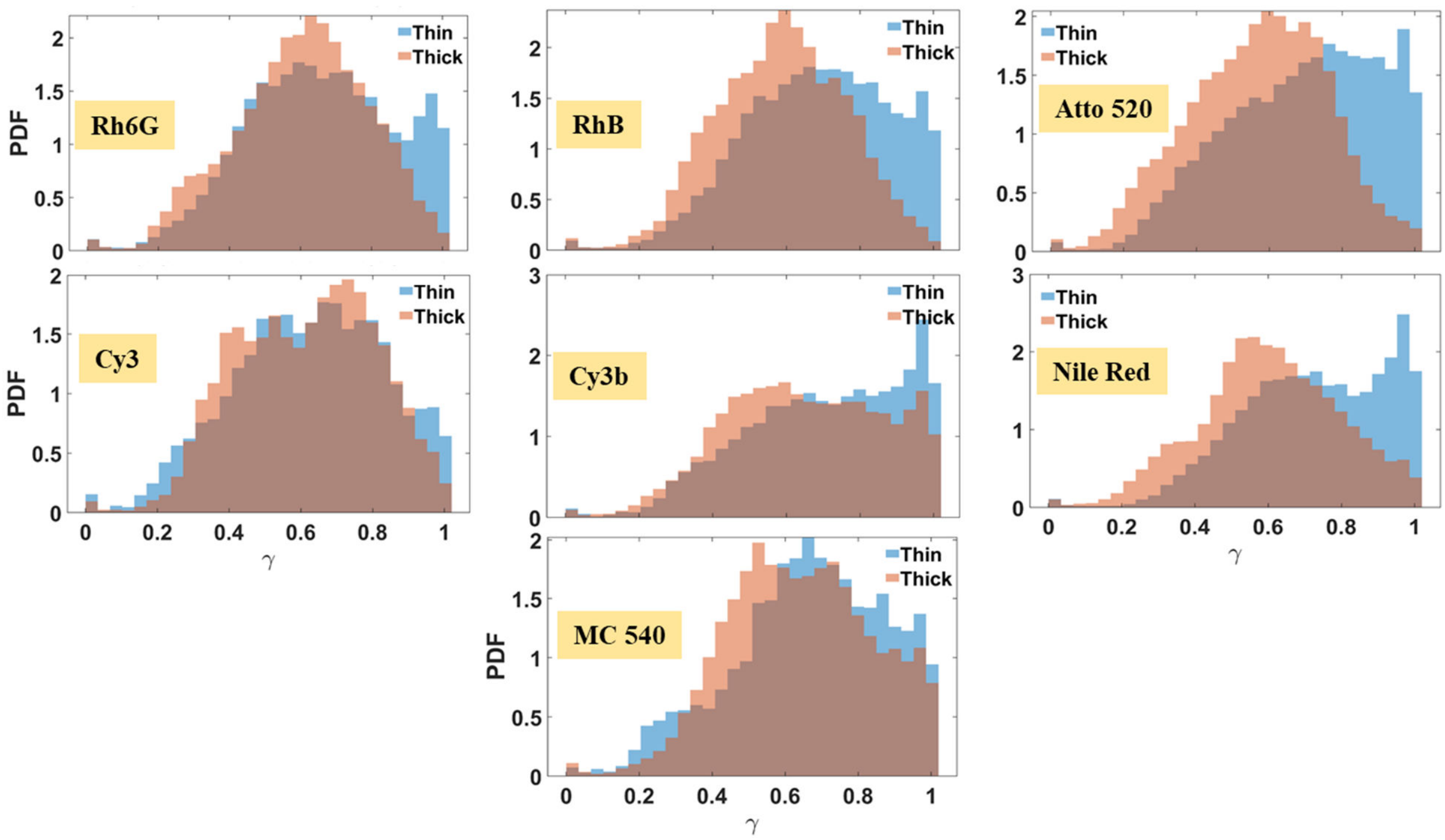

**Figure S19.** Comparison of γ distribution in thick and thin overlayers of PMMA (MW 350K).

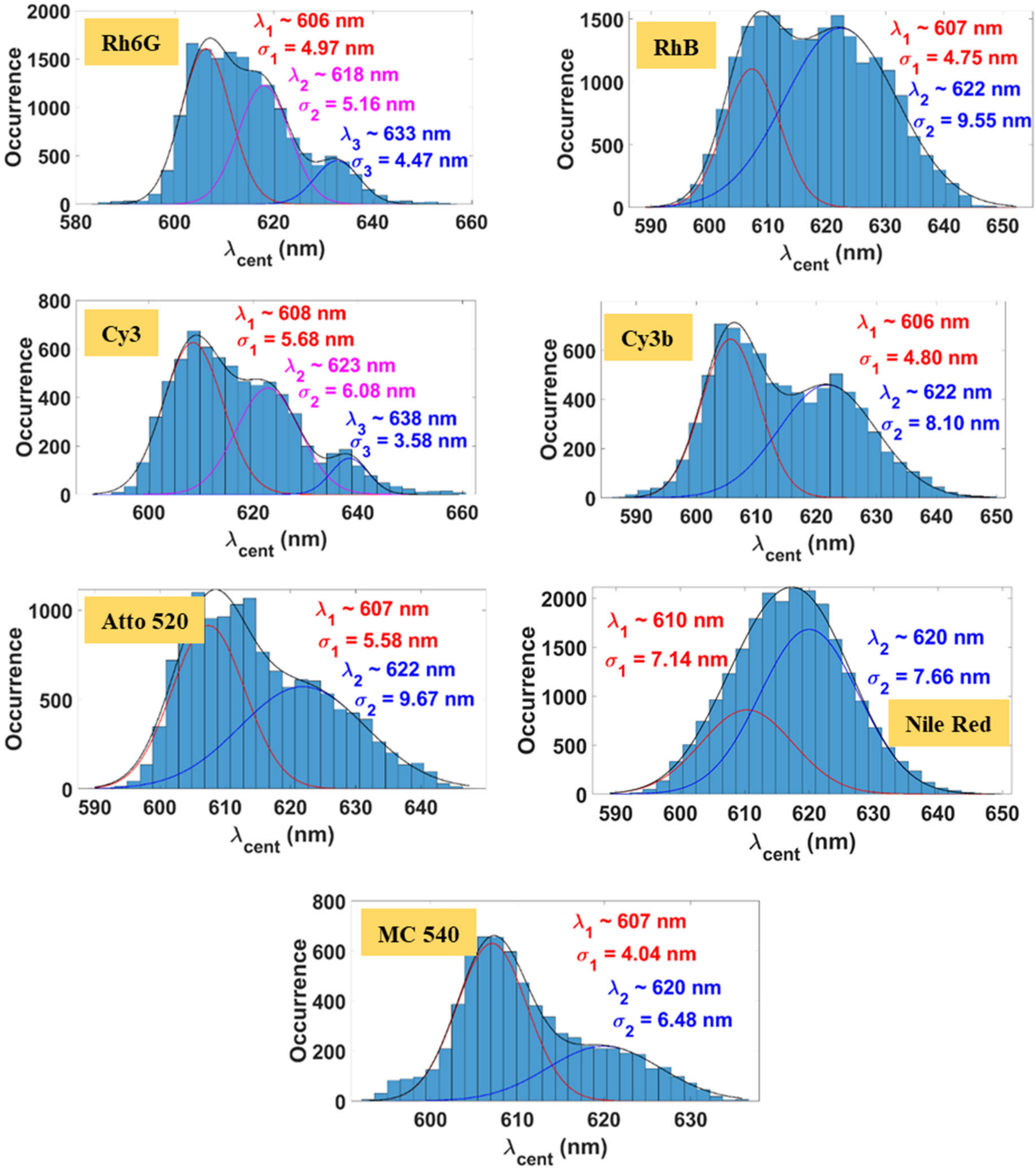

**Figure S20.** Distribution of $\lambda_c$ for all fluorophores with thin PMMA (MW 350K) overlayer. The multi-Gaussian fitting with parameters are also shown. Interestingly, we see distribution maxima at ~606 nm, ~620 nm & ~635 nm in all the dyes.

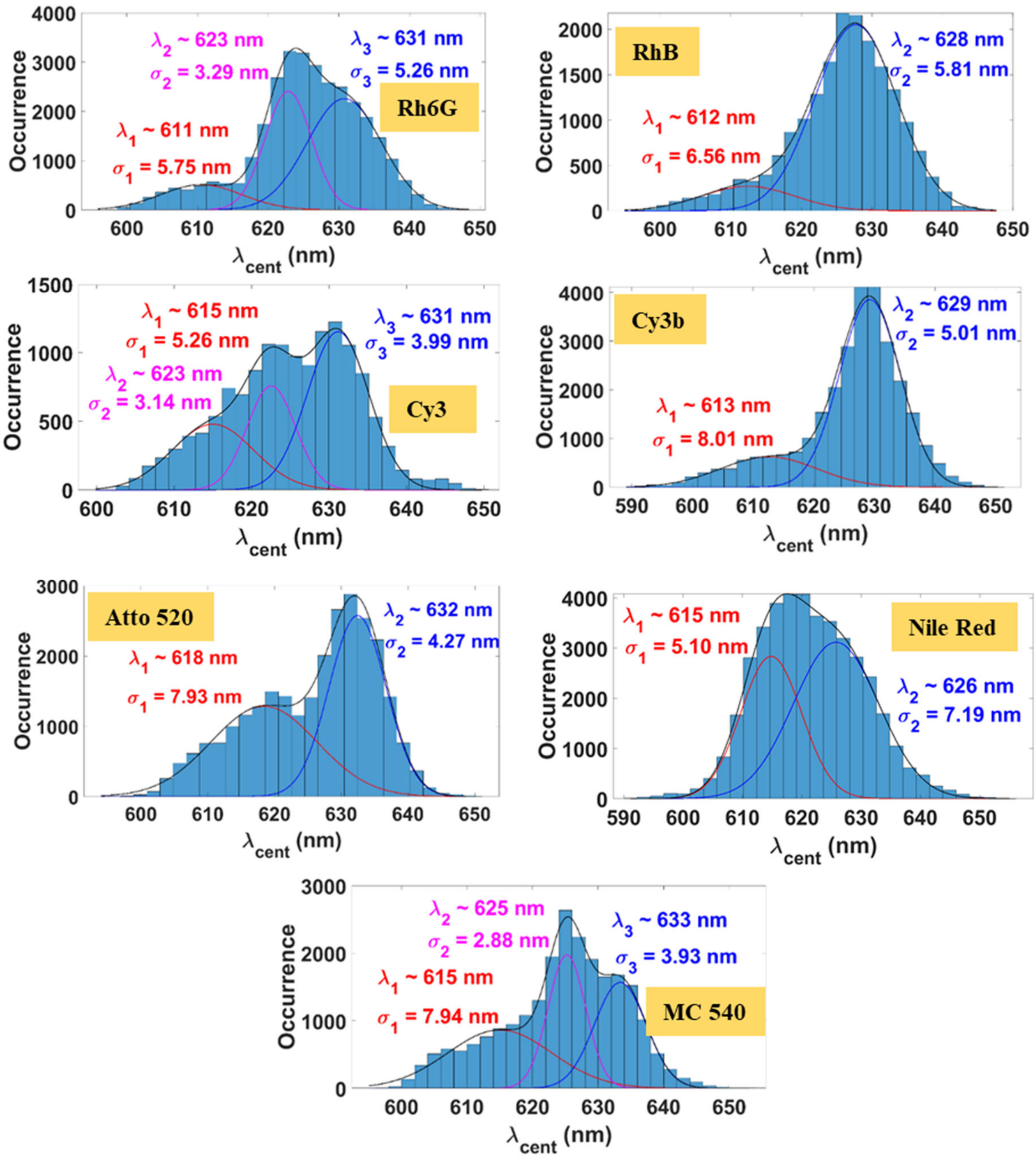

**Figure S21.** Distribution of $\lambda_c$ for all fluorophores with thick PMMA (MW 350K) overlayer. The multi-Gaussian fitting with parameters are also shown. In this case also, we see distribution maxima at ~615 nm, ~625 nm & ~635 nm for all the dyes. These values are slightly shifted towards higher wavelength than with thin overlayer case.

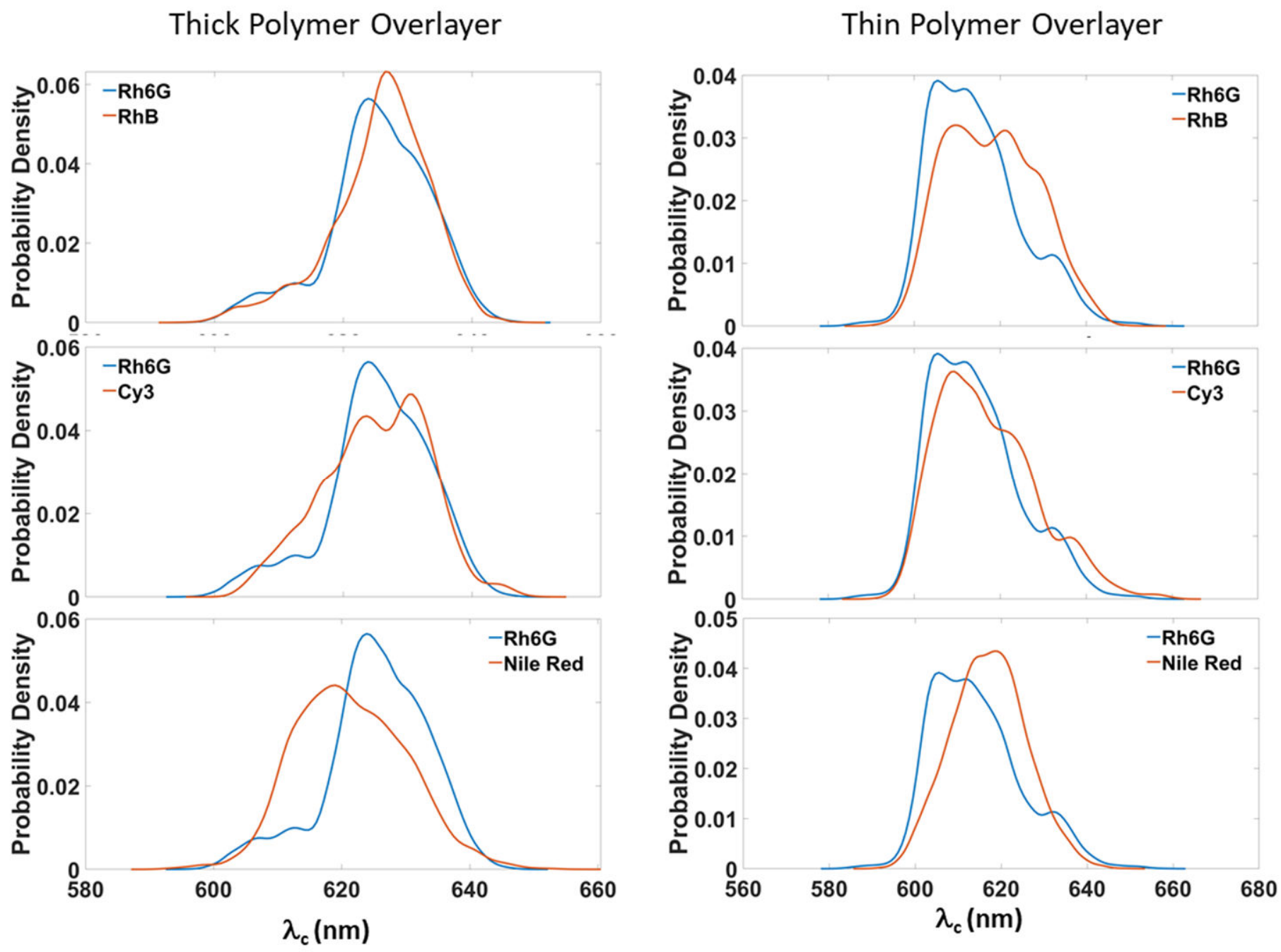

**Figure S22.** Two sample KS density comparison plots for different dye pairs with thick (left column) and thin (right column) overlayer configurations. The calculated test statistics (D) and p-values ($D < 0.05$, $p \ll 0.0001$) confirm that the spectral distributions of dye pairs are statistically distinct and independent in both regimes. One spectrum from each observed single molecule for a particular fluorophore, instead of multiple spectra from a SM spectral trajectory, were selected for this analysis.

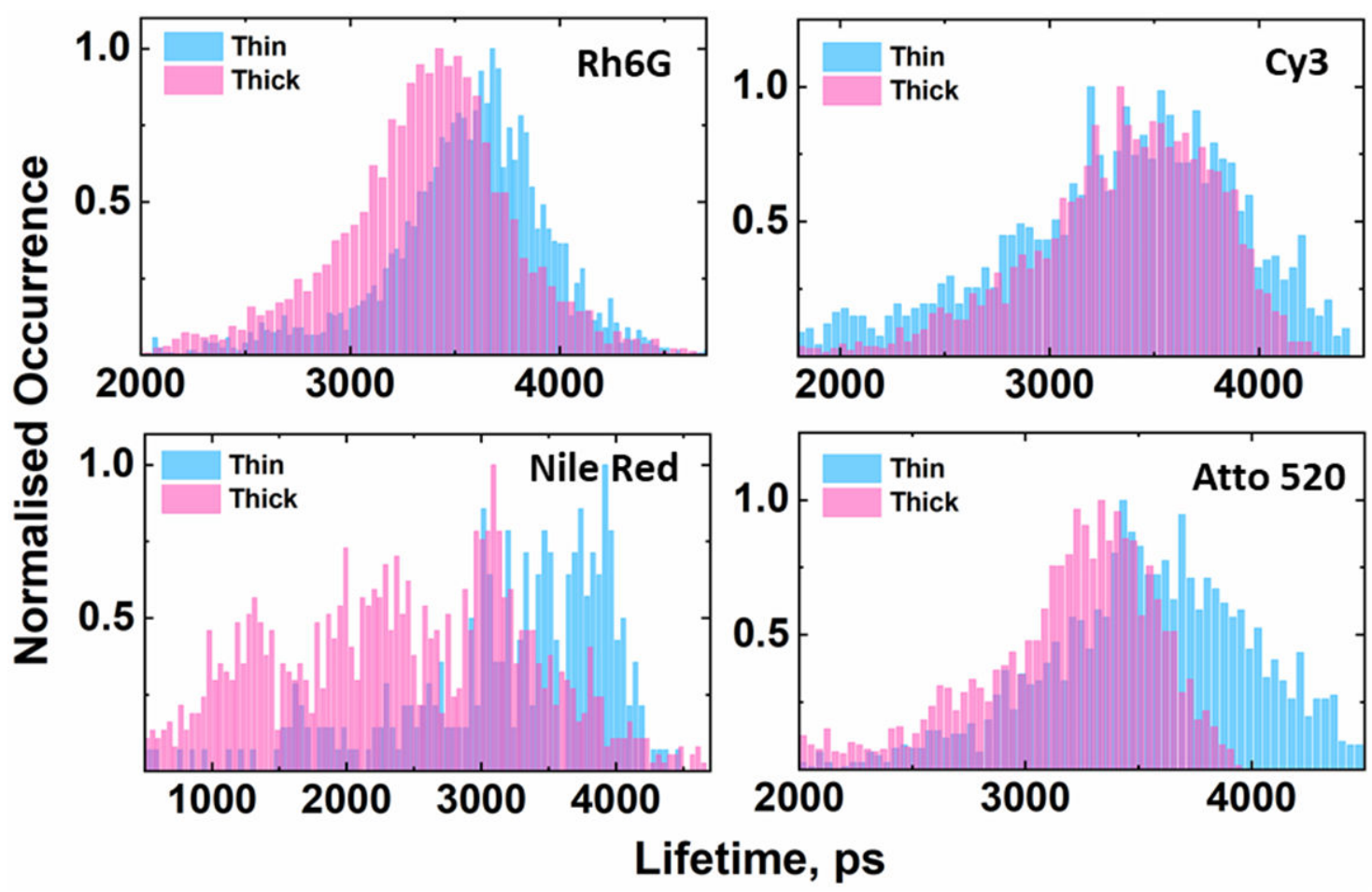

**Figure S23.** SM fluorescence lifetime distribution for thin and thick PMMA overlayers, measured with confocal fluorescence lifetime imaging microscopy (described earlier[25]). There is visible shift to lower values or narrowing of distribution with increase in thickness as expected from the role of electromagnetic boundary conditions in single molecule fluorescence lifetime studies of dyes embedded in thin films.[26]

## Section S4

## Cavity Induced Modification of Vibronic Emission via the Purcell Effect

When a fluorophore is embedded within an asymmetric thin film dielectric microcavity (such as a spin-coated PMMA matrix bound by glass and air boundaries), its photophysical properties diverge from standard free solution values. This alteration is driven by the Purcell effect,[27-29] which describes the modification of a quantum emitter's spontaneous emission rate and spectral profile as a direct consequence of altering the local density of optical states (LDOS).

### I. Spectral Re-shaping and the Cavity Pulling Effect

In free space or homogeneous solution, a fluorophore's emission spectrum is governed entirely by its intrinsic molecular coordinates, defined by the thermally resolved FC factors:

$$\left|\langle \nu_{upper} | \nu_{lower} \rangle\right|^2 = e^{-S} \frac{S^{|\nu_{upper}-\nu_{lower}|}}{|\nu_{upper} - \nu_{lower}|!} \qquad \text{(eq. S4)}$$

where S is the Huang-Rhys parameter describing electron-phonon coupling to a dominant intramolecular mode (e.g., 1350 $\text{cm}^{-1}$ for Rhodamine 6G/Atto 520,[30-32] 1205 $\text{cm}^{-1}$ for Cy3[33-36] and 1280 $\text{cm}^{-1}$ for Nile Red[37, 38]).[39] Once confined within a cavity of thickness $d$ and Finesse $\mathcal{F}$, the observed emission spectrum ($I_{cav}(\lambda)$) becomes the product of this intrinsic vibronic envelope ($I_{\text{int}}(\lambda)$) and the cavity's classical Fabry-Pérot transmission profile $H(\lambda)$:

$$I_{\text{cav}}(\lambda) = I_{\text{int}}(\lambda) \cdot H(\lambda) = I_{\text{int}}(\lambda) \cdot \left[1 + \mathcal{F} \sin^2\left(\frac{\pi \cdot 2n_{\text{eff}}d}{\lambda}\right)\right]^{-1} \qquad \text{(eq. S5)}$$

This boundary condition acts as an aggressive static filter. If a vibronic sideband (such as the $(0 \rightarrow 1)$ or $(0 \rightarrow 2)$ vibrational transitions) aligns with a cavity transmission node, it undergoes resonant enhancement. Conversely, off-resonance transitions are heavily suppressed. This geometric gating forces a cavity pulling effect, where the observed fluorescence peak shifts away from the literature zero-phonon line ($E_{00}$), depending purely on the film thickness $d$.

### II. Alteration of Intrinsic Linewidths (Γ) and Dephasing

The linewidth (Γ) of each vibronic sub-band, modelled as a temperature dependent Gaussian profile, $\Gamma(\text{T}) = \Gamma_0\sqrt{\text{T}/300}$, is dictated by radiative decay and non-radiative pure dephasing ($T_2^*$) timescales ($\Gamma \propto 1/T_2 = 1/(2T_1) + 1/T_2^*$), where $T_1$ and $T_2^*$ represents radiative lifetime and dephasing time.

By modifying the electromagnetic vacuum fluctuations at the position of the dipole, the microcavity alters the radiative lifetime ($T_1$). According to the Purcell factor $F_P$:

$$F_P = \frac{\Gamma_{\text{cav}}}{\Gamma_0} = \frac{3}{4\pi^2}\left(\frac{\lambda}{n}\right)^3 \frac{Q}{V_{\text{mode}}} \qquad \text{(eq. S6)}$$

where $Q$ is the quality factor and $V_{\text{mode}}$ is the mode volume. In weak coupling thin film configurations ($\mathcal{F} = 12$), $F_P$ typically ranges from slightly below unity (suppression) to modest enhancement ($\sim$ 1.5 to 3)

I. In zones where emission is enhanced, the radiative lifetime decreases ($T_1$ shortens), broadening the homogeneous lifetime limit of the line shape.

II. More visibly, the inhomogeneous broadening ($\Delta n = 0.042$) across the PMMA film convolutes with the Purcell-modulated LDOS, causing position dependent lifetime variations that break down simple, symmetric single molecule emission profiles.

**III. Impact on Polarized Dipole Vector Components**

The Richards-Wolf high NA vectorial integration engine relies on the collection of emitted fields. Free-space molecules project a uniform dipole field symmetry. Inside the cavity, the Purcell enhancement is highly anisotropic; it acts preferentially on dipole orientations ($\theta$) that couple directly into the supported waveguide or radiation modes of the PMMA thin film. Transverse dipoles ($\theta = 90°$) display altered coupling efficiencies compared to vertical dipoles ($\theta = 0°$). This spatial anisotropy modifies the weight of the horizontal ($E_x, E_y$) and longitudinal ($E_z$) fields projected through the phase-vortex mask, distorting the engineered donut PSF footprint based on the molecule's exact vertical depth ($z_{\text{pos}}$) inside the light sheet.

**IV. Points considered for simulation**

The intrinsic emission modelling accounts for the FC principle governing vibronic transitions (e.g. electronic state coupled to the prominent 1350 $cm^{-1}$ aromatic ring breathing mode of R6G):

**Thermal Weighting**: Uses a Boltzmann distribution to calculate the population of initial vibrational ground state sub-levels at temperature T (i.e. 300 K).

**FC Factor**: Evaluates transition probabilities using a Huang-Rhys parameter (S):

$$FC = e^{-S}\frac{S^{|\nu_{upper}-\nu_{lower}|}}{|\nu_{upper}-\nu_{lower}|!} \qquad \text{(eq. S7)}$$

**Line Shape Expansion**: Uses a temperature dependent Gaussian broadening function ($\Gamma \propto \sqrt{T}$) to replicate thermal broadening.

**Microcavity Boundary Modulation**

The simulation evaluates how a tiny optical microcavity alters emission intensity based on the Fabry-Pérot thin-film interference relation. It restricts or promotes wavelengths according to the cavity length and Finesse.

**PSF Engineering**

The simulation implements the electromagnetic diffraction integrals (Richards-Wolf theory) for high-numerical aperture lenses modified by a vortex phase mask.

## Section S5

**Principal Component Analysis (PCA):** Single-molecule fluorescence trajectory analysis reveals dynamic heterogeneity, conformational fluctuations, and discrete state transitions that are typically hidden in ensemble-averaged measurements. To systematically identify the dominant physical parameters driving these state shifts and to map the exact number of sub-states visited by a single molecule over time we performed Principal Component Analysis (PCA) coupled with unsupervised *K*-means clustering. PCA compresses the multidimensional parameter space $(\lambda_c, \theta, \gamma, \phi)$to identify which orthogonal variables account for the largest proportion of variance during the trajectory. Subsequently, *K*-means clustering on the reduced principal component space objectively classifies the discrete micro-states and trajectory jumps without prior manual thresholding bias.

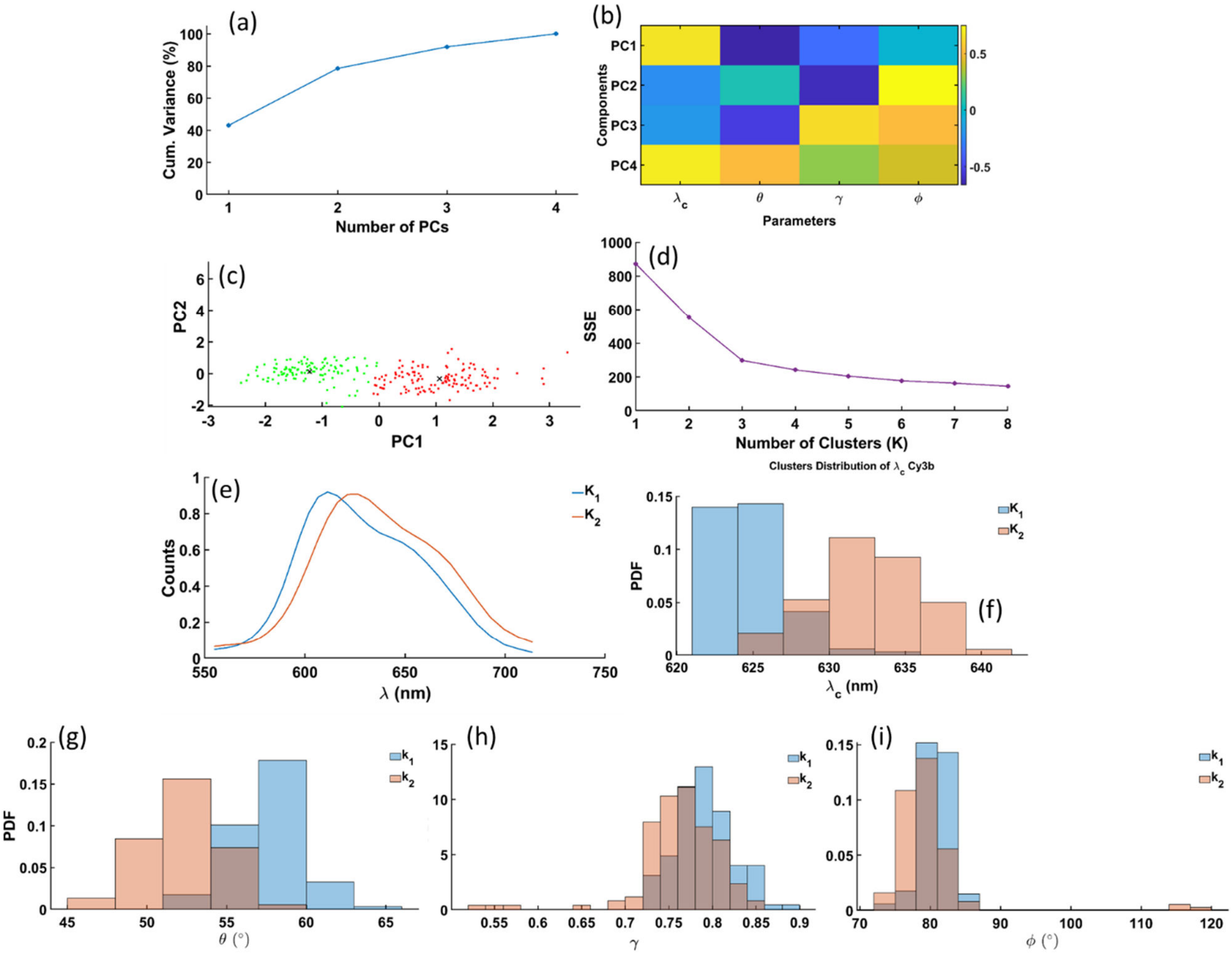

**Figure S24.** PCA of SM trajectory shown in **Figure S10**; Cy3b with thick PMMA (MW 350K) overlayer. (a) Cumulative explained variance as a function of the number of principal components (PCs), demonstrating that the first $N$ components capture over 90% of the total system variance.(b) PCA loading (weight) matrix illustrating the relative contribution of each physical parameter $(\lambda_c, \theta, \gamma, \phi)$ to the individual principal components, identifying the primary physical mechanisms driving spectral and orientation fluctuations. Unsupervised state identification and clustering via K-means: (c) Determination of the optimal number of clusters using the elbow method (Sum of Squared Errors vs. number of clusters). (d) Two-dimensional projection of the single-molecule trajectory onto the first two principal components (PC1 vs. PC2), highlighting distinct sub-states (clusters) and discrete state jumps visited during trajectory progression. Ensemble and structural parameter distributions across identified single-molecule sub-states: (e) Average emission spectra corresponding to each identified cluster, showing spectral shifts and small line-shape variations across states. Probability density distributions (PDF) of (f) central wavelength $(\lambda_c)$, (g) polar orientation angle $(\theta)$ and (h) wobble/constraint factor $(\gamma)$and (i) azimuthal angle $(\phi)$ segregated by cluster, revealing distinct physical configurations adopted by the single molecule as it progresses along its trajectory. Here $\theta$ and $\lambda_c$ shows distinctly different distributions for the two clusters, while in-plane angle and rotational constraint distributions remain nearly similar. For visual clarity we plotted only for the first two principal components.

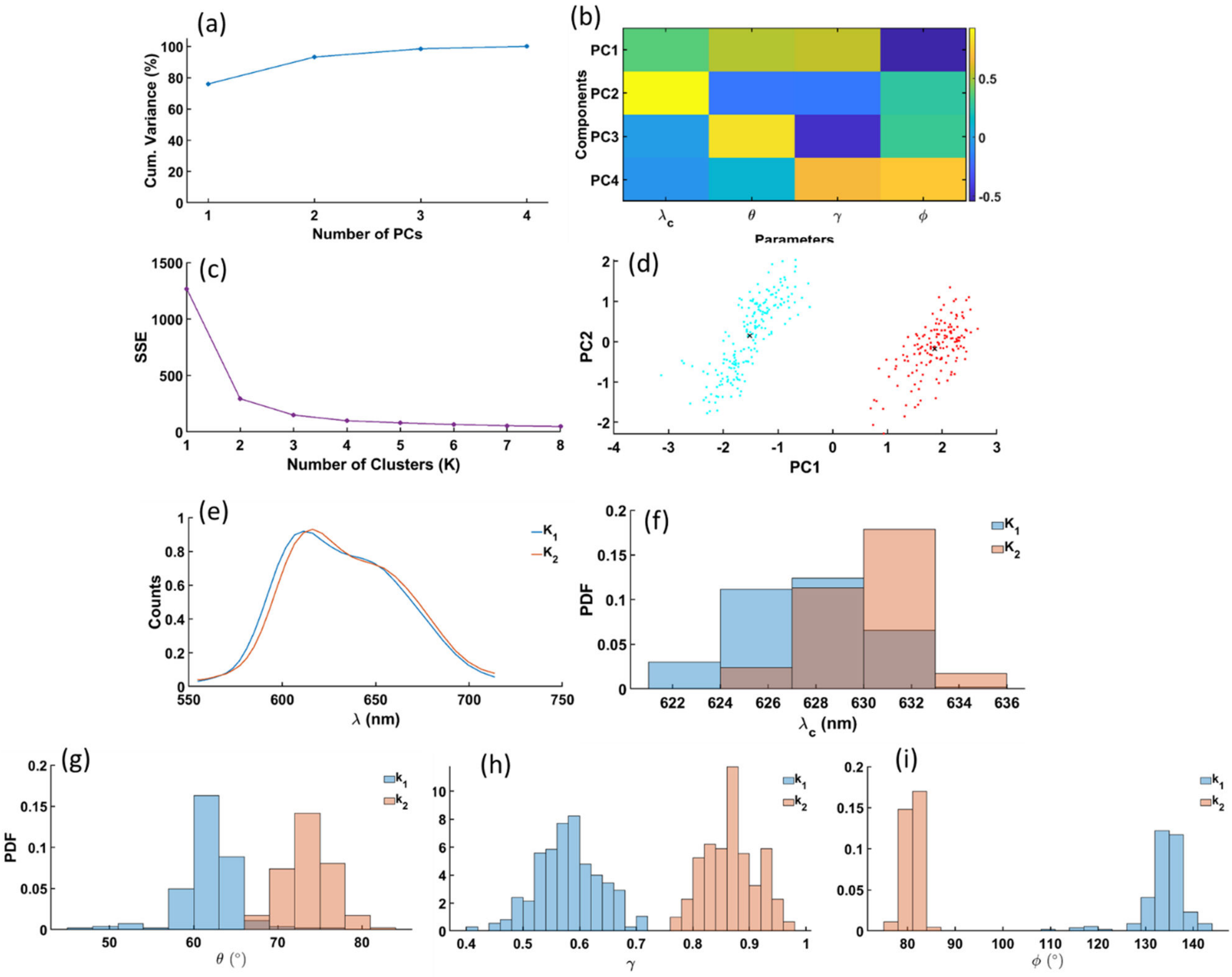

**Figure S25.** PCA of SM trajectory shown in **Figure S14**; NR with thick PMMA (MW 350K) overlayer. Description of figures are same as mentioned in **Figure S22**. In this SM trajectory TDM shows clearly distinct distributions for $\theta$, $\phi$ and $\gamma$ without substantial shift in $\lambda_c$ distribution.

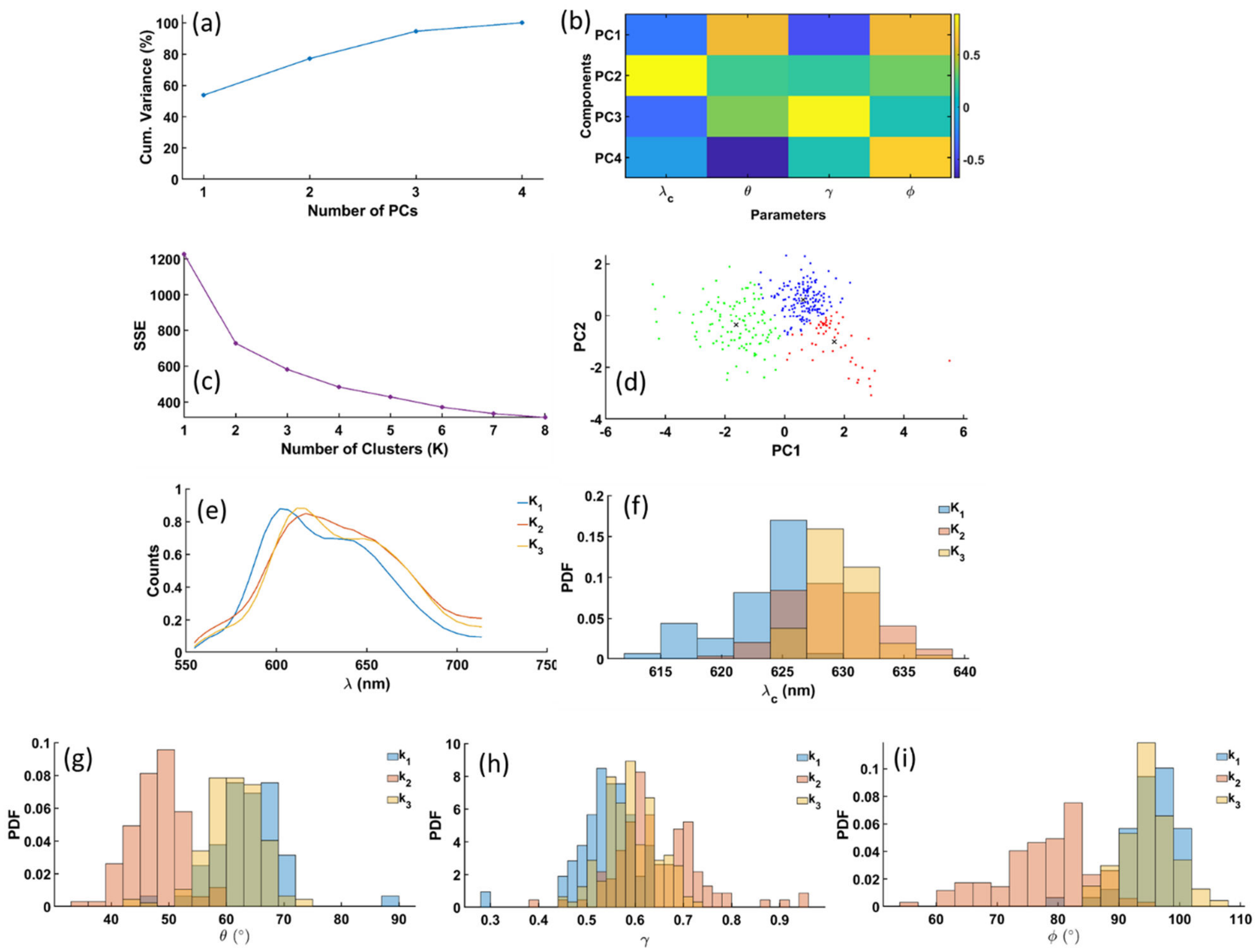

**Figure S26.** PCA of SM trajectory shown in **Figure S11**; MC540 with thick PMMA (MW 350K) overlayer. Description of figures are same as mentioned in **Figure S22**. Response from this SM indicates three distinct clusters of molecular behaviour for the trajectories.

## Section S6

**Large Scale TDM Reorientation:** The possibility for large scale TDM reorientation could be the consequence of chemical or physical interactions or intrinsic conformational transitions within the chromophore. PMMA is a basic polymer, where electron donor carbonyl oxygen of the ester group acts as the basic site, may involve in acid-base interaction with the fluorophores resulting in altered ionic states, e.g. Atto 520 or RhB may be in cation or zwitterion state, however both have same TDM orientation relative to xanthene frame (Section S7). Abrupt flipping of TDM owing to proton-transfer tautomerism is well documented for SM probes like hypercin, porphycenes, etc.[40, 41] However instead of true proton-transfer tautomerism, for the present bunch of dyes, lactone–zwitterion (chain↔ring) equilibria and cis–trans/photoisomerization seems more probable. In case of Rh6G the esterified carboxyl group prevents the usual rhodamine lactone-zwitterion equilibrium, but is likely for RhB (i.e. colored zwitterion ⇄ colourless lactone). But then RhB lactone form absorbs in UV region (~300–350 nm) and has a very weak, blue-shifted fluorescence ~450–500 nm.[42] In the present experiment, if formed, this will be a dark state like triplet. NR is not known for proton-transfer tautomerism but for pronounced solvatochromism. No conventional tautomerism is expected for Cy3 and MC540 but trans ⇄ cis photoisomerization of polymethine chain is possible. Cyanine-chain conformational/ isomerization behaviour is expected for Cy3 and MC540. Tautomerism is possible with Atto 520 between fluorescent form and colourless pseudo-base above pH ~7; the latter is again a dark invisible state for our setup. Therefore, contribution of dye conformational fluctuation towards TDM reorientation is expected to be negligible if not nil.

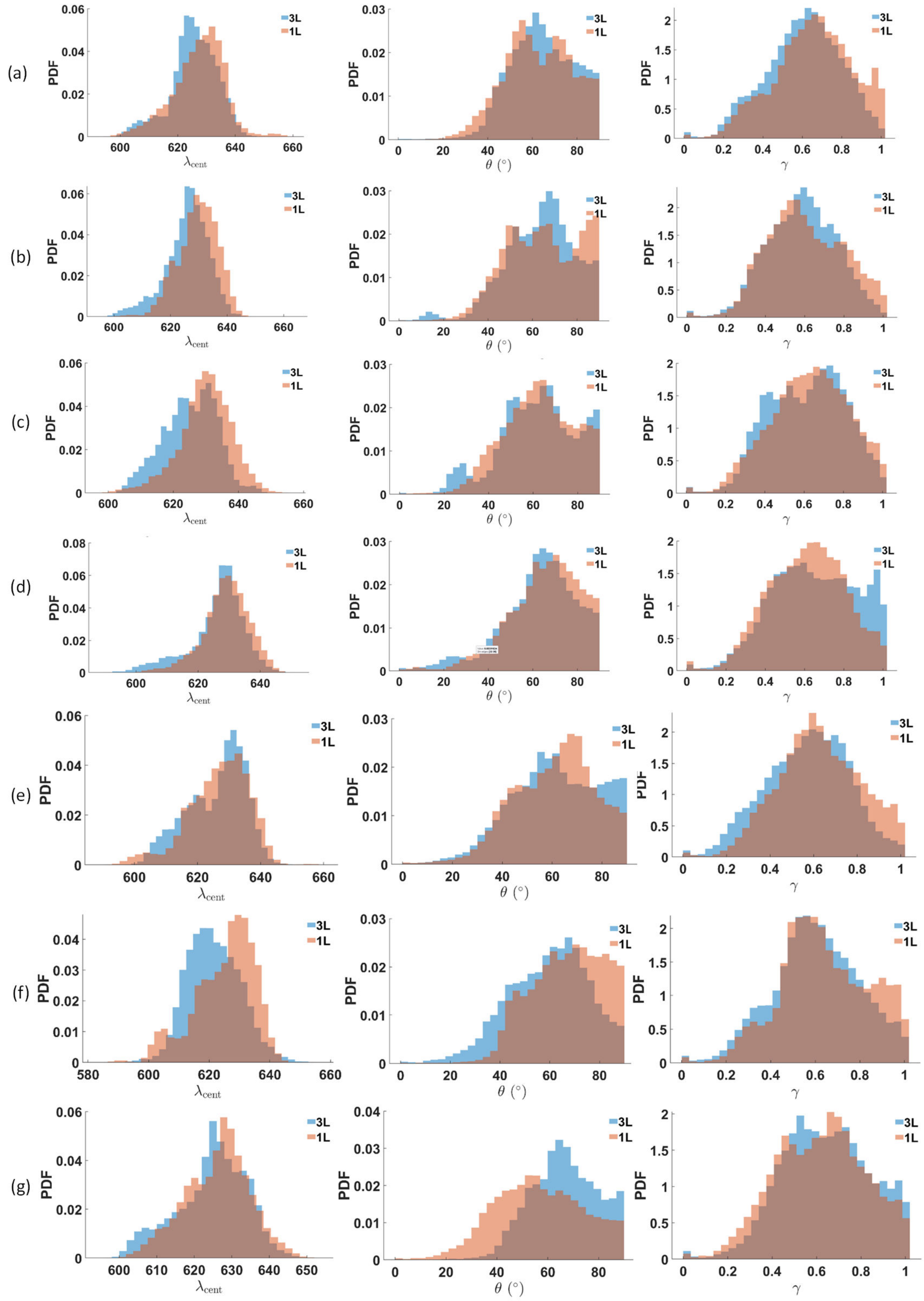

**Figure S27.** Distribution of parameters for thick PMMA overlayers of MW 100K (1L) & MW 350K (3L) molecular weights. Rh6G (a), RhB (b), Cy3 (c), Cy3b (d), Atto 520 (e), MC540 (f), and NR (g).

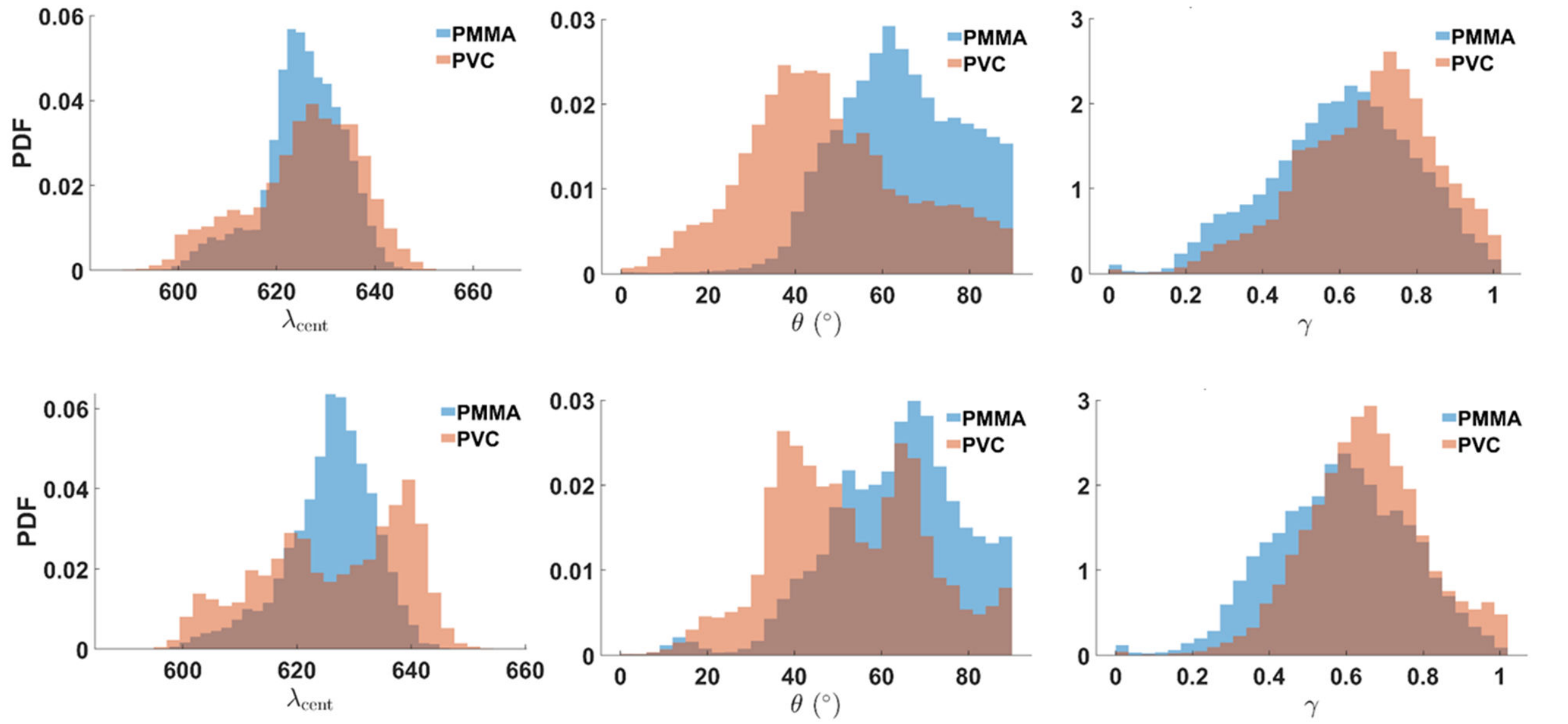

**Figure S28.** Comparison between PMMA (MW 350K) and PVC thick polymer overlayer. Top panel - Rh6G and bottom panel - RhB.

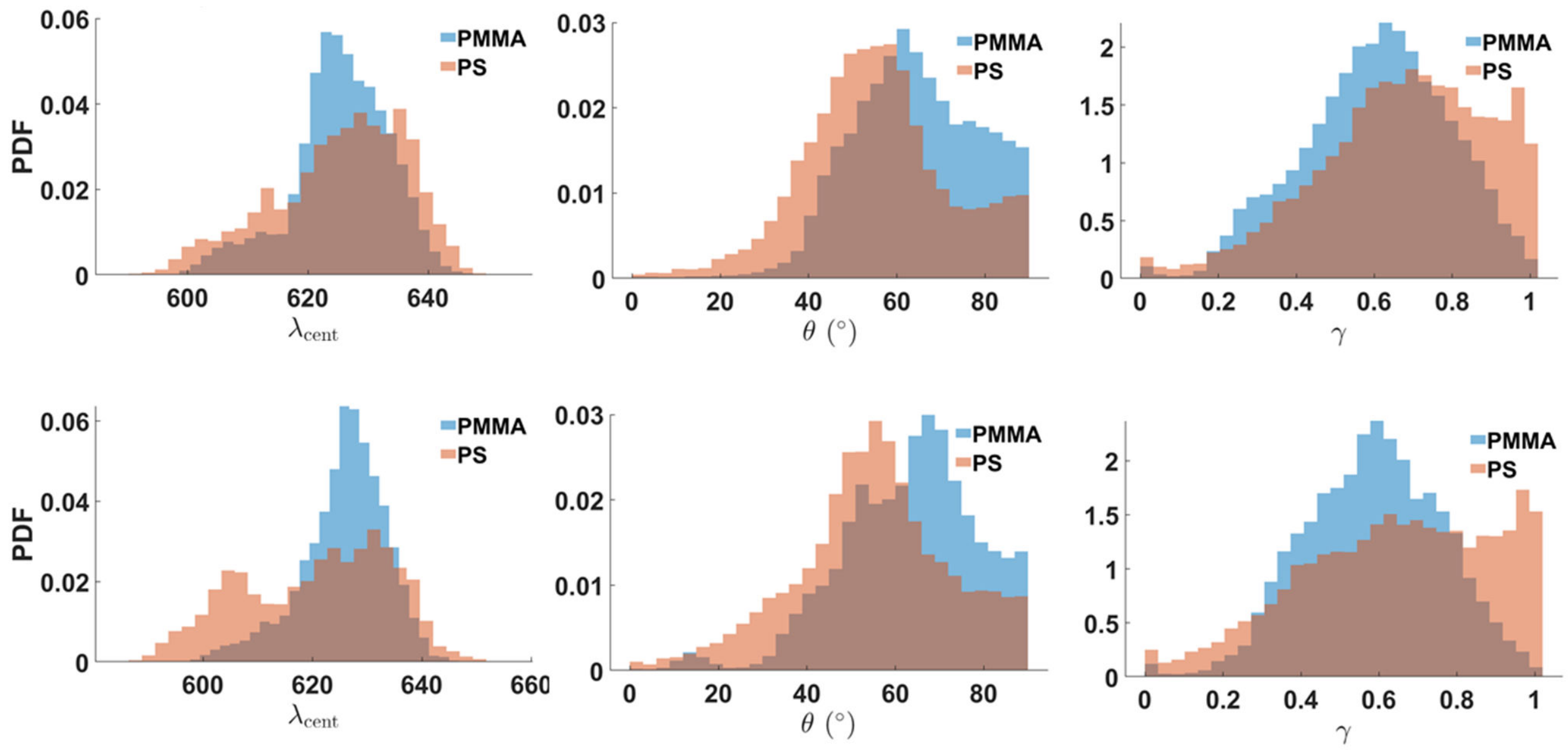

**Figure S29.** Comparison between PMMA (MW 350K) and PS thick polymer overlayer. Top panel - Rh6G and bottom panel - RhB.

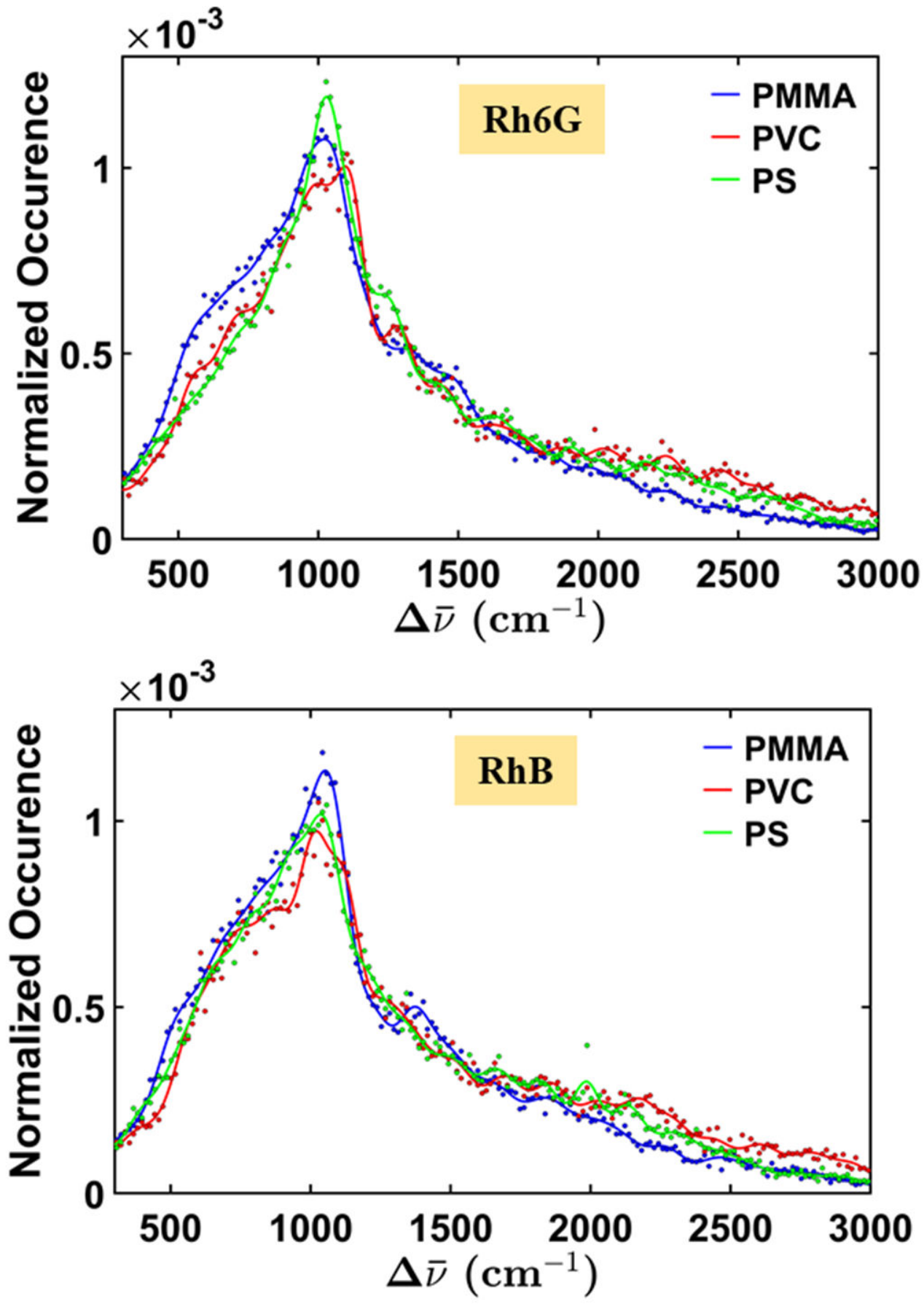

**Figure S30.** Distribution of vibronic band separations for thick overlayer of different polymers.

## Section S7

### Computational Methods and Intrinsic Dye Response

#### A. Methods

In order to separate intrinsic dye responses from polymer-matrix effects we used electronic structure calculations to identify if and which intramolecular degrees of freedom of the dyes can contribute to the experimentally observed single-molecule spectral diffusion and transition dipole reorientation. Specifically, we carried out density functional theory (DFT) for ground and excited state geometry optimizations and time-dependent density functional theory (TDDFT) calculations coupled with Franck-Condon (FC) and Herzberg-Teller (HT) vibronic analysis. The workflow began with geometry optimization and validation on the ground and bright excited surface. At each $S_0$ and $S_1$reference geometry we computed TDDFT single-point transition energies, oscillator strengths, and Cartesian transition dipole vectors, then defined a fixed right-handed molecular frame to describe the changes in dipole moment. Vibronic structure was treated within the harmonic approximation using an adiabatic-Hessian (AH) representation of the two-electronic-state potential energy surfaces, yielding Huang–Rhys factors and reorganization energies, while HT derivatives were obtained by symmetric finite-difference displacements and projected into the molecular frame to quantify longitudinal and transverse mode responses. We then employed a FC/HT threshold based filtering protocol to retain the most vibronically relevant and HT-active modes, after which selected coordinates were probed by multi-amplitude TDDFT displacements or relaxed torsional scans (for soft modes) to map the electronic energy surfaces, and TDM changes across sampled amplitudes. A detailed description of each these steps is provided below:

#### (I) Ground-state and Excited-state reference geometries

The ground-state electronic structure was optimised on the $S_0$ potential energy surface using Kohn-Sham density functional theory (DFT). The targeted lowest optically bright singlet excited state, $S_1$ was optimised using linear response time-dependent density functional theory (TDDFT). These calculations were performed using the CAM-B3LYP exchange-correlation functional, the def2-SVP basis set and a SMD continuum solvation ($\varepsilon_r = 24.85$) to include solvent effects. Further, we performed harmonic vibrational frequency calculations at each optimised $S_0$ and $S_1$ geometry and single-point TDDFT calculations at $S_0$ and $S_1$ reference geometries to obtain electronic transition properties. Based on these calculations we obtained the vertical $S_1 - S_0$ transition energy ($E_{\mathrm{vert}}(\mathrm{R}) = E_{S_1}(\mathrm{R}) - E_{S_0}(\mathrm{R})$) where $E_{S_1}(\mathrm{R})$ is the TDDFT energy of the selected bright state at geometry R and $E_{S_0}(\mathrm{R})$ is the ground-state energy at the same geometry. We also obtained the corresponding oscillator strength, and electric transition dipole moment vector $\mu_{01}(\mathrm{R}) = \langle \Psi_0(\mathrm{R}) | \hat{\mu} | \Psi_1(\mathrm{R}) \rangle$ where $\Psi_0$ and $\Psi_1$are the electronic wavefunctions of the ground and first bright singlet excited state, respectively, and $\hat{\mu}$ is the electric dipole operator. Here, for an electronic transition between the same pair of states, the absorption and emission transition moments are reciprocal $\mu_{10}(\mathrm{R}) = \mu_{01}(\mathrm{R})^* = \mu_{0,\mathrm{TDDFT}} = [\mu_x, \mu_y, \mu_z]^{\mathrm{T}} = \mu_0 = \mu$.The squared transition-moment magnitude contributes to the electric-dipole transition strength by $f_\mu = \frac{2}{3}\Delta E_{01}|\mu|^2$. For an isotropically averaged electric dipole transition, the transition moment norm is $|\mu| = \sqrt{\mu_x^2 + \mu_y^2 + \mu_z^2}$. The $f_\mu$ and $|\mu|$ report the strength of the electronic transition. Conversely, the vector direction of $\mu$ defines the intrinsic optical transition axis. The electronic transition moment extracted from Gaussian[43] is in an arbitrary Cartesian frame; therefore, we defined a fixed, right-handed molecular frame once from the optimised $S_1$ reference geometry. The unit vectors of this frame were the molecular long axis ($\hat{\mathrm{e}}_{\mathrm{long}}$). We defined the long axis from the chemically selected chromophore core and oriented it using chemically predefined anchor atoms. The short axis ($\hat{\mathrm{e}}_{\mathrm{short}}$) was generated from a chemically defined in-plane reference direction after projection orthogonal to the long axis, and the out-of-plane direction was then obtained by the right-handed cross product, $\hat{\mathrm{e}}_{\mathrm{oop}} = \hat{\mathrm{e}}_{\mathrm{long}} \times \hat{\mathrm{e}}_{\mathrm{short}}$. The molecular frame representation was calculated as,

$$\mu_{\mathrm{mol}} = \mathrm{R}_{\mathrm{mol}\leftarrow\mathrm{TDDFT}}\,\mu_{\mathrm{TDDFT}} = \begin{bmatrix} \mu_{\mathrm{long}} \\ \mu_{\mathrm{short}} \\ \mu_{\mathrm{oop}} \end{bmatrix}, where\ \mathrm{R}_{\mathrm{mol}\leftarrow\mathrm{TDDFT}} = \begin{bmatrix} \hat{\mathrm{e}}_{\mathrm{long}}^{\mathrm{T}} \\ \hat{\mathrm{e}}_{\mathrm{short}}^{\mathrm{T}} \\ \hat{\mathrm{e}}_{\mathrm{oop}}^{\mathrm{T}} \end{bmatrix} \qquad (\mathrm{eq.\,S8})$$

with projected components as $\mu_{\mathrm{long}} = \mu_{\mathrm{TDDFT}} \cdot \hat{\mathrm{e}}_{\mathrm{long}}$; $\mu_{\mathrm{short}} = \mu_{\mathrm{TDDFT}} \cdot \hat{\mathrm{e}}_{\mathrm{short}}$; and $\mu_{\mathrm{oop}} = \mu_{\mathrm{TDDFT}} \cdot \hat{\mathrm{e}}_{\mathrm{oop}}$. This rotation matrix ensures that any calculated change in the molecular frame transition moment reflects an intrinsic electronic response rather than a change in the coordinate axes used to describe it. A change predominantly along the long axis is expected primarily to modulate transition strength. In contrast, the short-axis and out-of-plane components are transverse to the principal optical axis and can produce intrinsic electronic transition-axis reorientation within or out of the molecular plane, respectively.

**(II) Vibronic Emission Model.**

Vibronic calculations were carried out within the harmonic approximation using an adiabatic-Hessian (AH) representation of the two-electronic-state potential energy surfaces. These calculations included both changes in normal mode character between the electronic states and the displacement of the equilibrium nuclear configuration upon electronic emission. The optimised $S_0$ and $S_1$ geometries, harmonic frequencies, and electronic-emission transition properties from the calculations were used as input to obtain vibrationally resolved emission properties in the time-dependent formulation (TD) at 300K within the *FCclasses3 3.0.4*package.[44] This is particularly useful for polyatomic fluorophores, for which a formally complete finite-TD Franck-Condon sum can contain an impractically large number of combination bands and overtones. We used these calculations to obtain mode-resolved vibronic descriptors, Huang-Rhys factors $S_k$ and reorganisation energies $\lambda_k$. For a normal mode $k$, $S_k$ reports the propensity of a mode to contribute to vibronic structure through nuclear displacement between the $S_1$ and $S_0$ potential energy surfaces as $S_k = 1/2\left(\Delta Q_k / Q_{\mathrm{zp},k}\right)^2$, where $\Delta Q_k$ is the equilibrium electronic state displacement projected on $k$ and $Q_{\mathrm{zp},k}$ is the zero-point amplitude of mode $k$. On the other hand, for a normal mode $k$, $\lambda_k$ identifies the energetic contribution to electronic state reorganisation as $\lambda_k = S_k \hbar \omega_k \; or \; \lambda_k = S_k hc\tilde{\nu}_k$ , where $\omega_k$ and $\tilde{\nu}_k$ are the angular frequency and wavenumber of mode *k*, respectively. A high $S_k$ does not necessarily imply a high $\lambda_k$, particularly for low-frequency modes. Conversely, a high frequency coordinate can contribute substantially to the energy even if its dimensionless displacement is moderate. Therefore, we preserved both descriptors from the vibronic emission model. This prevents the investigation of normal modes from being biased either towards low-frequency modes with large dimensionless amplitudes or towards high-frequency modes with large energetic weights alone.

**(III) Herzberg-Teller transition moment framework.**

The coordinate-dependent transition moment is expanded to first order as $\mu_{10}(\mathrm{Q}) = \mu_0 + \sum_k \left(\frac{\partial \mu_{10}}{\partial Q_k}\right)_{S_1} Q_k + O\left(Q_k^2\right)$, where the zeroth-order term $\mu_0 = \mu_{10}(Q = 0)$ is the Condon transition moment and gives the FC contribution. The FC approximation describes vibronic intensity borrowing through nuclear overlap (i.e., a mode with significant $S_k$ or $\lambda_k$ is vibrationally relevant), but it cannot describe a change in the electronic transition-moment vector induced by nuclear motion. In contrast, the first-order coefficient of mode *k*, $\mathrm{h}_k$ (the Herzberg-Teller (HT) transition-moment derivative) reports on the change in the transition moment vector. A mode with a substantial transverse HT derivative, in contrast, can change the direction of the effective transition moment even if its FC contribution is only moderate. Thus, HT coupling can redistribute vibronic intensity and alter the relative importance of emission channels, rather than simply adding an independent intensity correction. For an individual mode, the direction and magnitude of the HT derivative determine whether it primarily modulates intensity or can alter the effective transition axis. This is the basis for subsequently resolving longitudinal, short-axis transverse, and out-of-plane transverse responses. Thus, FC and HT terms offer complementary pictures of the electronic transitions. Within this FC/HT framework, the emission or absorption response contains three contributions $R = R_{\mathrm{FC}} + R_{\mathrm{HT}} + R_{\mathrm{FC-HT}}$, where $R_{\mathrm{FC}}$ arises from the constant transition moment, $R_{\mathrm{HT}}$ arises from the square of the coordinate-dependent term, and $R_{\mathrm{FC-HT}}$ is the interference term between the two contributions. Therefore, a mode can be strongly FC active but HT inactive, HT active but only weakly FC displaced, or simultaneously active in both channels. To connect the vibronic descriptors obtained from the FC-only emission *(AH-FC-TD-300K model in FCclasses3)* with the electronic transition-axis response, we evaluated the first-order nuclear-coordinate dependence of the transition dipole moment. For this purpose, we expanded the transition moment in the $S_1$ normal coordinates around $S_1$ reference geometry, where $Q_k$ is the dimensionless coordinate of normal mode *k*. The $S_1$ normal modes and mass-weighted Cartesian eigenvectors applied by *FCclasses3* were used to generate symmetric local displacements. For each normal mode *k*, the Cartesian geometry displacement was constructed using $\Delta \mathrm{R} = \sum_k Q_k \mathrm{L}_k$ or $\Delta \mathrm{R}_k(Q_k) = Q_k\sqrt{16.857762/\omega_k}\ \mathrm{L}_k$ , where $\mathrm{L}_k$ is the mass-weighted Cartesian eigenvector in $amu^{-1/2}$. The local HT finite difference calculation used a symmetric dimensionless displacement of $\delta Q = 0.02$. At both endpoint structures, we recorded TDDFT single-point calculations using the same electronic-structure-mode protocol as for the equilibrium reference. These endpoint vectors were projected onto the fixed molecular frame, $\mu^{\pm}_{k,\mathrm{mol}} = \mathrm{R}_{\mathrm{mol}\leftarrow\mathrm{TDDFT}}\mu^{\pm}_{k,\mathrm{TDDFT}}$, and $\mathrm{h}_k \approx \mu_k(+\delta Q) - \mu_k(-\delta Q)/2\delta Q$. The total local HT derivative magnitude, $|\mathrm{h}_k| = \sqrt{h^2_{k,\mathrm{long}} + h^2_{k,\mathrm{short}} + h^2_{k,\mathrm{oop}}}$ and projected HT components are $h_{k,\mathrm{long}} = (\mu_{\mathrm{long}}(+\delta Q) - \mu_{\mathrm{long}}(-\delta Q))/2\delta Q$ ; $h_{k,\mathrm{short}}$ and $h_{k,\mathrm{oop}}$. A large $h_{k,\mathrm{long}}$ describes a longitudinal response that can modulate the strength of the preexisting transition moment. A large $h_{k,\mathrm{short}}$ represents a transverse in-plane response that can rotate the intrinsic electronic transition axis within the molecular plane and a large $h_{k,\mathrm{oop}}$ represents a transverse out-of-plane response that can rotate the transition axis away from the molecular plane. Based on this, changes in the normalised directional fractions were recorded as $\Delta p_j(Q) = p_j(Q) - p_j(0),\ j \in \{\mathrm{long}, \mathrm{short}, \mathrm{oop}\}$.

**(IV) Candidate-mode selection and response-map design.**

The FC/HT analysis described above produced a mode-resolved descriptor space to select a computationally tractable set of normal modes which can then be examined for their contributions to the dye optical properties. We selected modes independently for each of the 10 dyes (including cation and zwitterion forms), using a common set of HT selection criteria. Specifically, we considered: *i)* FC displacement and reorganisation convergence; *ii)* molecular frame HT response; *iii)* brightness-gated effective channel orientation response; i*v)* persistence of these descriptors across predefined threshold grids; *v)* numerical HT resolution qualification; and *vi)* structural review for low-frequency modes. We then subjected the resulting candidate sets of modes to direct finite-amplitude TDDFT calculations and torsional scans (for low frequency modes), from which we assigned the final spectral, brightness, and transition-axis assignments. The mode selection criteria are described below:
<u>FC mode displacement and reorganization energies</u>: We defined the Huang-Rhys and mode-resolved reorganisation energy fractions for a set of modes $\mathcal{M}(T)$ as: $C_S(T) = \sum_{k\in\mathcal{M}(T)} S_k/\sum_k S_k\,, C_\lambda(T) =$ $\sum_{k\in\mathcal{M}(T)} \lambda_k/\sum_k \lambda_k$. We selected a set of modes $\mathcal{M}(T)$ which covered at least $C_S(T) \geq 0.95$, $C_\lambda(T) \geq 0.95$. We also checked that the subset of modes does not change significantly with small variations of this threshold.

_HT resolution thresholds_: We calculated the molecular-frame HT derivatives from finite differences of endpoint transition moments. We flagged modes whose apparent HT components were close to the empirical TDDFT output resolution (corresponding to dye specific HT component resolution increments of $2.5 \times 10^{-8} - 3.0 \times 10^{-7} a.u./Q$ , with vector norm resolution of $4.5 \times 10^{-8} - 5.0 \times 10^{-7} a.u./Q$). Such modes were considered only if they meet some or all of the other criteria as provided in this list.

_Transitions brightness thresholds_: The FC and HT descriptors were combined through the resolved one-quantum effective transition moment

$$\boldsymbol{\mu}_{\mathrm{eff},k} = \sqrt{S_k} \exp\left(-\frac{S_k}{2}\right)\left[\boldsymbol{\mu}_0 + \frac{\mathbf{h}_k}{\sqrt{2}}\right] \quad \text{(eq. S9)}$$

This provides a mode-specific, molecular-frame effective transition moment for the same resolved 0→$1_k$ vibronic transitions. Its relative brightness is $B_k \propto |\boldsymbol{\mu}_{\mathrm{eff},k}|^2$ with the normalised brightness fraction $b_k = B_k/\max_j B_j$. Here, brightness is normalised relative to the brightest resolved one quantum vibronic transition of the same dye, with $\max_j B_j$ evaluated over all mode specific transitions included in the analysis. Only transitions satisfying a normalized brightness threshold, $b_k \in \{0.10, 0.03, 0.01\}$ were considered in the effective axis and pairwise screening. The effective optical axis deviation of a transition from the equilibrium transition axis introduced along a mode is

$$\theta_{\mathrm{eff},k} = \cos^{-1}\left[\left|\frac{\boldsymbol{\mu}_{\mathrm{eff},k} \cdot \boldsymbol{\mu}_0}{|\boldsymbol{\mu}_{\mathrm{eff},k}||\boldsymbol{\mu}_0|}\right|\right] \quad \text{(eq. S10)}$$

$$0° \leq \theta_{\mathrm{eff},k} \leq 90°$$

and the pairwise effective axis separation of two vibronic transitions *i* and *j* along a mode is,

$$\theta_{\mathrm{eff},ij} = \cos^{-1}\left[\left|\frac{\boldsymbol{\mu}_{\mathrm{eff},i} \cdot \boldsymbol{\mu}_{\mathrm{eff},j}}{|\boldsymbol{\mu}_{\mathrm{eff},i}||\boldsymbol{\mu}_{\mathrm{eff},j}|}\right|\right] \quad \text{(eq. S11)}$$

$$0° \leq \theta_{\mathrm{eff},ij} \leq 90°$$

We set thresholds of $\theta_{\mathrm{eff},k} \in \{0.5°, 1°, 2°\}$ , $\theta_{\mathrm{eff},ij} \in \{1°, 2°, 5°\}$. These thresholds were not considered as experimental angular detection limits. Modes which change the transition axis above these thresholds were further examined using direct finite-amplitude and torsional calculations.

*Persistence Rule:* For each mode, persistence was quantified as the fraction of tested settings for which that mode remained above the thresholds defined above. This approach distinguished a mode that remains relevant under multiple criteria from a mode that only satisfies a single criterion. we used the following four persistence thresholds:

$$\text{P1: } b_k \geq 0.03 \text{ and } P_{k,\text{orientation}} \geq 0.50 \quad \text{(eq. S12)}$$

$$\text{P2: } b_k \geq 0.03 \text{ and } P_{k,\text{pairwise}} \geq 0.50 \quad \text{(eq. S13)}$$

$$\text{P3: } \frac{\text{FC}}{\text{HR}_k} \geq 0.005, P_{k,\text{HT}} \geq 0.80, \max(r_{k,\text{short}}, r_{k,\text{oop}}) \geq 0.15 \quad \text{(eq. S14)}$$

$$\text{P4: } \frac{\text{FC}}{\text{HR}_k} \geq 0.01 \text{ and } b_k \geq 0.03 \quad \text{(eq. S15)}$$

Here, $r_{k,\text{short}}$ and $r_{k,\text{oop}}$ denote the normalised local short axis and out-of-plane HT response components, respectively. P1 and P2 identify bright channels with significant single or pairwise effective axis impact. P3 identifies modes with jointly strong FC/HR relevance, persistent HT response and a substantial local transverse component. P4 retains a strong FC brightness reference class when P1-P3 are not met.

**(V) Normal-coordinate and Conformational-coordinate treatment.**

Following the FC/HT-based candidate mode identifications, we separated them according to their frequencies and examined their atomic displacement character:

*High frequency/non-soft normal modes:* Normal modes at or above the threshold $\tilde{\nu}_k \geq 300\ cm^{-1}$ were treated as locally harmonic coordinates around the optimised $S_1$ reference geometry. For a selected normal mode k, the finite-amplitude Cartesian displacement was generated from the optimised $S_1$ geometry as $\mathrm{R}_k(Q_k) = \mathrm{R}_{S_1}^{\mathrm{eq}} + \Delta\mathrm{R}_k(Q_k)$. Modes in this category are further categorised into primary and secondary sets. Primary modes were selected based on persistent bright channel axis or pairwise axis screening evidence, or strong FC relevance combined with robust, above resolution HT directional susceptibility. Secondary modes were above resolution, non-soft coordinates supported by at least one of the FC-HT, effective-axis, pairwise-axis or FC intensity screening criteria but not meeting the primary-selection criteria, and were evaluated using a reduced symmetric amplitude screen. Primary modes were evaluated using a symmetric mass-weighted, multi-amplitude grid of amplitudes, $Q/Q_{\text{rms}} \in \{0, \pm0.5, \pm1.0, \pm1.5, \pm2.0\}$. For a harmonic oscillator of frequency $\tilde{\nu}_k$ at temperature *T*, the dimensionless thermal coordinate amplitude used here is $Q_{\text{rms},k} = [1/2 \coth(hc\tilde{\nu}_k/2k_BT)]^{1/2}$. The $Q = 0$ structure was a common $S_1$ equilibrium reference. We then performed TDDFT single-point calculations at each displaced geometry to probe the local electronic susceptibility of the bright transition. The direct responses extracted at each displaced geometry were $\Delta E_{\text{vert}}(Q)$ or $\Delta\tilde{\nu}(Q)$, $\Delta f_{\text{rel}}(Q)$, and $\theta_{\text{axis}}(Q)$ as given below

$$\Delta E_{\text{vert}}(Q_k) = E_{\text{vert}}(Q_k) - E_{\text{vert}}(0) \quad \text{(eq. S16)}$$

$$\Delta f_{\text{rel}}(Q_k) = f(Q_k)/f(0) - 1 \quad \text{(eq. S17)}$$

$$\theta_{\text{axis}}(Q_k) = \cos^{-1}\left[\left|\frac{\boldsymbol{\mu}_{\text{mol}}(Q_k)\cdot\boldsymbol{\mu}_{0,\text{mol}}}{|\boldsymbol{\mu}_{\text{mol}}(Q_k)||\boldsymbol{\mu}_{0,\text{mol}}|}\right|\right] \quad \text{(eq. S18)}$$

For each primary mode, the final response metrics from the full sampled grid are $\max_Q|\Delta E_{\text{vert}}(Q)|$, $\max_Q|\Delta\tilde{\nu}(Q)|$, $\max_Q\theta_{\text{axis}}(Q)$, $\max_Q|\Delta f_{\text{rel}}(Q)|$, $\max_Q|\Delta\mu_{\text{rel}}(Q)|$, $\max_Q|\Delta p_{\text{long}}(Q)|$, $\max_Q|\Delta p_{\text{short}}(Q)|$, and $\max_Q|\Delta p_{\text{oop}}(Q)|$. Secondary modes were treated by a lower-cost symmetric grid of amplitudes $Q_{\text{rms}}/Q_{\text{rms}} \in \{0, \pm 1.0\}$. The secondary calculation also uses symmetric positive and negative displacements and extracts the same fundamental observables.

*Low frequency/soft normal modes( $\tilde{\nu}_k < 300\ cm^{-1}$)*: In these cases, a single Cartesian normal-coordinate displacement may inadequately represent the physically relevant large-amplitude conformational degree of freedom. The harmonic approximation is not suitable for flexible and curvilinear coordinates such as torsions. We divided such modes into two categories as described below

*Category A*, *torsional and torsion-coupled coordinates:* This set includes soft coordinates whose displacement was dominated by a specific dihedral deformation. We treated these coordinates using bidirectional, one-dimensional constrained relaxed $S_1$ torsional scans rather than interpreting them solely as frozen harmonic displacements. For a selected torsional coordinate (dihedral, *D*), scan points were generated in both directions about the equilibrium structure $\phi = \phi_0 + \Delta\phi$ to span an energy range of a few $k_BT$. At each constrained value of $\phi$, the geometry was relaxed on the $S_1$ surface while retaining the specified dihedral constraint. Here, each scan geometry was first rigidly aligned to the equilibrium $S_1$ reference using a chemically predeclared set of alignment atoms by Kabsch alignment, thereafter the TDDFT transition moment at each scan coordinate $\phi$ is correspondingly rotated and then projected into the same equilibrium-defined molecular frame used for the harmonic normal mode calculations. We used a second independent alignment atom set as a sensitivity calculation to test the sensitivity of the induced internal transition-axis response to the choice of rigid-core alignment atoms. For every retained torsional scan point, vertical energy, oscillator strength, transition axis, and molecular frame directional response were computed as:

$$E_{span}^{\mathrm{D}} = \max_D \mathrm{E}_{vert}(\phi) - \min_D \mathrm{E}_{vert}(\phi) \qquad \text{(eq. S19)}$$

$$\text{range of } f(\phi) = \frac{f(\varphi)}{f_{\mathrm{e}}} \qquad \text{(eq. S20)}$$

$$Rf^{\mathrm{D}} = \frac{\max_{\varphi \in D} f(\varphi)}{\min_{\varphi \in D} f(\varphi)} \qquad \text{(eq. S21)}$$

$$\max_{\varphi \in D} \boldsymbol{\theta}_{\mathbf{axis,e}} = \max_{\varphi \in D} \cos^{-1} \left[ \left| \frac{\boldsymbol{\mu}_{\mathrm{mol}}(\phi) \cdot \boldsymbol{\mu}_{0,\mathrm{mol}}}{|\boldsymbol{\mu}_{\mathrm{mol}}(\phi)||\boldsymbol{\mu}_{0,\mathrm{mol}}|} \right| \right] \qquad \text{(eq. S22)}$$

$$\Delta p_j(\phi) = p_j(\phi) - p_j(\phi_0), j \in \{\text{long, short, oop}\} \qquad \text{(eq. S23)}$$

To quantify the largest intrinsic axis distinction sampled within one torsional family, pairwise transition axis separations were calculated as:

$$max\varphi_{\mathrm{i}}, \varphi_{\mathrm{j}} = \max_{\varphi \in D} \cos^{-1} \left[ \left| \frac{\mu_{\mathrm{mol}}(\phi_i) \cdot \mu_{\mathrm{mol}}(\phi_j)}{|\mu_{\mathrm{mol}}(\phi_i)| \, |\mu_{\mathrm{mol}}(\phi_j)|} \right| \right] \qquad \text{(eq. S24)}$$

The equilibrium reference angle indicates whether a particular constrained conformation differs from the equilibrium $S_1$ structure. The pairwise angle indicates whether any of two sampled conformations along a set of torsional degree of freedom possess different intrinsic electronic transition axes. The latter is a relevant descriptor to test the possibility of a conformational switching mechanism.

*Category B, collective non-torsional or mixed coordinates:* Soft modes that remained predominantly collective, non-conformational, or not adequately represented by a single torsion were assigned to this category. They were evaluated on the full set of symmetric amplitude grids: $\frac{Q}{Q_{\mathrm{rms}}} \in \{0, \pm 0.5, \pm 1.0, \pm 1.5, \pm 2.0\}$.The same workflow protocol of phase-corrected, fixed r-frame TDM analysis, vertical energy, oscillator strength response, and directional fraction analysis used for primary non-soft normal modes was also applied to category B modes. Therefore, the normal mode results characterise local response near the $S_1$ minimum, and the torsional results characterise response within a validated sampled conformational interval.

Taken together the analysis separates local vibronic/electronic susceptibility from finite amplitude conformational susceptibility. The workflow here determines whether a dye's intrinsic response is more naturally associated with local vibronic or electronic energy modulation, local TDM strength modulation, local short-axis or out-of-plane electronic-axis susceptibility, collective low-frequency motion, or a coordinate that jointly modulates the vertical energy and intrinsic transition axis.

## B. Computational Results:

### (I) Rh6G (Cation)

For Rh6G cation, the calculated responses reveal both coupled and comparatively selective intramolecular channels. Two closely spaced coordinates near 1600 cm$^{-1}$ involve xanthene core stretching/compression coupled to amino group motion, with different participation of the two amino substituted sides. Despite their similar frequencies and qualitative displacement character, their optical responses differ; one predominantly couples electronic gap and transition strength modulation, whereas the other couples gap modulation to a short axis transition deviation of approximately 0.51°. Their maximum absolute vertical gap changes are approximately 800 cm$^{-1}$ and 520 cm$^{-1}$, respectively. A collective core deformation coordinate near 780 cm$^{-1}$ produces concurrent gap, strength and axis responses, but its axis deviation remains approximately 0.26°. Rh6G also exhibits low frequency axis selective local responses with reported equilibrium axis deviations reaching approximately 0.7°. These normal coordinate responses must be distinguished from the constrained relaxed torsional results which incorporate relaxation of the remaining molecular coordinates. Within the sampled torsional domains, xanthene-amino torsion produces an approximately 1.33° equilibrium axis deviation with an out of plane response whereas xanthene-phenyl torsion produces and approximately 1.51° deviation with a short axis response. The latter also modulates the vertical gap over a span of approximately 810 cm$^{-1}$ and changes oscillator strength by a maximum to minimum ration of 1.09. In contrast the amino torsional family enhances in strength relative to the equilibrium reference but exhibits little strength variation within its sampled domain.

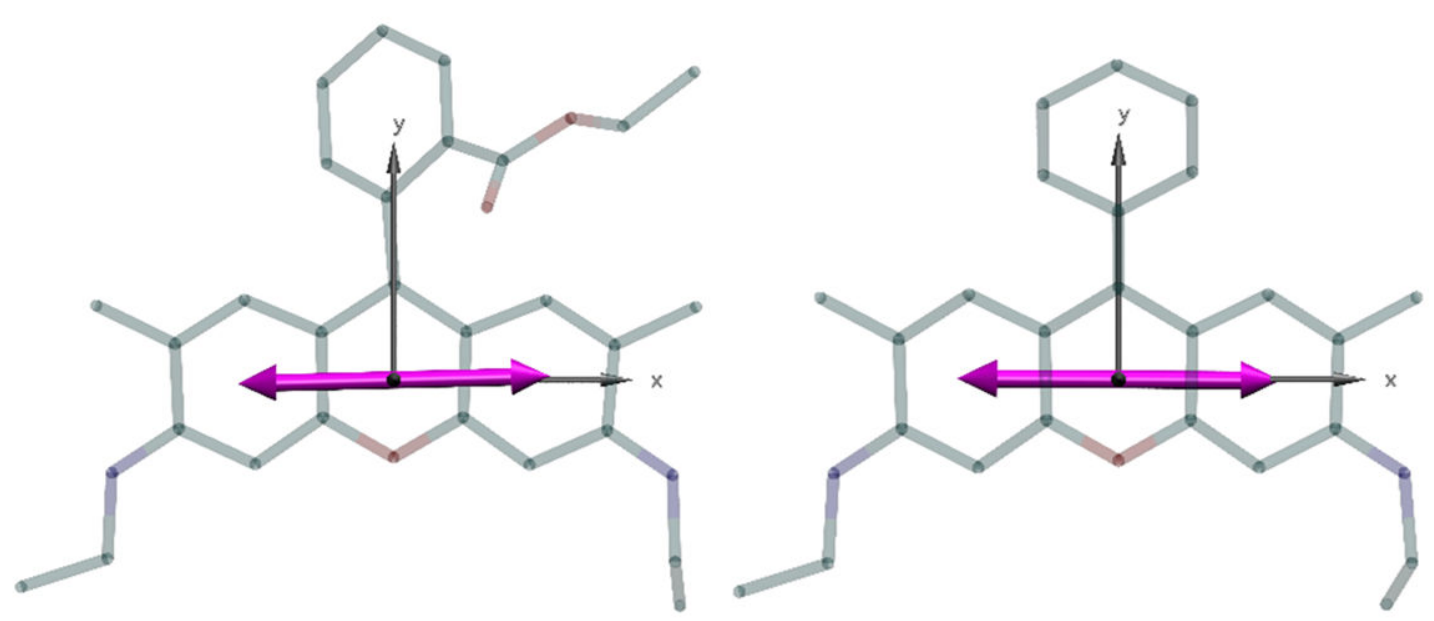

**Figure S31.** TDM direction does not change due to alteration in xanthene-phenyl dihedral angle.

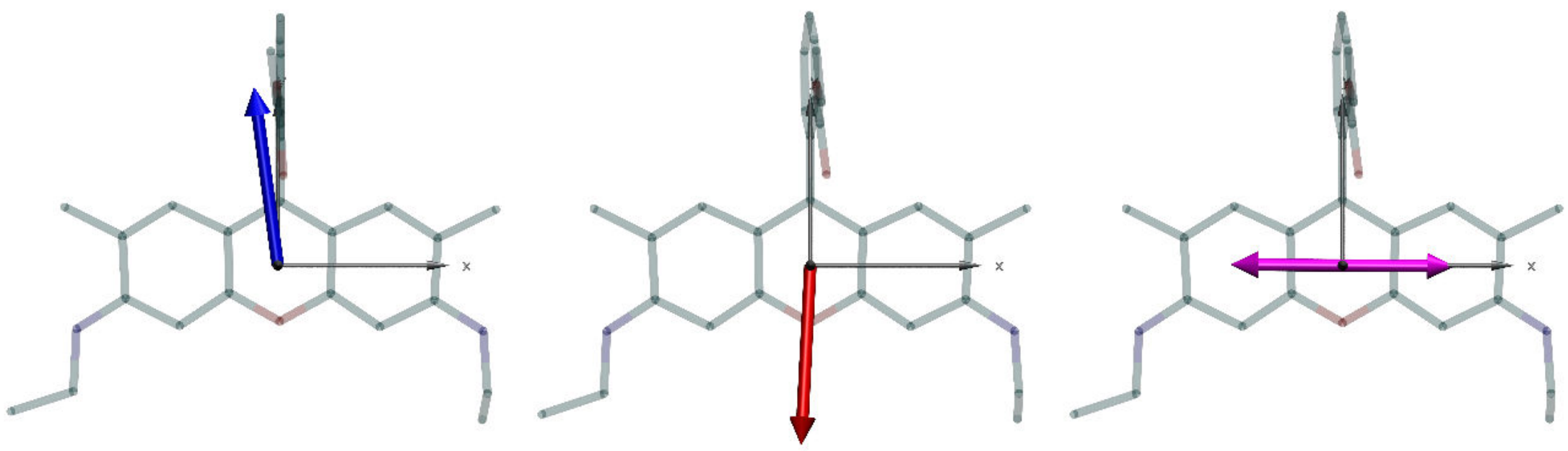

**Figure S32.** (Left and Centre) Dipole moment vectors for ground $(\boldsymbol{\mu}_{S_0})$ and first excited state $(\boldsymbol{\mu}_{S_1})$ of Rh6G cation. (Right) Emission Transition Dipole Moment vector $(\boldsymbol{\mu}_{\mathbf{em}})$.

**Table S2.** Modes with Huang-Rhys factors $(\mathbf{S_k})$ and reorganization energies $(\boldsymbol{\lambda}_\mathbf{k})$ ranked in increasing order of frequency. We list only modes that collectively account for at least 95% of the total FC contributions *(Inclusion threshold:* $C_S(T) \geq 0.95$ *and* $C_\lambda(T) \geq 0.95$*).*

| **k** | $\tilde{\nu}_\mathbf{k}(\mathbf{cm^{-1}})$ | $\mathbf{S_k}$ | $\boldsymbol{\lambda}_\mathbf{k}(\mathbf{meV})$ |
|---|---|---|---|
| 1 | 7.54 | 0.8822 | 0.8248 |
| 2 | 27.92 | 0.0129 | 0.0446 |
| 4 | 41.04 | 0.0067 | 0.0339 |
| 5 | 43.20 | 0.1381 | 0.7394 |
| 6 | 51.18 | 0.0022 | 0.0138 |
| 7 | 54.31 | 0.0278 | 0.1875 |
| 8 | 69.15 | 0.0181 | 0.1551 |
| 9 | 87.98 | 0.0090 | 0.0978 |
| 11 | 107.88 | 0.0012 | 0.0155 |
| 12 | 114.70 | 0.0546 | 0.7771 |
| 13 | 119.81 | 0.0107 | 0.1583 |
| 14 | 134.68 | 0.0098 | 0.1634 |
| 15 | 140.20 | 0.0080 | 0.1397 |
| 16 | 145.63 | 0.0466 | 0.8417 |
| 17 | 154.71 | 0.0040 | 0.0777 |
| 18 | 162.87 | 0.0135 | 0.2721 |
| 19 | 174.18 | 0.0038 | 0.0821 |
| 20 | 186.14 | 0.0683 | 1.5753 |
| 21 | 216.41 | 0.0079 | 0.2117 |
| 23 | 224.32 | 0.0029 | 0.0801 |
| 24 | 239.06 | 0.0916 | 2.7148 |
| 25 | 246.45 | 0.0305 | 0.9323 |
| 27 | 262.95 | 0.0040 | 0.1293 |
| 28 | 271.08 | 0.0121 | 0.4079 |
| 29 | 276.48 | 0.0010 | 0.0342 |
| 30 | 293.60 | 0.0141 | 0.5121 |
| 31 | 301.30 | 0.0585 | 2.1867 |
| 33 | 329.59 | 0.0116 | 0.4759 |
| 34 | 340.08 | 0.0071 | 0.2986 |
| 35 | 373.85 | 0.0540 | 2.5023 |
| 36 | 380.57 | 0.0016 | 0.0741 |
| 37 | 397.32 | 0.0010 | 0.0491 |
| 38 | 413.86 | 0.1121 | 5.7529 |
| 39 | 416.56 | 0.0037 | 0.1904 |
| 40 | 443.37 | 0.0087 | 0.4791 |
| 42 | 465.49 | 0.0027 | 0.1560 |
| 43 | 472.32 | 0.0015 | 0.0879 |
| 46 | 514.93 | 0.0235 | 1.5018 |
| 47 | 537.48 | 0.0127 | 0.8441 |
| 50 | 571.25 | 0.0033 | 0.2305 |
| 51 | 584.12 | 0.0007 | 0.0504 |
| 53 | 625.04 | 0.0416 | 3.2260 |
| 54 | 630.65 | 0.0006 | 0.0505 |
| 55 | 682.47 | 0.0016 | 0.1393 |
| 57 | 710.59 | 0.0004 | 0.0332 |
| 58 | 721.79 | 0.0064 | 0.5699 |
| 63 | 769.11 | 0.0031 | 0.2952 |
| 64 | 781.95 | 0.0309 | 2.9960 |
| 65 | 802.26 | 0.0015 | 0.1512 |
| 69 | 841.59 | 0.0006 | 0.0630 |
| 73 | 887.23 | 0.0004 | 0.0455 |
| 76 | 918.05 | 0.0009 | 0.1062 |
| 89 | 1083.73 | 0.0096 | 1.2939 |
| 90 | 1090.12 | 0.0007 | 0.1000 |
| 93 | 1123.52 | 0.0015 | 0.2083 |
| 95 | 1145.45 | 0.0050 | 0.7167 |
| 96 | 1150.14 | 0.0006 | 0.0878 |
| 98 | 1170.41 | 0.0004 | 0.0539 |
| 99 | 1176.88 | 0.0003 | 0.0389 |
| 103 | 1206.63 | 0.0170 | 2.5453 |
| 104 | 1234.44 | 0.0064 | 0.9730 |
| 108 | 1289.78 | 0.0015 | 0.2374 |
| 113 | 1321.87 | 0.0021 | 0.3397 |
| 114 | 1342.72 | 0.0049 | 0.8164 |
| 116 | 1362.25 | 0.0008 | 0.1430 |
| 117 | 1373.64 | 0.0050 | 0.8586 |
| 118 | 1379.33 | 0.0098 | 1.6780 |
| 119 | 1382.26 | 0.0049 | 0.8348 |
| 121 | 1399.72 | 0.0004 | 0.0636 |
| 122 | 1403.88 | 0.0005 | 0.0816 |
| 123 | 1410.22 | 0.0017 | 0.3049 |
| 124 | 1416.64 | 0.0050 | 0.8731 |
| 126 | 1435.68 | 0.0021 | 0.3723 |
| 132 | 1450.44 | 0.0003 | 0.0476 |
| 134 | 1452.94 | 0.0002 | 0.0381 |
| 139 | 1484.97 | 0.0003 | 0.0641 |
| 141 | 1493.63 | 0.0003 | 0.0573 |
| 142 | 1496.86 | 0.0005 | 0.0993 |
| 145 | 1544.89 | 0.0003 | 0.0557 |
| 146 | 1550.99 | 0.0002 | 0.0453 |
| 147 | 1570.83 | 0.0073 | 1.4151 |
| 148 | 1600.90 | 0.0308 | 6.1225 |
| 149 | 1603.64 | 0.0138 | 2.7385 |
| 150 | 1612.36 | 0.0105 | 2.0987 |
| 151 | 1659.80 | 0.0016 | 0.3394 |
| 152 | 1691.18 | 0.0058 | 1.2167 |
| 153 | 1718.70 | 0.0067 | 1.4318 |
| 155 | 1806.72 | 0.0004 | 0.0891 |
| 159 | 3059.28 | 0.0001 | 0.0516 |
| 161 | 3063.79 | 0.0003 | 0.1281 |
| 163 | 3086.03 | 0.0005 | 0.1928 |
| 164 | 3095.80 | 0.0003 | 0.1182 |
| 178 | 3218.72 | 0.0009 | 0.3618 |
| 181 | 3228.35 | 0.0007 | 0.2979 |
| 183 | 3250.98 | 0.0004 | 0.1470 |
| 184 | 3251.00 | 0.0003 | 0.1317 |
| 186 | 3606.08 | 0.0007 | 0.2921 |

**Table S3.** Low and High frequency normal mode finite amplitude contributions to maximum changes in oscillator strength ($\mathbf{\Delta f_{rel}}$) and transition dipole magnitude ($\mathbf{\Delta \mu_{rel}}$). Here, modes are ranked in decreasing order of oscillator strength modulation *(Inclusion thresholds:* $\Delta f_{rel} \geq 0.02$ *or* $\Delta\mu_{rel} \geq 0.02$ *,* $\Delta\tilde{\nu} < 500\ cm^{-1}$ *and* $\theta_{axis} < 0.25^o$*).*

| **k** | $\mathbf{\tilde{\nu}_k (cm^{-1})}$ | $\max_Q \lvert\Delta f_{rel}(Q)\rvert$ | $\max_Q \lvert\Delta \mu_{rel}(Q)\rvert$ |
|---|---|---|---|
| 153 | 1718.70 | 0.0569 | 0.0305 |
| 53 | 625.04 | 0.0347 | 0.0070 |
| 147 | 1570.83 | 0.0246 | 0.0154 |
| 31 | 301.30 | 0.0240 | 0.0136 |
| 46 | 514.93 | 0.0204 | 0.0032 |

**Table S4.** Low and High frequency normal mode finite amplitude contributions to maximum transition dipole axis deviations ($\mathbf{\theta_{axis}}$), along with dominant directional responses and oscillator strength changes ($\mathbf{\Delta f_{rel}}$). Here, modes are ranked in decreasing order of axis deviation *(Inclusion thresholds:* $\theta_{axis} \geq 0.25^o$ *and* $\Delta\tilde{\nu} < 500\ cm^{-1}$*, Mode 8 additionally meets strength modulation* $\Delta f_{rel}) \geq 0.02$ *or* $\Delta\mu_{rel} \geq 0.02$*).*

| **k** | $\mathbf{\tilde{\nu}_k (cm^{-1})}$ | $\max_Q \lvert\Delta f_{rel}(Q)\rvert$ | $\max_Q \theta_{axis}(Q)$ | **Dominant Direction** |
|---|---|---|---|---|
| 4 | 41.04 | 0.0123 | 0.7058 | mixed |
| 16 | 145.63 | 0.0175 | 0.6989 | short |
| 8 | 69.15 | 0.0293 | 0.3328 | short |
| 17 | 154.71 | 0.0084 | 0.3254 | short |
| 30 | 293.60 | 0.0048 | 0.3018 | short |
| 23 | 224.32 | 0.0088 | 0.3001 | mixed |
| 116 | 1362.25 | 0.0032 | 0.2657 | short |

**Table S5.** Modes with large vertical electronic gap, along with their contributions to changes in oscillator strength ($\mathbf{\Delta f_{rel}}$) transition dipole moment ($\mathbf{\Delta \mu_{rel}}$) and transition dipole axis deviations ($\mathbf{\theta_{axis}}$), with dominant directional responses. Here, modes are ranked in decreasing order of oscillator strength modulation *(Inclusion thresholds:* $\Delta\tilde{\nu} \geq 500\ cm^{-1}$*,* $\Delta f_{rel} \geq 0.02$ *or* $\Delta\mu_{rel} \geq 0.02$ *, and* $\theta_{axis} \geq 0.25^o$*).*

| **k** | $\mathbf{\tilde{\nu}_k (cm^{-1})}$ | $\max_Q \lvert\Delta \tilde{\nu}(Q)\rvert$ $\mathbf{(cm^{-1})}$ | $\max_Q \lvert\Delta f_{rel}(Q)\rvert$ | $\max_Q \lvert\Delta \mu_{rel}(Q)\rvert$ | $\max_Q \theta_{axis}(Q)$ | **Dominant Direction** |
|---|---|---|---|---|---|---|
| 64 | 781.95 | 560.55 | 0.0217 | 0.0039 | 0.2592 | short |
| 148 | 1600.90 | 796.06 | 0.0215 | 0.0139 | <0.25 | below threshold |
| 149 | 1603.64 | 517.00 | 0.0114 | 0.0088 | 0.5062 | short |

**Table S6.** Torsional modes giving significant oscillator strength enhancement $(\mathbf{f(\varphi)/f_e})$ and/or changes in transition dipole axis deviations $(\boldsymbol{\theta_{axis}})$ with dominant directional responses. Here, modes are ranked in decreasing transition dipole axis deviations with respect to reference geometry axis *(Inclusion thresholds:* $\theta_{axis} \geq 1.0^o$, $\max_{\varphi \in D} f(\varphi) / \min_{\varphi \in D} f(\varphi) \geq 1.05$*, enhancement/suppression relative to 1.0* $(f(\varphi)/f_e)$*).*

| $\boldsymbol{\phi_D}$ | $\boldsymbol{\Delta\phi_D}$**(°)** | $\mathbf{\frac{f(\varphi)}{f_e}}$ | $\mathbf{\frac{\max_{\varphi \in D} f(\varphi)}{\min_{\varphi \in D} f(\varphi)}}$ | $\mathbf{\max_{Q} \theta_{axis,e}}$ | **Espan**$^{D}$ | **max**$\boldsymbol{\varphi_i, \varphi_j}$ | **Dominant Direction** |
|---|---|---|---|---|---|---|---|
| xanthene-phenyl | -40 to +40 | 1.1904 to 1.2987 | 1.0910 | 1.5143 | 805.74 | 1.6586 | short |
| xanthene-amino | -30 to +30 | 1.2878 to 1.2980 | 1.0079 | 1.3332 | 128.24 | 1.5759 | oop |
| phenyl-carboxyl | -40 to +40 | 1.2628 to 1.2994 | 1.0290 | 0.4860 | 284.71 | 0.6411 | mixed (below threshold) |
| amino-ethyl | -20 to +40 | 1.2898 to 1.3178 | 1.0217 | 0.4769 | 119.37 | 0.5450 | mixed (below threshold) |

## (II) RhB (Cation)

For RhB cation, a xanthene core stretching/compression coordinate near 1600 $cm^{-1}$ similarly produces pronounces gap modulation, reaching approximately 1290 $cm^{-1}$ together with transition strength changes. Collective low frequency coordinates in the approximately 30-80 $cm^{-1}$ region described qualitatively as butterfly like framework and ring breathing like motion coupled to amino-ethyl substituent, active coupled optical responses but produce sub-degree axis deviations. Among the sampled relaxed torsions, xanthene-phenyl showed 1.27° deviation and a ratio of 1.09. The amino-ethyl and phenyl-carboxyl torsions remain axis subthreshold within their sampled domains, although their continuous responses remain informative.

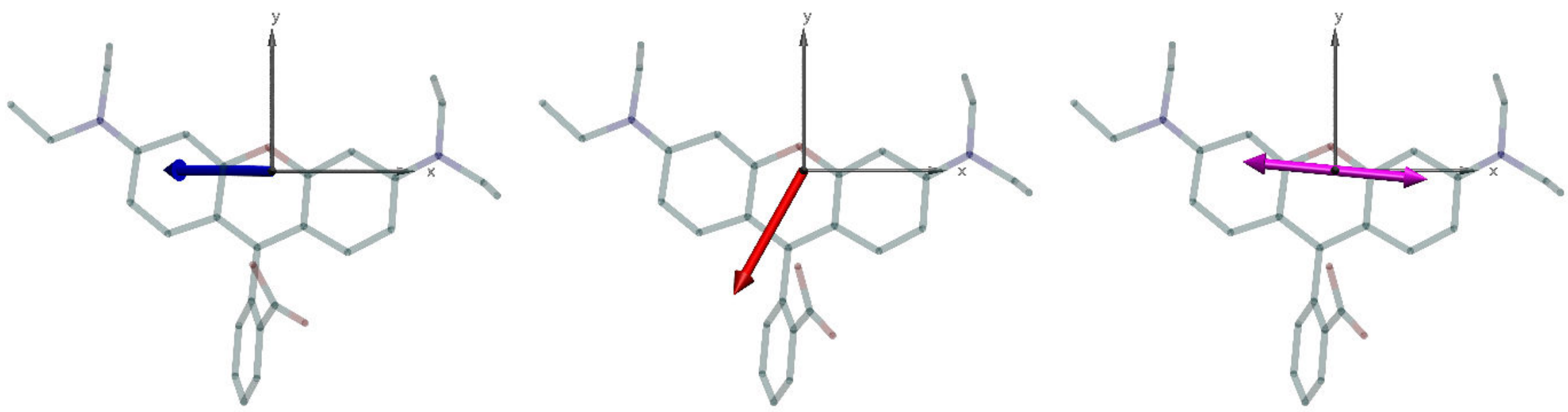

**Figure S33.** (Left and Centre) Dipole moment vectors for ground ($\boldsymbol{\mu}_{S_0}$) and first excited state ($\boldsymbol{\mu}_{S_1}$) for RhB cation. (Right) Emission Transition Dipole Moment vector ($\boldsymbol{\mu}_{em}$)

**Table S7.** Modes with Huang-Rhys factors ($\mathbf{S_k}$) and reorganization energies ($\boldsymbol{\lambda}_k$) ranked in increasing order of frequency. We list only modes that collectively account for at least 95% of the total FC contributions *(Inclusion threshold:* $C_S(T) \geq 0.95$ *and* $C_\lambda(T) \geq 0.95$*).*

| $\mathbf{k}$ | $\tilde{\nu}_k(cm^{-1})$ | $\mathbf{S_k}$ | $\lambda_k(meV)$ |
|---|---|---|---|
| 1 | 24.85 | 0.2340 | 0.7209 |
| 2 | 25.02 | 0.0418 | 0.1298 |
| 3 | 33.32 | 0.3457 | 1.4283 |
| 4 | 43.33 | 0.8883 | 4.7720 |
| 5 | 53.66 | 0.2610 | 1.7362 |
| 6 | 64.25 | 0.0807 | 0.6425 |
| 7 | 67.87 | 1.1116 | 9.3537 |
| 8 | 73.61 | 0.0663 | 0.6047 |
| 9 | 79.99 | 1.7311 | 17.1678 |
| 10 | 92.53 | 1.0464 | 12.0051 |
| 11 | 104.44 | 0.4687 | 6.0696 |
| 12 | 119.51 | 0.0160 | 0.2376 |
| 14 | 136.01 | 0.0174 | 0.2933 |
| 15 | 142.13 | 0.0092 | 0.1621 |
| 17 | 169.49 | 0.0050 | 0.1057 |
| 19 | 207.50 | 0.0637 | 1.6401 |
| 21 | 225.18 | 0.0661 | 1.8446 |
| 23 | 244.70 | 0.0386 | 1.1698 |
| 25 | 262.94 | 0.0309 | 1.0059 |
| 26 | 268.66 | 0.0158 | 0.5257 |
| 28 | 310.91 | 0.0044 | 0.1705 |
| 30 | 317.55 | 0.0105 | 0.4152 |
| 33 | 392.27 | 0.0041 | 0.1985 |
| 34 | 393.86 | 0.0062 | 0.3034 |
| 35 | 402.64 | 0.0183 | 0.9138 |
| 36 | 419.60 | 0.0123 | 0.6390 |
| 37 | 436.39 | 0.0079 | 0.4279 |
| 38 | 455.95 | 0.0112 | 0.6327 |
| 40 | 487.61 | 0.0028 | 0.1680 |
| 41 | 506.31 | 0.0100 | 0.6263 |
| 43 | 517.04 | 0.0319 | 2.0428 |
| 44 | 523.97 | 0.0558 | 3.6268 |
| 45 | 534.71 | 0.0083 | 0.5531 |
| 47 | 574.23 | 0.1793 | 12.7643 |
| 48 | 596.34 | 0.0241 | 1.7842 |
| 49 | 600.22 | 0.0065 | 0.4852 |
| 50 | 608.32 | 0.0719 | 5.4195 |
| 51 | 610.43 | 0.0113 | 0.8516 |
| 52 | 639.27 | 0.0271 | 2.1465 |
| 55 | 667.58 | 0.0641 | 5.3042 |
| 57 | 714.36 | 0.0618 | 5.4724 |
| 63 | 793.42 | 0.0032 | 0.3119 |
| 64 | 804.00 | 0.0123 | 1.2268 |
| 66 | 815.19 | 0.1366 | 13.8026 |
| 68 | 837.79 | 0.0019 | 0.1923 |
| 70 | 842.96 | 0.0065 | 0.6837 |
| 71 | 846.44 | 0.0327 | 3.4365 |
| 85 | 1046.22 | 0.0017 | 0.2232 |
| 86 | 1057.19 | 0.0018 | 0.2341 |
| 88 | 1079.89 | 0.0012 | 0.1546 |
| 92 | 1116.18 | 0.0456 | 6.3147 |
| 94 | 1118.25 | 0.0022 | 0.3100 |
| 96 | 1149.63 | 0.0014 | 0.1955 |
| 97 | 1154.85 | 0.1415 | 20.2643 |
| 98 | 1163.45 | 0.0115 | 1.6551 |
| 102 | 1216.00 | 0.1190 | 17.9471 |
| 104 | 1226.46 | 0.0011 | 0.1637 |
| 105 | 1228.72 | 0.0048 | 0.7254 |
| 109 | 1285.43 | 0.0023 | 0.3691 |
| 110 | 1302.06 | 0.0019 | 0.3129 |
| 111 | 1321.26 | 0.0027 | 0.4458 |
| 112 | 1329.63 | 0.0018 | 0.2923 |
| 113 | 1332.44 | 0.0032 | 0.5301 |
| 114 | 1342.62 | 0.0083 | 1.3738 |
| 116 | 1363.29 | 0.0113 | 1.9153 |
| 117 | 1370.06 | 0.0039 | 0.6614 |
| 118 | 1375.14 | 0.0018 | 0.2996 |
| 119 | 1379.41 | 0.0106 | 1.8166 |
| 120 | 1386.17 | 0.0014 | 0.2385 |
| 121 | 1390.70 | 0.0110 | 1.9042 |
| 122 | 1394.39 | 0.0026 | 0.4492 |
| 137 | 1470.10 | 0.0017 | 0.3131 |
| 139 | 1491.33 | 0.0010 | 0.1807 |
| 145 | 1534.02 | 0.0010 | 0.1930 |
| 147 | 1549.29 | 0.0021 | 0.3972 |
| 149 | 1571.58 | 0.0012 | 0.2265 |
| 150 | 1601.35 | 0.0471 | 9.3446 |
| 155 | 1835.03 | 0.1205 | 27.4168 |
| 181 | 3237.65 | 0.0005 | 0.2100 |
| 186 | 3731.97 | 0.1503 | 69.5434 |

**Table S8.** Low and High frequency normal mode finite amplitude contributions to maximum transition dipole axis deviations $(\boldsymbol{\theta_{axis}})$, along with dominant directional responses and oscillator strength changes $(\boldsymbol{\Delta f_{rel}})$. Here, modes are ranked in decreasing order of axis deviation *(Inclusion thresholds:* $\theta_{axis} \geq 0.25^{o}$ *and* $\Delta\tilde{\nu} < 500\ cm^{-1}$, *Mode 43 additionally meets strength modulation* $\Delta f_{rel}) \geq 0.02$ *or* $\Delta\mu_{rel} \geq 0.02$*).*

| k | $\tilde{\nu}_k(\mathbf{cm^{-1}})$ | $\max_Q \lvert\Delta f_{rel}(Q)\rvert$ | $\max_Q \theta_{axis}(Q)$ | **Dominant Direction** |
|---|---|---|---|---|
| 38 | 455.95 | 0.0170 | 0.0032 | short |
| 50 | 608.32 | 0.0136 | 0.0040 | short |
| 36 | 419.60 | 0.0076 | 0.0066 | short |
| 43 | 517.04 | 0.0229 | 0.0018 | short |

**Table S9.** Modes with large vertical electronic gap, along with their contributions to changes in oscillator strength $(\boldsymbol{\Delta f_{rel}})$, transition dipole moment $(\boldsymbol{\Delta\mu_{rel}})$ and transition dipole axis deviations $(\boldsymbol{\theta_{axis}})$, with dominant directional responses. Here, modes are ranked in decreasing order of oscillator strength modulation *(Inclusion thresholds:* $\Delta\tilde{\nu} \geq 500\ cm^{-1}, \Delta f_{rel} \geq 0.02$ *or* $\Delta\mu_{rel} \geq 0.02$ *, and* $\theta_{axis} \geq 0.25^{o}$*).*

| k | $\tilde{\nu}_k(\mathbf{cm^{-1}})$ | $\max_Q \lvert\Delta\tilde{\nu}(Q)\rvert$ $(\mathbf{cm^{-1}})$ | $\max_Q \lvert\Delta f_{rel}(Q)\rvert$ | $\max_Q \lvert\Delta\mu_{rel}(Q)\rvert$ | $\max_Q \theta_{axis}(Q)$ | **Dominant Direction** |
|---|---|---|---|---|---|---|
| 150 | 1601.35 | 1290.48 | 0.0281 | 0.0298 | 0.1104 | below threshold |
| 8 | 73.61 | 571.84 | 0.0259 | 0.0120 | 0.6092 | short |
| 3 | 33.32 | 994.48 | 0.0032 | 0.0230 | 0.3455 | short |

**Table S10.** Torsional modes giving significant oscillator strength enhancement $(f(\varphi)/f_e)$ and/or changes in transition dipole axis deviations $(\theta_{axis})$ with dominant directional responses. Here, modes are ranked in decreasing transition dipole axis deviations with respect to reference geometry axis *(Inclusion thresholds:* $\theta_{axis} \geq 1.0^{o}$, $\max_{\varphi\in D} f(\varphi)/\min_{\varphi\in D} f(\varphi) \geq 1.05$, *enhancement/suppression relative to 1.0* $(f(\varphi)/f_e)$*).*

| $\boldsymbol{\phi_D}$ | $\boldsymbol{\Delta\phi_D}$(°) | $\frac{f(\varphi)}{f_e}$ | $\frac{\max_{\varphi\in D} f(\varphi)}{\min_{\varphi\in D} f(\varphi)}$ | $\max_Q \theta_{axis}$ | **Espan**[D] | **max**$\boldsymbol{\varphi_i, \varphi_j}$ | **Dominant Direction** |
|---|---|---|---|---|---|---|---|
| Xanthene–phenyl | -30 to +60 | 0.9362–1.0192 | 1.0887 | 1.2667 | 491.19 | 1.2862 | short |
| Phenyl–carboxyl | -40 to +40 | 0.9980–1.0174 | 1.0194 | 0.8374 | 1038.03 | 0.8531 | below threshold |
| Left xanthene–amino | -30 to +40 | 0.9970–1.0054 | 1.0084 | 0.7339 | 232.28 | 1.2460 | below threshold |
| Right xanthene–amino | -40 to +30 | 0.9931–1.0010 | 1.0079 | 0.4560 | 172.60 | 0.8843 | below threshold |
| Left N-ethyl | -20 to +20 | 0.9916–1.0153 | 1.0239 | 0.4329 | 11.29 | 0.6642 | below threshold |
| Right N-ethyl | -30 to +50 | 0.9964–1.0156 | 1.0192 | 0.4127 | 87.10 | 0.7670 | below threshold |

### (III) RhB (Zwitterion)

The RhB zwitterion displays a different response pattern. Collective low frequency rocking and flapping coordinates in the approximately 30-70 $cm^{-1}$ region jointly modulate the gap, strength and transition axis, with short axis deviations approaching 1.9°. Its relaxed torsional families also show more pronounced sampled responses than those reported for the cation; equilibrium axis deviation span approximately 2.1 – 3.1°, accompanied by strength modulation rations of approximately 1.06 – 1.10. The xanthene-phenyl family exhibits a maximum pairwise axis separation of approximately 4.45°. These results establish charge state dependent intrinsic response patterns within RhB, without implying that the zwitterion is responsible for observations attributed experimentally to the cation.

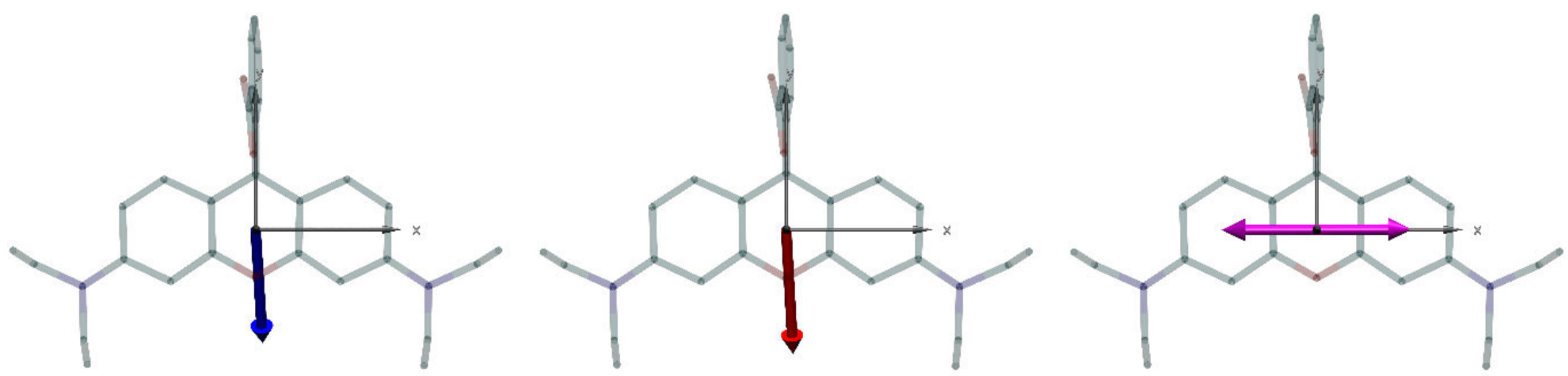

**Figure S34.** (Left and Centre) Dipole moment vectors for ground ($\boldsymbol{\mu_{S_0}}$) and first excited state ($\boldsymbol{\mu_{S_1}}$) for RhB zwitterion. (Right) Emission Transition Dipole Moment vector ($\boldsymbol{\mu_{em}}$).

**Table S11.** Modes with Huang-Rhys factors **($S_k$)** and reorganization energies **($\lambda_k$)** ranked in increasing order of frequency. We list only modes that collectively account for at least 95% of the total FC contributions *(Inclusion threshold: $C_S(T) \geq 0.95$ and $C_\lambda(T) \geq 0.95$).*

| k | $\tilde{\nu}_k(cm^{-1})$ | $S_k$ | $\lambda_k(meV)$ |
|---|---|---|---|
| 1 | 19.67 | 0.0353 | 0.0861 |
| 2 | 31.83 | 0.1156 | 0.4562 |
| 3 | 34.69 | 0.0067 | 0.0290 |
| 4 | 42.91 | 0.0128 | 0.0681 |
| 5 | 46.34 | 0.0018 | 0.0104 |
| 6 | 57.24 | 0.0438 | 0.3113 |
| 7 | 61.84 | 0.0098 | 0.0753 |
| 8 | 65.64 | 0.0317 | 0.2583 |
| 9 | 74.99 | 0.0867 | 0.8064 |
| 10 | 88.90 | 0.0180 | 0.1993 |
| 11 | 94.30 | 0.0210 | 0.2460 |
| 12 | 105.54 | 0.0239 | 0.3128 |
| 13 | 113.52 | 0.0394 | 0.5546 |
| 14 | 132.17 | 0.0115 | 0.1899 |
| 15 | 137.52 | 0.0203 | 0.3469 |
| 16 | 139.73 | 0.0011 | 0.0201 |
| 17 | 154.04 | 0.1737 | 3.3180 |
| 18 | 189.44 | 0.1075 | 2.5258 |
| 19 | 193.97 | 0.2547 | 6.1267 |
| 20 | 215.61 | 0.0526 | 1.4071 |
| 21 | 218.56 | 0.0034 | 0.0923 |
| 22 | 227.16 | 0.0326 | 0.9198 |
| 23 | 240.04 | 0.1354 | 4.0323 |
| 24 | 240.85 | 0.1084 | 3.2398 |
| 25 | 254.23 | 0.0168 | 0.5310 |
| 26 | 262.46 | 0.0294 | 0.9599 |
| 28 | 278.59 | 0.0899 | 3.1062 |
| 29 | 306.60 | 0.0195 | 0.7430 |
| 32 | 350.43 | 0.0357 | 1.5540 |
| 33 | 354.96 | 0.0204 | 0.8984 |
| 34 | 395.38 | 0.0999 | 4.8978 |
| 35 | 397.65 | 0.0015 | 0.0769 |
| 37 | 440.21 | 0.0121 | 0.6657 |
| 39 | 465.84 | 0.0033 | 0.1910 |
| 40 | 466.49 | 0.0114 | 0.6597 |
| 41 | 496.50 | 0.0248 | 1.5318 |
| 42 | 510.09 | 0.0006 | 0.0411 |
| 43 | 517.26 | 0.0029 | 0.1875 |
| 45 | 569.48 | 0.0097 | 0.6907 |
| 46 | 571.32 | 0.0009 | 0.0705 |
| 47 | 593.12 | 0.0021 | 0.1581 |
| 49 | 597.31 | 0.0125 | 0.9326 |
| 51 | 638.49 | 0.0370 | 2.9315 |
| 53 | 658.48 | 0.0009 | 0.0777 |
| 56 | 723.65 | 0.0079 | 0.7141 |
| 59 | 749.50 | 0.0210 | 1.9519 |
| 60 | 765.45 | 0.0050 | 0.4782 |
| 62 | 799.09 | 0.0025 | 0.2571 |
| 63 | 802.29 | 0.0003 | 0.0369 |
| 64 | 803.10 | 0.0003 | 0.0392 |
| 66 | 826.42 | 0.0005 | 0.0524 |
| 67 | 832.59 | 0.0029 | 0.3014 |
| 68 | 844.29 | 0.0069 | 0.7308 |
| 75 | 949.46 | 0.0005 | 0.0650 |
| 77 | 957.06 | 0.0004 | 0.0549 |
| 78 | 960.78 | 0.0004 | 0.0483 |
| 79 | 971.22 | 0.0011 | 0.1357 |
| 84 | 1048.15 | 0.0011 | 0.1507 |
| 85 | 1050.73 | 0.0006 | 0.0867 |
| 88 | 1095.10 | 0.0004 | 0.0656 |
| 89 | 1098.30 | 0.0011 | 0.1629 |
| 92 | 1107.44 | 0.0012 | 0.1719 |
| 95 | 1151.39 | 0.0033 | 0.4742 |
| 97 | 1160.06 | 0.0005 | 0.0805 |
| 101 | 1222.07 | 0.0010 | 0.1569 |
| 102 | 1224.85 | 0.0015 | 0.2385 |
| 103 | 1225.68 | 0.0008 | 0.1237 |
| 104 | 1234.82 | 0.0014 | 0.2296 |
| 108 | 1297.81 | 0.0002 | 0.0394 |
| 109 | 1311.69 | 0.0007 | 0.1196 |
| 110 | 1316.08 | 0.0040 | 0.6577 |
| 111 | 1327.92 | 0.0013 | 0.2289 |
| 113 | 1338.80 | 0.0002 | 0.0470 |
| 114 | 1371.05 | 0.0025 | 0.4253 |
| 115 | 1374.02 | 0.0028 | 0.4823 |
| 116 | 1380.10 | 0.0010 | 0.1845 |
| 119 | 1391.36 | 0.0172 | 2.9818 |
| 121 | 1398.89 | 0.0005 | 0.0893 |
| 122 | 1401.70 | 0.0005 | 0.0929 |
| 125 | 1416.25 | 0.0005 | 0.0894 |
| 126 | 1442.61 | 0.0014 | 0.2661 |
| 131 | 1450.89 | 0.0004 | 0.0820 |
| 135 | 1472.26 | 0.0003 | 0.0549 |
| 137 | 1486.74 | 0.0007 | 0.1394 |
| 140 | 1502.90 | 0.0006 | 0.1230 |
| 142 | 1531.39 | 0.0002 | 0.0567 |
| 145 | 1551.88 | 0.0017 | 0.3275 |
| 147 | 1579.46 | 0.0078 | 1.5391 |
| 148 | 1601.29 | 0.0326 | 6.4904 |
| 149 | 1654.92 | 0.0002 | 0.0455 |
| 150 | 1685.47 | 0.0004 | 0.0841 |
| 151 | 1711.78 | 0.0088 | 1.8864 |
| 152 | 1728.90 | 0.0021 | 0.4565 |

**Table S12.** Low and High frequency normal mode finite amplitude contributions to maximum changes in oscillator strength **($\Delta f_{rel}$)** and transition dipole magnitude **($\Delta\mu_{rel}$).** Here, modes are ranked in decreasing order of oscillator strength modulation *(Inclusion thresholds:* $\Delta f_{rel} \geq 0.02$ *or* $\Delta\mu_{rel} \geq 0.02$ , $\Delta\tilde{\nu} < 500\ cm^{-1}$ *and* $\theta_{axis} < 0.25^{o}$*).*

| **k** | $\tilde{\nu}_k(\mathbf{cm^{-1}})$ | $\max_{Q}\lvert\Delta f_{rel}(Q)\rvert$ | $\max_{Q}\lvert\Delta\mu_{rel}(Q)\rvert$ |
|---|---|---|---|
| 151 | 1711.7800 | 0.1438 | 0.0765 |
| 8 | 65.6400 | 0.0480 | 0.0216 |
| 17 | 154.0400 | 0.0463 | 0.0205 |
| 51 | 638.4900 | 0.0240 | 0.0013 |
| 147 | 1579.4600 | 0.0233 | 0.0189 |

**Table S13.** Modes with large vertical electronic gap, along with their contributions to changes in oscillator strength **($\Delta f_{rel}$),** transition dipole moment **($\Delta\mu_{rel}$)** and transition dipole axis deviations **($\theta_{axis}$)**, with dominant directional responses. Here, modes are ranked in decreasing order of oscillator strength modulation *(Inclusion thresholds:* $\Delta\tilde{\nu} \geq 500\ cm^{-1}$, $\Delta f_{rel} \geq 0.02$ *or* $\Delta\mu_{rel} \geq 0.02$ , *and* $\theta_{axis} \geq 0.25^{o}$*).*

| **k** | $\tilde{\nu}_k(\mathbf{cm^{-1}})$ | $\max_{Q}\lvert\Delta\tilde{\nu}(Q)\rvert$ $(\mathbf{cm^{-1}})$ | $\max_{Q}\lvert\Delta f_{rel}(Q)\rvert$ | $\max_{Q}\lvert\Delta\mu_{rel}(Q)\rvert$ | $\max_{Q}\theta_{axis}(Q)$ | **Dominant Direction** |
|---|---|---|---|---|---|---|
| 7 | 61.8400 | 716.22 | 0.0575 | 0.0135 | 1.9341 | short |
| 3 | 34.6900 | 784.77 | 0.0545 | 0.0104 | 1.3926 | short |
| 148 | 1601.2900 | 1090.46 | 0.0379 | 0.0194 | 0.0516 | below threshold |
| 49 | 597.3100 | 658.14 | 0.0338 | 0.0136 | 0.0953 | below threshold |
| 2 | 31.8300 | 972.70 | 0.0238 | 0.0098 | 0.8799 | short |
| 34 | 395.3800 | 561.36 | 0.0152 | 0.0106 | 0.0525 | below threshold |
| 9 | 74.9900 | 672.66 | 0.0133 | 0.0083 | 0.2899 | oop |

**Table S14.** Torsional modes giving significant oscillator strength enhancement ($f(\varphi)/f_e$) and/or changes in transition dipole axis deviations ($\theta_{axis}$) with dominant directional responses. Here, modes are ranked in decreasing transition dipole axis deviations with respect to reference geometry axis *(Inclusion thresholds:* $\theta_{axis} \geq 1.0^o$, $\max_{\varphi \in D} f(\varphi) / \min_{\varphi \in D} f(\varphi) \geq 1.05$*, enhancement/suppression relative to 1.0* $(f(\varphi)/f_e)$*).*

| $\phi_D$ | $\Delta\phi_D$(°) | $\frac{f(\varphi)}{f_e}$ | $\frac{\max_{\varphi \in D} f(\varphi)}{\min_{\varphi \in D} f(\varphi)}$ | $\max_Q \theta_{axis}$ | $Espan^D$ | $\max \varphi_i, \varphi_j$ | Dominant Direction |
|---|---|---|---|---|---|---|---|
| Xanthene-amino | -40 to +40 | 0.9352–0.9999 | 1.0692 | 3.0855 | 555.71 | 3.2072 | short |
| N-ethyl | -50 to +40 | 0.9329–1.0087 | 1.0813 | 2.5593 | 496.83 | 2.5274 | short |
| Xanthene-phenyl | -50 to +60 | 0.8920–0.9811 | 1.0999 | 2.3542 | 901.72 | 4.4537 | short |
| Phenyl-carboxylate | -60 to +70 | 0.9415–1.0010 | 1.0632 | 2.1365 | 671.85 | 2.3457 | short |

## (IV) Atto 520 (Cation)

The Atto520 cation exhibits distinct spectral sensitivity from soft coordinate axis sensitivity. A core stretching/compression coordinate near 1600 $cm^{-1}$ coupled to peripheral hydrogen and amino group motion, produces a maximum absolute gap change of approximately 970 $cm^{-1}$ together with strength modulation. A soft flapping coordinate near 40 $cm^{-1}$ instead gives an out of plane axis response of approximately 1.19$^{o}$ with mixed directional character and strength ratios close to unity. The sampled xanthene-pendant torsion modulates strength and the vertical gap but remains below the equilibrium axis threshold, illustrating that conformationally driven spectral changes need not entail substantial intrinsic reorientations.

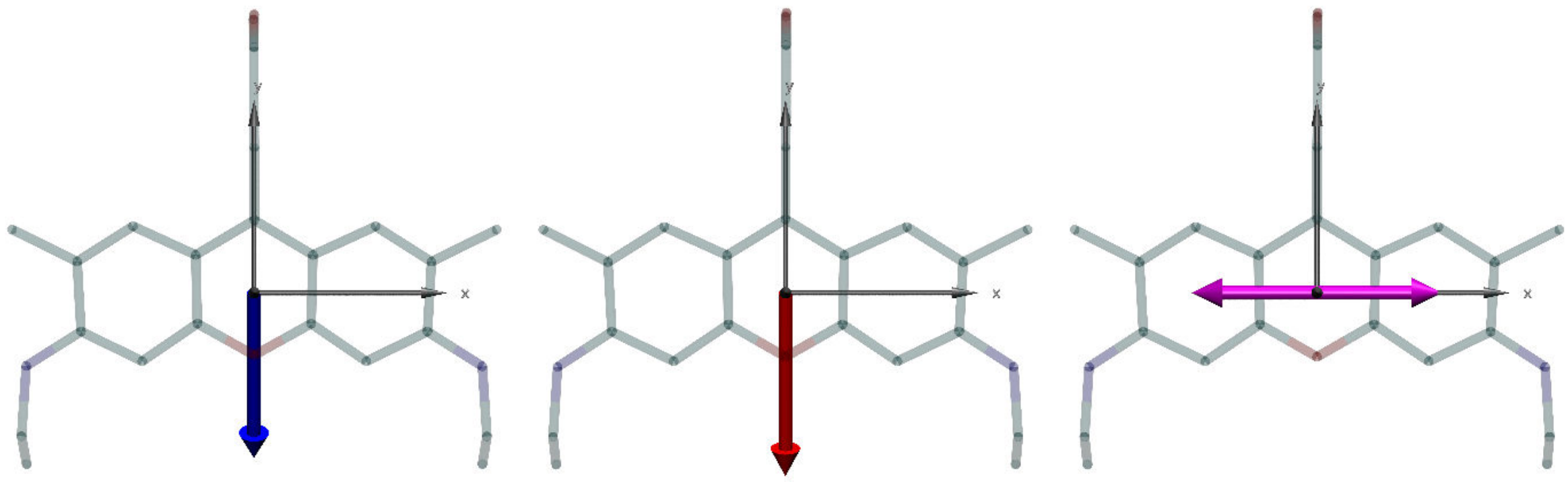

**Figure S35.** (Left and Centre) Dipole moment vectors for ground ($\boldsymbol{\mu}_{\mathbf{S_0}}$) and first excited state ($\boldsymbol{\mu}_{\mathbf{S_1}}$) of Atto 520 cation. (Right) Emission Transition Dipole Moment vector ($\boldsymbol{\mu}_{\mathbf{em}}$).

**Table S15.** Modes with Huang-Rhys factors $(\mathbf{S_k})$ and reorganization energies $(\boldsymbol{\lambda}_{\mathbf{k}})$ ranked in increasing order of frequency. We list only modes that collectively account for at least 95% of the total FC contributions *(Inclusion threshold:* $C_S(T) \geq 0.95$ *and* $C_\lambda(T) \geq 0.95$*).*

| $k$ | $\tilde{\nu}_k(cm^{-1})$ | $S_k$ | $\lambda_k(meV)$ |
|---|---|---|---|
| 1 | 21.32 | 0.0114 | 0.0303 |
| 2 | 29.87 | 0.0128 | 0.0474 |
| 3 | 38.69 | 0.1233 | 0.5918 |
| 4 | 42.83 | 0.0133 | 0.0707 |
| 6 | 56.49 | 0.0154 | 0.1079 |
| 8 | 86.31 | 0.0356 | 0.3810 |
| 10 | 112.23 | 0.0534 | 0.7436 |
| 14 | 176.79 | 0.0057 | 0.1259 |
| 15 | 178.09 | 0.0156 | 0.3449 |
| 16 | 181.53 | 0.0778 | 1.7527 |
| 19 | 220.80 | 0.0365 | 1.0000 |
| 23 | 253.53 | 0.0018 | 0.0596 |
| 25 | 309.00 | 0.0524 | 2.0108 |
| 27 | 337.89 | 0.0010 | 0.0455 |
| 28 | 353.94 | 0.0640 | 2.8085 |
| 30 | 412.37 | 0.0205 | 1.0517 |
| 32 | 453.93 | 0.0272 | 1.5328 |
| 34 | 471.77 | 0.0033 | 0.1959 |
| 37 | 534.75 | 0.0011 | 0.0778 |
| 39 | 546.11 | 0.0749 | 5.0747 |
| 40 | 555.17 | 0.0008 | 0.0591 |
| 41 | 559.26 | 0.0048 | 0.3363 |
| 43 | 568.98 | 0.0185 | 1.3099 |
| 44 | 589.18 | 0.0046 | 0.3395 |
| 51 | 728.29 | 0.0109 | 0.9920 |
| 54 | 770.44 | 0.0044 | 0.4261 |
| 56 | 794.67 | 0.0231 | 2.2813 |
| 65 | 929.51 | 0.0005 | 0.0611 |
| 66 | 935.33 | 0.0006 | 0.0782 |
| 70 | 1049.48 | 0.0006 | 0.0854 |
| 72 | 1052.14 | 0.0029 | 0.3890 |
| 76 | 1098.59 | 0.0016 | 0.2293 |
| 80 | 1152.04 | 0.0056 | 0.8042 |
| 82 | 1202.26 | 0.0025 | 0.3774 |
| 84 | 1206.69 | 0.0060 | 0.9053 |
| 85 | 1207.91 | 0.0135 | 2.0315 |
| 88 | 1272.01 | 0.0005 | 0.0820 |
| 92 | 1313.78 | 0.0023 | 0.3810 |
| 95 | 1350.44 | 0.0060 | 1.0116 |
| 97 | 1373.25 | 0.0090 | 1.5366 |
| 98 | 1374.49 | 0.0046 | 0.7937 |
| 99 | 1379.73 | 0.0049 | 0.8444 |
| 101 | 1397.73 | 0.0005 | 0.1023 |
| 103 | 1408.71 | 0.0047 | 0.8240 |
| 104 | 1411.86 | 0.0005 | 0.0979 |
| 105 | 1427.93 | 0.0016 | 0.2913 |
| 107 | 1435.34 | 0.0018 | 0.3224 |
| 112 | 1448.24 | 0.0010 | 0.1852 |
| 119 | 1494.69 | 0.0009 | 0.1697 |
| 123 | 1572.44 | 0.0087 | 1.6978 |
| 124 | 1597.41 | 0.0403 | 8.0010 |
| 125 | 1598.11 | 0.0005 | 0.1047 |
| 126 | 1609.74 | 0.0138 | 2.7581 |
| 127 | 1718.19 | 0.0067 | 1.4407 |
| 129 | 1853.78 | 0.0003 | 0.0875 |
| 137 | 3097.51 | 0.0002 | 0.1075 |
| 155 | 3606.21 | 0.0004 | 0.2228 |

**Table S16.** Low and High frequency normal mode finite amplitude contributions to maximum changes in oscillator strength ($\mathbf{\Delta f_{rel}}$) and transition dipole magnitude ($\mathbf{\Delta \mu_{rel}}$). Here, modes are ranked in decreasing order of oscillator strength modulation *(Inclusion thresholds:* $\Delta f_{rel} \geq 0.02$ *or* $\Delta \mu_{rel} \geq 0.02$ , $\Delta\tilde{\nu} < 500\ cm^{-1}$ *and* $\theta_{axis} < 0.25^{o}$*).*

| k | $\tilde{\nu}_k(\mathbf{cm^{-1}})$ | $\max_Q \lvert \Delta f_{rel}(Q) \rvert$ | $\max_Q \lvert \Delta \mu_{rel}(Q) \rvert$ |
|---|---|---|---|
| 127 | 1718.1900 | 0.1267 | 0.0686 |
| 8 | 86.3100 | 0.0249 | 0.0051 |
| 51 | 728.2900 | 0.0216 | 0.0047 |

**Table S17.** Low and High frequency normal mode finite amplitude contributions to maximum transition dipole axis deviations ($\boldsymbol{\theta_{axis}}$), along with dominant directional responses and oscillator strength changes ($\mathbf{\Delta f_{rel}}$). Here, modes are ranked in decreasing order of axis deviation *(Inclusion thresholds:* $\theta_{axis} \geq 0.25^{o}$ *and* $\Delta\tilde{\nu} < 500\ cm^{-1}$, *Mode 8 additionally meets strength modulation* $\Delta f_{rel}) \geq 0.02$ *or* $\Delta \mu_{rel} \geq 0.02$*).*

| k | $\tilde{\nu}_k(\mathbf{cm^{-1}})$ | $\max_Q \lvert \Delta f_{rel}(Q) \rvert$ | $\max_Q \theta_{axis}(Q)$ | **Dominant Direction** |
|---|---|---|---|---|
| 4 | 42.8300 | 0.0112 | 0.0018 | oop |
| 3 | 38.6900 | 0.0219 | 0.0078 | short |
| 9 | 102.9500 | 0.0072 | 0.0031 | oop |
| 24 | 278.3700 | 0.0012 | 0.0006 | short |
| 13 | 152.8200 | 0.0035 | 0.0005 | oop |

**Table S18.** Modes with large vertical electronic gap, along with their contributions to changes in oscillator strength ($\mathbf{\Delta f_{rel}}$), transition dipole moment ($\mathbf{\Delta \mu_{rel}}$) and transition dipole axis deviations ($\boldsymbol{\theta_{axis}}$), with dominant directional responses. Here, modes are ranked in decreasing order of oscillator strength modulation *(Inclusion thresholds:* $\Delta\tilde{\nu} \geq 500\ cm^{-1}$, $\Delta f_{rel} \geq 0.02$ *or* $\Delta \mu_{rel} \geq 0.02$ , *and* $\theta_{axis} \geq 0.25^{o}$*).*

| k | $\tilde{\nu}_k(\mathbf{cm^{-1}})$ | $\max_Q \lvert \Delta\tilde{\nu}(Q) \rvert$ $(\mathbf{cm^{-1}})$ | $\max_Q \lvert \Delta f_{rel}(Q) \rvert$ | $\max_Q \lvert \Delta \mu_{rel}(Q) \rvert$ | $\max_Q \theta_{axis}(Q)$ | **Dominant Direction** |
|---|---|---|---|---|---|---|
| 123 | 1572.44 | 597.65 | 0.0577 | 0.0341 | 0.0109 | below threshold |
| 39 | 546.11 | 633.14 | 0.0308 | 0.0017 | 0.0191 | below threshold |
| 43 | 568.98 | 541.19 | 0.0275 | 0.0020 | 0.0284 | below threshold |
| 124 | 1597.41 | 967.05 | 0.0235 | 0.0203 | 0.0486 | below threshold |
| 126 | 1609.74 | 506.51 | 0.0060 | 0.0106 | 0.0062 | below threshold |

**Table S19.** Torsional modes giving significant oscillator strength enhancement ($f(\varphi)/f_e$) and/or changes in transition dipole axis deviations ($\theta_{axis}$) with dominant directional responses. Here, modes are ranked in decreasing transition dipole axis deviations with respect to reference geometry axis *(Inclusion thresholds:* $\theta_{axis} \geq 1.0^o$, $\max_{\varphi \in D} f(\varphi) / \min_{\varphi \in D} f(\varphi) \geq 1.05$, *enhancement/suppression relative to 1.0* $(f(\varphi)/f_e)$*)*.

| $\phi_D$ | $\Delta\phi_D(°)$ | $\frac{f(\varphi)}{f_e}$ | $\frac{\max_{\varphi \in D} f(\varphi)}{\min_{\varphi \in D} f(\varphi)}$ | $\max_Q \theta_{axis}$ | $Espan^D$ | $\max \varphi_i, \varphi_j$ | Dominant Direction |
|---|---|---|---|---|---|---|---|
| xanthene-amino (right-end) | -40 to +40 | 0.9894–1.0012 | 1.0119 | 1.1951 | 114.53 | 1.6861 | mixed |
| xanthene-amino (left-end) | -40 to +40 | 0.9906–1.0010 | 1.0105 | 1.1893 | 97.59 | 1.6225 | mixed |
| xanthene-pendant | -40 to +50 | 0.9238–0.9994 | 1.0819 | 0.5928 | 1052.55 | 1.1507 | below threshold |
| amino-ethyl (right-end) | -40 to +90 | 0.9864–1.0186 | 1.0326 | 0.5568 | 87.10 | 1.1011 | below threshold |
| amino-ethyl (left-end) | -90 to +50 | 0.9887–1.0185 | 1.0302 | 0.4640 | 83.88 | 0.7684 | below threshold |
| Pendant-carboxyl | -60 to 100 | 0.9923–0.9996 | 1.0073 | 0.1048 | 183.89 | 0.1550 | below threshold |

## (V) Atto 520 (Zwitterion)

For Atto520 zwitterion, neighbouring low frequency collective deformations demonstrate particularly clear response selectively. A butterfly like coordinate near 40 $cm^{-1}$ produces strong gap and strength modulation, whereas a nearby flapping coordinate produces an axis selective, out of lane response approaching 1.70°. The relaxed amino torsions yield approximately 1.2-1.3° equilibrium axis deviations within little within domain strength modulation. Unlike RhB, therefore Atto does not show a general enhancement of all torsional response channels upon changing from cation to zwitterion. Therefore, the charge state effect is coordinate and dye dependent rather than a uniform consequence of zwitterionic character.

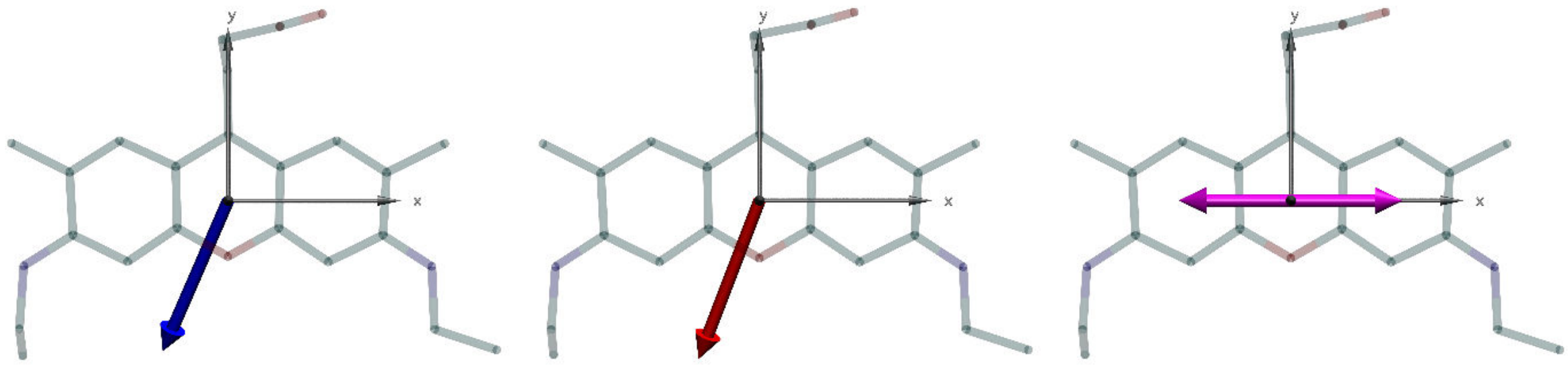

**Figure S36.** (Left and Centre) Dipole moment vectors for ground ($\boldsymbol{\mu}_{S_0}$) and first excited state ($\boldsymbol{\mu}_{S_1}$) of Atto 520 zwitterion. (Right) Emission Transition Dipole Moment vector ($\boldsymbol{\mu}_{em}$).

**Table S20.** Modes with Huang-Rhys factors ($\mathbf{S_k}$) and reorganization energies ($\boldsymbol{\lambda}_\mathbf{k}$) ranked in increasing order of frequency. We list only modes that collectively account for at least 95% of the total FC contributions *(Inclusion threshold:* $C_S(T) \geq 0.95$ *and* $C_\lambda(T) \geq 0.95$*).*

| k | $\tilde{\nu}_k(cm^{-1})$ | $S_k$ | $\lambda_k(meV)$ |
|---|---|---|---|
| 1 | 26.40 | 0.0309 | 0.1013 |
| 2 | 35.08 | 0.2298 | 0.9997 |
| 3 | 39.83 | 0.1485 | 0.7334 |
| 4 | 42.61 | 0.1715 | 0.9059 |
| 5 | 56.00 | 0.2725 | 1.8920 |
| 6 | 68.15 | 0.0139 | 0.1173 |
| 8 | 95.78 | 0.0167 | 0.1980 |
| 9 | 98.64 | 0.1235 | 1.5110 |
| 10 | 103.10 | 0.0792 | 1.0124 |
| 11 | 131.60 | 0.0819 | 1.3364 |
| 12 | 137.81 | 0.0276 | 0.4712 |
| 13 | 157.34 | 0.0065 | 0.1261 |
| 14 | 160.60 | 0.0121 | 0.2402 |
| 15 | 177.87 | 0.0013 | 0.0287 |
| 16 | 187.88 | 0.0039 | 0.0916 |
| 17 | 200.16 | 0.0010 | 0.0250 |
| 18 | 208.69 | 0.0024 | 0.0609 |
| 19 | 219.19 | 0.0032 | 0.0872 |
| 20 | 224.48 | 0.0034 | 0.0958 |
| 21 | 257.41 | 0.0493 | 1.5740 |
| 22 | 262.23 | 0.0012 | 0.0394 |
| 24 | 269.62 | 0.0138 | 0.4599 |
| 25 | 294.61 | 0.0332 | 1.2143 |
| 26 | 316.43 | 0.0035 | 0.1362 |
| 27 | 334.79 | 0.0074 | 0.3052 |
| 29 | 377.18 | 0.0667 | 3.1212 |
| 30 | 391.73 | 0.0091 | 0.4398 |
| 31 | 418.83 | 0.0105 | 0.5443 |
| 32 | 445.21 | 0.0466 | 2.5721 |
| 33 | 465.71 | 0.0048 | 0.2767 |
| 34 | 475.19 | 0.0042 | 0.2504 |
| 35 | 477.67 | 0.0111 | 0.6572 |
| 36 | 486.29 | 0.0053 | 0.3218 |
| 38 | 501.36 | 0.0904 | 5.6182 |
| 39 | 529.28 | 0.0075 | 0.4916 |
| 40 | 539.11 | 0.0163 | 1.0893 |
| 41 | 565.99 | 0.0038 | 0.2697 |
| 43 | 592.35 | 0.0148 | 1.0868 |
| 44 | 609.05 | 0.0009 | 0.0697 |
| 47 | 697.67 | 0.0009 | 0.0805 |
| 48 | 708.31 | 0.0029 | 0.2516 |
| 49 | 715.32 | 0.0074 | 0.6529 |
| 51 | 743.72 | 0.0009 | 0.0792 |
| 52 | 762.85 | 0.0043 | 0.4037 |
| 54 | 789.32 | 0.0386 | 3.7751 |
| 55 | 811.41 | 0.0004 | 0.0393 |
| 62 | 928.01 | 0.0004 | 0.0459 |
| 65 | 958.96 | 0.0004 | 0.0448 |
| 66 | 966.31 | 0.0012 | 0.1454 |
| 71 | 1055.37 | 0.0014 | 0.1886 |
| 72 | 1060.93 | 0.0005 | 0.0638 |
| 73 | 1082.94 | 0.0004 | 0.0538 |
| 74 | 1096.66 | 0.0005 | 0.0642 |
| 75 | 1102.62 | 0.0004 | 0.0596 |
| 77 | 1149.22 | 0.0007 | 0.0999 |
| 78 | 1151.74 | 0.0032 | 0.4613 |
| 81 | 1199.44 | 0.0002 | 0.0349 |
| 83 | 1209.34 | 0.0203 | 3.0485 |
| 84 | 1229.02 | 0.0011 | 0.1699 |
| 89 | 1309.70 | 0.0006 | 0.0897 |
| 90 | 1319.21 | 0.0016 | 0.2695 |
| 91 | 1327.48 | 0.0013 | 0.2191 |
| 92 | 1331.36 | 0.0016 | 0.2579 |
| 93 | 1348.81 | 0.0040 | 0.6606 |
| 94 | 1370.75 | 0.0008 | 0.1287 |
| 95 | 1375.43 | 0.0052 | 0.8876 |
| 96 | 1379.04 | 0.0124 | 2.1229 |
| 97 | 1381.75 | 0.0002 | 0.0321 |
| 99 | 1399.93 | 0.0014 | 0.2426 |
| 100 | 1410.08 | 0.0031 | 0.5333 |
| 101 | 1416.67 | 0.0019 | 0.3251 |
| 103 | 1433.12 | 0.0014 | 0.2559 |
| 104 | 1435.21 | 0.0003 | 0.0475 |
| 106 | 1444.75 | 0.0005 | 0.0923 |
| 108 | 1450.01 | 0.0004 | 0.0662 |
| 109 | 1450.64 | 0.0011 | 0.1961 |
| 110 | 1454.18 | 0.0002 | 0.0361 |
| 112 | 1460.07 | 0.0003 | 0.0455 |
| 114 | 1476.71 | 0.0003 | 0.0526 |
| 115 | 1481.16 | 0.0002 | 0.0394 |
| 116 | 1483.94 | 0.0002 | 0.0349 |
| 117 | 1494.95 | 0.0003 | 0.0485 |
| 119 | 1531.20 | 0.0004 | 0.0708 |
| 121 | 1570.29 | 0.0084 | 1.6261 |
| 122 | 1597.40 | 0.0444 | 8.7883 |
| 124 | 1613.23 | 0.0066 | 1.3179 |
| 126 | 1718.73 | 0.0116 | 2.4685 |
| 127 | 1728.04 | 0.0002 | 0.0463 |
| 128 | 3041.03 | 0.0001 | 0.0345 |
| 129 | 3041.55 | 0.0008 | 0.3082 |
| 130 | 3042.45 | 0.0002 | 0.0755 |
| 131 | 3054.51 | 0.0015 | 0.5511 |
| 134 | 3076.44 | 0.0012 | 0.4722 |
| 139 | 3122.38 | 0.0001 | 0.0548 |
| 147 | 3165.76 | 0.0001 | 0.0369 |
| 152 | 3606.48 | 0.0002 | 0.0723 |
| 153 | 3611.94 | 0.0004 | 0.1910 |

**Table S21.** Low and High frequency normal mode finite amplitude contributions to maximum changes in oscillator strength ($\mathbf{\Delta f_{rel}}$) and transition dipole magnitude ($\boldsymbol{\Delta\mu_{rel}}$). Here, modes are ranked in decreasing order of oscillator strength modulation *(Inclusion thresholds:* $\Delta f_{rel} \geq 0.02$ *or* $\Delta\mu_{rel} \geq 0.02$ *,* $\Delta\tilde{\nu} < 500\ cm^{-1}$ *and* $\theta_{axis} < 0.25^o$*).*

| **k** | $\tilde{\nu}_k(\mathbf{cm^{-1}})$ | $\max_Q\lvert\Delta f_{rel}(Q)\rvert$ | $\max_Q\lvert\Delta\mu_{rel}(Q)\rvert$ |
|---|---|---|---|
| 126 | 1718.73 | 0.0574 | 0.0326 |
| 38 | 501.36 | 0.0257 | 0.0022 |
| 32 | 445.21 | 0.0217 | 0.0021 |

**Table S22.** Low and High frequency normal mode finite amplitude contributions to maximum transition dipole axis deviations ($\boldsymbol{\theta_{axis}}$), along with dominant directional responses and oscillator strength changes ($\mathbf{\Delta f_{rel}}$). Here, modes are ranked in decreasing order of axis deviation *(Inclusion thresholds:* $\theta_{axis} \geq 0.25^o$ *and* $\Delta\tilde{\nu} < 500\ cm^{-1}$*, Mode 8 additionally meets strength modulation* $\Delta f_{rel}) \geq 0.02$ *or* $\Delta\mu_{rel} \geq 0.02$*).*

| **k** | $\tilde{\nu}_k(\mathbf{cm^{-1}})$ | $\max_Q\lvert\Delta f_{rel}(Q)\rvert$ | $\max_Q\theta_{axis}(Q)$ | **Dominant Direction** |
|---|---|---|---|---|
| 4 | 42.61 | 0.0042 | 1.6979 | oop |
| 10 | 103.10 | 0.0068 | 0.7528 | oop |
| 8 | 95.78 | 0.0132 | 0.7178 | oop |
| 52 | 762.85 | 0.0059 | 0.6655 | short |
| 7 | 83.70 | 0.0043 | 0.4842 | short |
| 40 | 539.11 | 0.0017 | 0.4379 | short |
| 14 | 160.60 | 0.0057 | 0.4215 | short |
| 22 | 262.23 | 0.0063 | 0.4105 | short |
| 25 | 294.61 | 0.0052 | 0.3174 | short |
| 31 | 418.83 | 0.0128 | 0.3038 | mixed |
| 13 | 157.34 | 0.0062 | 0.2963 | short |
| 124 | 1613.23 | 0.0058 | 0.2720 | short |
| 74 | 1096.66 | 0.0028 | 0.2699 | short |
| 41 | 565.99 | 0.0017 | 0.2694 | short |
| 27 | 334.79 | 0.0114 | 0.2629 | short |
| 20 | 224.48 | 0.0079 | 0.2522 | oop |

**Table S23.** Modes with large vertical electronic gap, along with their contributions to changes in oscillator strength **($\Delta f_{rel}$)**, transition dipole moment **($\Delta\mu_{rel}$)** and transition dipole axis deviations **($\theta_{axis}$)**, with dominant directional responses. Here, modes are ranked in decreasing order of oscillator strength modulation *(Inclusion thresholds: $\Delta\tilde{\nu} \geq 500\ cm^{-1}, \Delta f_{rel} \geq 0.02$ or $\Delta\mu_{rel} \geq 0.02$ , and $\theta_{axis} \geq 0.25^o$).*

| **k** | **$\tilde{\nu}_k(cm^{-1})$** | **$\max_Q\lvert\Delta\tilde{\nu}(Q)\rvert$ $(cm^{-1})$** | **$\max_Q\lvert\Delta f_{rel}(Q)\rvert$** | **$\max_Q\lvert\Delta\mu_{rel}(Q)\rvert$** | **$\max_Q\theta_{axis}(Q)$** | **Dominant Direction** |
|---|---|---|---|---|---|---|
| 121 | 1570.29 | 571.04 | 0.0616 | 0.0359 | 0.3751 | short |
| 3 | 39.83 | 1845.39 | 0.0536 | 0.0239 | 0.7046 | oop |
| 43 | 592.35 | 567.81 | 0.0406 | 0.0084 | 0.0256 | below threshold |
| 2 | 35.08 | 666.21 | 0.0381 | 0.0049 | 1.2029 | short |
| 1 | 26.40 | 573.46 | 0.0287 | 0.0057 | 2.3826 | short |
| 122 | 1597.40 | 1037.22 | 0.0256 | 0.0221 | 0.1039 | below threshold |
| 54 | 789.32 | 511.35 | 0.0183 | 0.0038 | 0.1577 | below threshold |

**Table S24.** Torsional modes giving significant oscillator strength enhancement ($f(\varphi)/f_e$) and/or changes in transition dipole axis deviations ($\theta_{axis}$) with dominant directional responses. Here, modes are ranked in decreasing transition dipole axis deviations with respect to reference geometry axis *(Inclusion thresholds:* $\theta_{axis} \geq 1.0^o$, $\max_{\varphi\in D} f(\varphi) / \min_{\varphi\in D} f(\varphi) \geq 1.05$*, enhancement/suppression relative to 1.0* $(f(\varphi)/f_e)$*).*

| $\phi_D$ | $\Delta\phi_D(°)$ | $\frac{f(\varphi)}{f_e}$ | $\frac{\max_{\varphi\in D} f(\varphi)}{\min_{\varphi\in D} f(\varphi)}$ | $\max_Q \theta_{axis}$ | Espan[D] | $\max\varphi_i, \varphi_j$ | Dominant Direction |
|---|---|---|---|---|---|---|---|
| xanthene-amino (left-end) | -40 to +40 | 0.9927–1.0011 | 1.0084 | 1.3030 | 79.8489 | 1.5407 | mixed |
| xanthene-amino (right-end) | -40 to +40 | 0.9942–1.0013 | 1.0072 | 1.2346 | 55.6523 | 1.7974 | mixed |
| xanthene-pendant | -40 to +40 | 0.9886–1.0073 | 1.0189 | 0.8057 | 240.3532 | 0.7973 | below threshold |
| amino-ethyl (right-end) | -70 to +40 | 0.9934–1.0201 | 1.0269 | 0.7640 | 82.2686 | 0.9931 | below threshold |
| N-ethyl (left-end) | -60 to +40 | 0.9993–1.0207 | 1.0213 | 0.4942 | 81.4620 | 0.6062 | below threshold |
| Pendant-carboxylate | -50 to +70 | 0.9856–1.0149 | 1.0298 | 0.3616 | 335.5266 | 0.4772 | below threshold |

### (VI) Cy3 (anion)

For Cy3 anion, collective butterfly like and flapping deformations in the calculated 15-30 $cm^{-1}$ region jointly perturb the vertical electronic gap, transition strength, and transition axis. Two related butterfly like coordinates distinguished by their distributions of peripheral displacement yield maximum equilibrium axis deviations of approximately 4.34° and 1.47°, with short axis and mixed directional response, respectively. A flapping deformation involving opposing motion of the molecular halves produces and approximately 5.44° out of plane deviation. For Cy3 anion, collective butterfly like and flapping deformations in the calculated 15-30 $cm^{-1}$ region jointly perturb the vertical electronic gap, transition strength, and transition axis. Two related butterfly like coordinates distinguished by their distributions of peripheral displacement yield maximum equilibrium axis deviations of approximately 4.34° and 1.47°, with short axis and mixed directional response, respectively. A flapping deformation involving opposing motion of the molecular halves produces an approximately 5.44° out of plane deviation. The Cy3 response is not uniformly coupled across all soft coordinates. An in-plane seesaw like deformation near 75 $cm^{-1}$ produces an approximately 2.43° short axis deviation with comparatively little transition strength modulation. An asymmetric, torsion like deformation near 40 $cm^{-1}$, concentrated on one side of the molecule and coupled to the conjugated bridge, instead produces concurrent gap, strength, and out of plane axis responses. Higher frequency Cy3 coordinates provide additional channels. Stretching/compression of the terminal fused heterocyclic framework near 1500-1520 $cm^{-1}$ predominantly couples vertical gap and strength modulation, with a maximum absolute gap change of approximately 1525 $cm^{-1}$. Heterocyclic framework deformation near 1690 $cm^{-1}$ likewise processes substantial spectral strength response. In plane wagging coupled to the central conjugated bridge near 1420 $cm^{-1}$ actives all three response channels, although its axis deviation is smaller, approximately 0.44°. The relaxed Cy3 scans examine local twisting along two cis-trans torsional coordinates of the conjugate chain, each sampled within ±30° of the reference equilibrium. Within these windows, both families produce oscillator strength ratios of approximately 1.07, while their equilibrium axis deviations remain approximately 0.6°-0.64°. The sampled torsions therefore provide evidence for local conformational modulation of transition strength, accompanied by smaller axis changes. Larger angle sampling is required to determine the same coordinates develop stronger gap or axis responses, or access additional conformational states.

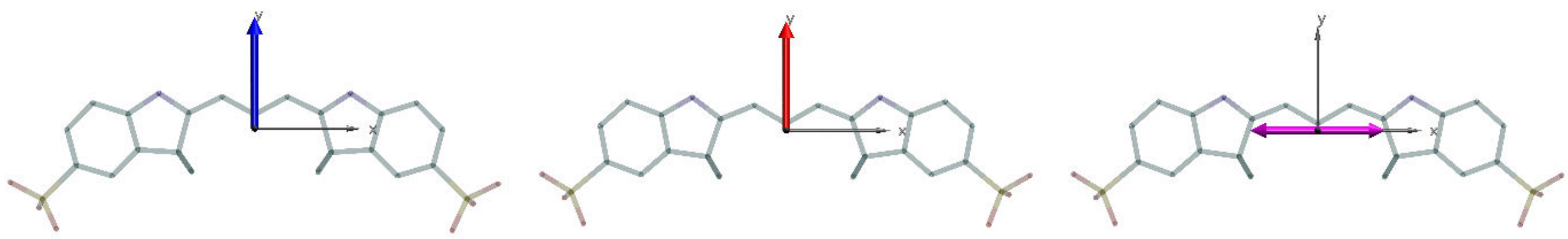

**Figure S37.** (Left and Centre) Dipole moment vectors for ground ($\boldsymbol{\mu}_{S_0}$) and first excited state ($\boldsymbol{\mu}_{S_1}$) of Cy3 anion. (Right) Emission Transition Dipole Moment vector ($\boldsymbol{\mu}_{em}$).

**Table S25.** Modes with Huang-Rhys factors **($\mathbf{S_k}$)** and reorganization energies **($\boldsymbol{\lambda_k}$)** ranked in increasing order of frequency. We list only modes that collectively account for at least 95% of the total FC contributions *(Inclusion threshold: $C_S(T) \geq 0.95$ and $C_\lambda(T) \geq 0.95$)*.

| k | $\tilde{\nu}_k(cm^{-1})$ | $S_k$ | $\lambda_k(meV)$ |
|---|---|---|---|
| 1 | 11.79 | 2.9384 | 4.2953 |
| 2 | 17.64 | 2.7417 | 5.9963 |
| 3 | 20.75 | 3.1244 | 8.0380 |
| 4 | 27.66 | 0.0257 | 0.0881 |
| 5 | 39.26 | 0.2040 | 0.9931 |
| 6 | 42.05 | 0.0553 | 0.2881 |
| 7 | 64.59 | 0.0213 | 0.1708 |
| 12 | 128.85 | 0.1285 | 2.0535 |
| 13 | 132.30 | 0.2763 | 4.5329 |
| 15 | 156.62 | 0.0141 | 0.2731 |
| 25 | 272.69 | 0.0053 | 0.1778 |
| 27 | 281.64 | 0.0047 | 0.1641 |
| 30 | 317.77 | 0.0308 | 1.2116 |
| 33 | 344.55 | 0.1293 | 5.5234 |
| 34 | 346.10 | 0.0041 | 0.1748 |
| 36 | 382.10 | 0.0379 | 1.7969 |
| 37 | 403.80 | 0.0047 | 0.2350 |
| 47 | 563.13 | 0.0039 | 0.2736 |
| 48 | 569.40 | 0.0027 | 0.1903 |
| 49 | 572.58 | 0.0251 | 1.7833 |
| 50 | 596.45 | 0.0427 | 3.1553 |
| 57 | 672.38 | 0.1536 | 12.8025 |
| 59 | 690.72 | 0.0026 | 0.2222 |
| 61 | 731.72 | 0.0038 | 0.3413 |
| 70 | 870.38 | 0.0018 | 0.1993 |
| 72 | 914.90 | 0.0037 | 0.4156 |
| 73 | 914.96 | 0.0083 | 0.9439 |
| 79 | 970.20 | 0.0088 | 1.0530 |
| 86 | 1039.63 | 0.0367 | 4.7275 |
| 87 | 1079.79 | 0.0025 | 0.3340 |
| 88 | 1081.86 | 0.0017 | 0.2323 |
| 89 | 1131.39 | 0.0041 | 0.5727 |
| 90 | 1135.32 | 0.0185 | 2.6074 |
| 91 | 1144.71 | 0.0026 | 0.3741 |
| 92 | 1146.63 | 0.0132 | 1.8800 |
| 96 | 1195.57 | 0.0014 | 0.2058 |
| 99 | 1216.79 | 0.0095 | 1.4289 |
| 100 | 1220.86 | 0.0021 | 0.3201 |
| 101 | 1223.81 | 0.0021 | 0.3208 |
| 102 | 1234.11 | 0.0009 | 0.1341 |
| 104 | 1259.64 | 0.0304 | 4.7499 |
| 106 | 1279.50 | 0.0064 | 1.0130 |
| 108 | 1292.55 | 0.0400 | 6.4079 |
| 110 | 1336.45 | 0.0065 | 1.0734 |
| 112 | 1368.84 | 0.0297 | 5.0414 |
| 116 | 1390.89 | 0.0129 | 2.2198 |
| 117 | 1407.54 | 0.0012 | 0.2070 |
| 118 | 1418.33 | 0.0399 | 7.0108 |
| 119 | 1420.58 | 0.0167 | 2.9413 |
| 123 | 1450.36 | 0.0010 | 0.1847 |
| 124 | 1451.53 | 0.0032 | 0.5697 |
| 125 | 1454.39 | 0.0011 | 0.2027 |
| 126 | 1460.05 | 0.0014 | 0.2522 |
| 128 | 1477.44 | 0.0009 | 0.1654 |
| 132 | 1518.98 | 0.0764 | 14.3843 |
| 134 | 1615.13 | 0.0020 | 0.4020 |
| 138 | 1690.78 | 0.0236 | 4.9433 |
| 142 | 3065.07 | 0.0006 | 0.2431 |
| 148 | 3159.56 | 0.0004 | 0.1454 |
| 156 | 3226.60 | 0.0003 | 0.1361 |
| 158 | 3234.32 | 0.0005 | 0.2146 |
| 159 | 3236.20 | 0.0005 | 0.1925 |
| 161 | 3580.94 | 0.0011 | 0.5051 |

**Table S26.** Low and High frequency normal mode finite amplitude contributions to maximum changes in oscillator strength ($\mathbf{\Delta f_{rel}}$) and transition dipole magnitude ($\mathbf{\Delta \mu_{rel}}$)**.** Here, modes are ranked in decreasing order of oscillator strength modulation *(Inclusion thresholds:* $\Delta f_{rel} \geq 0.02$ *or* $\Delta\mu_{rel} \geq 0.02$ *,* $\Delta\tilde{\nu} < 500\ cm^{-1}$ *and* $\theta_{axis} < 0.25^{o}$*).*

| **k** | $\tilde{\nu}_k(\mathbf{cm^{-1}})$ | $\max_Q \lvert \Delta f_{rel}(Q) \rvert$ | $\max_Q \lvert \Delta\mu_{rel}(Q) \rvert$ |
|---|---|---|---|
| 110 | 1336.45 | 0.0320 | 0.0203 |
| 13 | 132.30 | 0.0299 | 0.0144 |
| 33 | 344.55 | 0.0242 | 0.0033 |
| 116 | 1390.89 | 0.0232 | 0.0062 |
| 36 | 382.10 | 0.0207 | 0.0045 |

**Table S27.** Low and High frequency normal mode finite amplitude contributions to maximum transition dipole axis deviations ($\mathbf{\theta_{axis}}$), along with dominant directional responses and oscillator strength changes ($\mathbf{\Delta f_{rel}}$). Here, modes are ranked in decreasing order of axis deviation *(Inclusion thresholds:* $\theta_{axis} \geq 0.25^{o}$ *and* $\Delta\tilde{\nu} < 500\ cm^{-1}$*, Mode 8 additionally meets strength modulation* $\Delta f_{rel}) \geq 0.02$ *or* $\Delta\mu_{rel} \geq 0.02$*).*

| **k** | $\tilde{\nu}_k(\mathbf{cm^{-1}})$ | $\max_Q \lvert \Delta f_{rel}(Q) \rvert$ | $\max_Q \theta_{axis}(Q)$ | **Dominant Direction** |
|---|---|---|---|---|
| 8 | 74.37 | 0.0102 | 2.4320 | short |
| 10 | 103.72 | 0.0138 | 0.6063 | oop |
| 119 | 1420.58 | 0.0221 | 0.5440 | short |
| 15 | 156.62 | 0.0046 | 0.5009 | short |
| 9 | 94.38 | 0.0141 | 0.4741 | mixed |
| 7 | 64.59 | 0.0137 | 0.2853 | short |
| 117 | 1407.54 | 0.0041 | 0.2601 | short |
| 102 | 1234.11 | 0.0020 | 0.2551 | short |

**Table S28.** Modes with large vertical electronic gap, along with their contributions to changes in oscillator strength ($\mathbf{\Delta f_{rel}}$)**,** transition dipole moment ($\mathbf{\Delta \mu_{rel}}$) and transition dipole axis deviations ($\mathbf{\theta_{axis}}$), with dominant directional responses. Here, modes are ranked in decreasing order of oscillator strength modulation *(Inclusion thresholds:* $\Delta\tilde{\nu} \geq 500\ cm^{-1}, \Delta f_{rel} \geq 0.02$ *or* $\Delta\mu_{rel} \geq 0.02$ *, and* $\theta_{axis} \geq 0.25^{o}$*).*

| **k** | $\tilde{\nu}_k(\mathbf{cm^{-1}})$ | $\max_Q \lvert \Delta\tilde{\nu}(Q) \rvert$ $(\mathbf{cm^{-1}})$ | $\max_Q \lvert \Delta f_{rel}(Q) \rvert$ | $\max_Q \lvert \Delta\mu_{rel}(Q) \rvert$ | $\max_Q \theta_{axis}(Q)$ | **Dominant Direction** |
|---|---|---|---|---|---|---|
| 2 | 17.64 | 4090.84 | 0.2911 | 0.0843 | 4.3356 | short |
| 3 | 20.75 | 1692.95 | 0.1295 | 0.0388 | 1.4658 | mixed |
| 4 | 27.66 | 1405.82 | 0.0867 | 0.0123 | 5.4364 | oop |
| 138 | 1690.78 | 1274.35 | 0.0510 | 0.0533 | 0.0186 | below threshold |
| 57 | 672.38 | 805.74 | 0.0438 | 0.0036 | 0.0305 | below threshold |
| 118 | 1418.33 | 651.69 | 0.0376 | 0.0040 | 0.4374 | short |
| 6 | 42.05 | 705.73 | 0.0339 | 0.0037 | 1.0070 | oop |
| 132 | 1518.98 | 1525.19 | 0.0189 | 0.0273 | 0.0523 | below threshold |
| 108 | 1292.55 | 567.00 | 0.0113 | 0.0078 | 0.0485 | below threshold |
| 50 | 596.45 | 700.89 | 0.0089 | 0.0121 | 0.0124 | below threshold |
| 112 | 1368.84 | 824.29 | 0.0070 | 0.0198 | 0.0557 | below threshold |

**Table S29.** Torsional modes giving significant oscillator strength enhancement ($f(\varphi)/f_e$) and/or changes in transition dipole axis deviations ($\theta_{axis}$) with dominant directional responses. Here, modes are ranked in decreasing transition dipole axis deviations with respect to reference geometry axis *(Inclusion thresholds:* $\theta_{axis} \geq 1.0^o$, $\max_{\varphi \in D} f(\varphi) / \min_{\varphi \in D} f(\varphi) \geq 1.05$, *enhancement/suppression relative to 1.0* $(f(\varphi)/f_e)$*)*.

| $\boldsymbol{\phi_D}$ | $\boldsymbol{\Delta\phi_D(°)}$ | $\boldsymbol{\frac{f(\varphi)}{f_e}}$ | $\boldsymbol{\frac{\max_{\varphi \in D} f(\varphi)}{\min_{\varphi \in D} f(\varphi)}}$ | $\boldsymbol{\max_{Q} \theta_{axis}}$ | $\mathbf{Espan^D}$ | $\boldsymbol{\max \varphi_i, \varphi_j}$ | **Dominant Direction** |
|---|---|---|---|---|---|---|---|
| cis-trans C-C bond (right-end) | -30 to +30 | 0.9232-0.9917 | 1.0742 | 0.6357 | 391.9855 | 1.1894 | oop |
| cis-trans C-C bond (left-end) | -30 to +30 | 0.9232-0.9908 | 1.0732 | 0.6105 | 440.3787 | 1.2116 | oop |

## (VII) Cy3b (anion)

Cy3b anion also retains pronounced susceptibility to selected collective deformations despite its capped bridge. A flapping coordinate near 20 $cm^{-1}$ produces an approximately 5.60 ° out of plane axis deviation, together with a maximum absolute vertical gap change of approximately 2150 $cm^{-1}$ and substantial transition strength modulation. Neighbouring breathing like and in plane wagging coordinates in approximately 25-35 $cm^{-1}$ region, coupled respectively to methyl and methyl/sulfonate substituent motion, activate concurrent gap, strength and axis responses, but their axis deviations retain to 0.6°. A separate flapping coordinate near 85 $cm^{-1}$ produces an axis selective short axis response of approximately 1.84°. These results show that the capped framework remains optically responsive to collective internal deformation. In the higher frequency responses, related bridge/core stretching compression coordinates near 1500 $cm^{-1}$ differ in their left and right-side participation. Both couple gap modulation to short axis reorientation, with maximum axis deviations of approximately 1.58° and 0.79°, respectively. A nearby deformation near 1530 $cm^{-1}$, concentrated predominantly on one terminal fused ring framework, instead coupled gap, strength, axis response with a smaller axis of deviation of approximately 0.36°. Heterocyclic framework stretching/compression near 1690 $cm^{-1}$ provides another coupled channel. Related structural motions therefore need not have equivalent optical effects: their fragments participation influences how the response is distributed between transition strength and direction. The investigated Cy3b bridge coordinate is a cis-trans torsion within the capped framework. Its retained domain is approximately -10° to +15 ° around the reference conformation, over which the equilibrium axis deviation reaches approximately 0.51° and the oscillator strength ratio approximately 1.03. The vertical gap varies over a span of approximately 404 $cm^{-1}$ demonstrating a finite local spectral response. These observations support a locally gap-sensitive torsional channel with smaller accompanying strength and axis changes. Extended sampling will be needed to separate the effects of structural constraint from those of angular coverage and to determine whether additional torsion dependent response channels emerge. Therefore, soft molecular deformations can couple gap modulation to substantial intrinsic axis change, while local cis trans twisting can modulate transition strength or the vertical gap. The results therefore motivate both broader torsional sampling and explicit dye-matrix calculations.

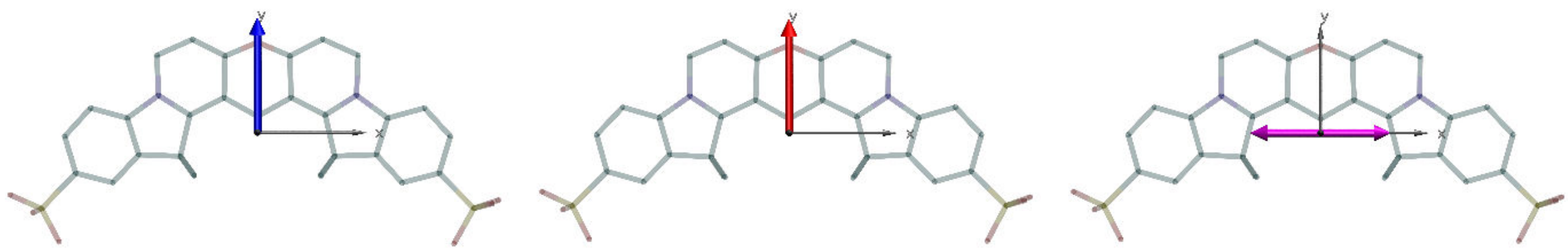

**Figure S38.** (Left and Centre) Dipole moment vectors for ground ($\boldsymbol{\mu}_{S_0}$) and first excited state ($\boldsymbol{\mu}_{S_1}$) of Cy3b anion. (Right) Emission Transition Dipole Moment vector ($\boldsymbol{\mu}_{em}$).

**Table S30.** Modes with Huang-Rhys factors ($\mathbf{S_k}$) and reorganization energies ($\boldsymbol{\lambda}_\mathbf{k}$) ranked in increasing order of frequency. We list only modes that collectively account for at least 95% of the total FC contributions *(Inclusion threshold:* $C_S(T) \geq 0.95$ *and* $C_\lambda(T) \geq 0.95$*).*

| k | $\tilde{\nu}_k(cm^{-1})$ | $S_k$ | $\lambda_k(meV)$ |
|---|---|---|---|
| 2 | 25.82 | 0.5896 | 1.8873 |
| 3 | 30.49 | 0.4966 | 1.8771 |
| 4 | 31.47 | 1.3400 | 5.2282 |
| 5 | 42.03 | 0.0108 | 0.0562 |
| 6 | 45.35 | 0.0470 | 0.2640 |
| 8 | 74.65 | 0.0148 | 0.1369 |
| 10 | 97.31 | 0.0054 | 0.0654 |
| 11 | 104.14 | 0.0141 | 0.1825 |
| 13 | 127.31 | 0.2872 | 4.5336 |
| 15 | 139.84 | 0.0054 | 0.0943 |
| 16 | 156.81 | 0.0110 | 0.2145 |
| 17 | 169.94 | 0.0051 | 0.1065 |
| 18 | 172.26 | 0.0045 | 0.0971 |
| 21 | 210.38 | 0.0048 | 0.1254 |
| 25 | 241.25 | 0.0179 | 0.5369 |
| 26 | 251.34 | 0.0030 | 0.0944 |
| 34 | 320.88 | 0.0038 | 0.1511 |
| 35 | 329.78 | 0.0029 | 0.1206 |
| 36 | 336.30 | 0.0166 | 0.6917 |
| 37 | 341.69 | 0.1133 | 4.7980 |
| 40 | 379.16 | 0.0021 | 0.0980 |
| 42 | 388.46 | 0.0348 | 1.6751 |
| 43 | 420.60 | 0.0033 | 0.1707 |
| 44 | 424.57 | 0.0017 | 0.0908 |
| 46 | 477.84 | 0.0333 | 1.9745 |
| 53 | 534.16 | 0.0082 | 0.5447 |
| 56 | 572.74 | 0.0076 | 0.5366 |
| 57 | 574.16 | 0.0354 | 2.5234 |
| 58 | 587.41 | 0.0151 | 1.0997 |
| 59 | 588.14 | 0.0035 | 0.2539 |
| 60 | 608.32 | 0.0065 | 0.4926 |
| 61 | 617.57 | 0.0074 | 0.5677 |
| 63 | 640.72 | 0.0080 | 0.6346 |
| 68 | 685.55 | 0.0222 | 1.8846 |
| 69 | 711.94 | 0.0221 | 1.9518 |
| 74 | 802.22 | 0.0043 | 0.4327 |
| 75 | 832.79 | 0.0038 | 0.3941 |
| 77 | 846.68 | 0.0042 | 0.4402 |
| 80 | 857.08 | 0.0016 | 0.1661 |
| 84 | 942.43 | 0.0074 | 0.8674 |
| 91 | 986.38 | 0.0018 | 0.2159 |
| 100 | 1069.37 | 0.0013 | 0.1726 |
| 108 | 1144.63 | 0.0025 | 0.3501 |
| 109 | 1151.26 | 0.0197 | 2.8171 |
| 111 | 1161.16 | 0.0011 | 0.1639 |
| 112 | 1175.87 | 0.0006 | 0.0925 |
| 114 | 1181.21 | 0.0047 | 0.6929 |
| 119 | 1219.35 | 0.0089 | 1.3518 |
| 120 | 1220.41 | 0.0007 | 0.1006 |
| 122 | 1226.75 | 0.0032 | 0.4901 |
| 125 | 1265.04 | 0.0107 | 1.6801 |
| 127 | 1276.65 | 0.0008 | 0.1269 |
| 131 | 1329.49 | 0.0021 | 0.3380 |
| 132 | 1330.87 | 0.0060 | 0.9974 |
| 133 | 1333.41 | 0.0007 | 0.1087 |
| 134 | 1352.28 | 0.0582 | 9.7517 |
| 135 | 1365.21 | 0.0062 | 1.0443 |
| 137 | 1378.31 | 0.0047 | 0.8064 |
| 141 | 1396.81 | 0.0047 | 0.8138 |
| 142 | 1397.10 | 0.0014 | 0.2359 |
| 143 | 1400.49 | 0.0024 | 0.4111 |
| 144 | 1401.19 | 0.0015 | 0.2530 |
| 146 | 1428.01 | 0.0012 | 0.2122 |
| 148 | 1434.73 | 0.0433 | 7.7040 |
| 157 | 1462.20 | 0.0017 | 0.3009 |
| 159 | 1464.85 | 0.0026 | 0.4655 |
| 161 | 1476.53 | 0.0190 | 3.4759 |
| 163 | 1503.48 | 0.0100 | 1.8640 |
| 164 | 1505.87 | 0.0344 | 6.4164 |
| 165 | 1528.38 | 0.0006 | 0.1058 |
| 166 | 1529.89 | 0.0101 | 1.9131 |
| 172 | 1686.35 | 0.0192 | 4.0100 |
| 173 | 2977.64 | 0.0006 | 0.2305 |
| 175 | 3057.56 | 0.0007 | 0.2597 |
| 176 | 3058.12 | 0.0010 | 0.3650 |
| 201 | 3284.83 | 0.0003 | 0.1314 |

**Table S31.** Low and High frequency normal mode finite amplitude contributions to maximum changes in oscillator strength **($\Delta f_{rel}$)** and transition dipole magnitude **($\Delta \mu_{rel}$).** Here, modes are ranked in decreasing order of oscillator strength modulation *(Inclusion thresholds:* $\Delta f_{rel} \geq 0.02$ *or* $\Delta \mu_{rel} \geq 0.02$ , $\Delta\tilde{\nu} < 500\ cm^{-1}$ *and* $\theta_{axis} < 0.25^o$*).*

| **k** | $\tilde{\nu}_k(cm^{-1})$ | $\max_Q\lvert\Delta f_{rel}(Q)\rvert$ | $\max_Q\lvert\Delta \mu_{rel}(Q)\rvert$ |
|---|---|---|---|
| 37 | 341.69 | 0.0296 | 0.0036 |
| 68 | 685.55 | 0.0226 | 0.0022 |
| 137 | 1378.31 | 0.0219 | 0.0077 |
| 42 | 388.46 | 0.0205 | 0.0036 |

**Table S32.** Low and High frequency normal mode finite amplitude contributions to maximum transition dipole axis deviations **($\theta_{axis}$)**, along with dominant directional responses and oscillator strength changes **($\Delta f_{rel}$)**. Here, modes are ranked in decreasing order of axis deviation *(Inclusion thresholds:* $\theta_{axis} \geq 0.25^o$ *and* $\Delta\tilde{\nu} < 500\ cm^{-1}$, *Mode 8 additionally meets strength modulation* $\Delta f_{rel}) \geq 0.02$ *or* $\Delta \mu_{rel} \geq 0.02$*).*

| **k** | $\tilde{\nu}_k(cm^{-1})$ | $\max_Q\lvert\Delta f_{rel}(Q)\rvert$ | $\max_Q\theta_{axis}(Q)$ | **Dominant Direction** |
|---|---|---|---|---|
| 9 | 85.48 | 0.0039 | 1.8430 | short |
| 6 | 45.35 | 0.0202 | 1.2292 | oop |
| 165 | 1528.38 | 0.0041 | 0.5569 | short |
| 18 | 172.26 | 0.0049 | 0.5084 | short |
| 132 | 1330.87 | 0.0037 | 0.5009 | short |
| 8 | 74.65 | 0.0149 | 0.4918 | mixed |
| 20 | 207.08 | 0.0053 | 0.3936 | short |
| 10 | 97.31 | 0.0069 | 0.3857 | short |
| 108 | 1144.63 | 0.0010 | 0.3756 | short |
| 11 | 104.14 | 0.0193 | 0.3271 | short |
| 59 | 588.14 | 0.0017 | 0.3090 | short |
| 109 | 1151.26 | 0.0041 | 0.2580 | short |
| 131 | 1329.49 | 0.0005 | 0.2505 | short |

**Table S33.** Modes with large vertical electronic gap, along with their contributions to changes in oscillator strength ($\mathbf{\Delta f_{rel}}$), transition dipole moment ($\mathbf{\Delta \mu_{rel}}$) and transition dipole axis deviations ($\mathbf{\theta_{axis}}$), with dominant directional responses. Here, modes are ranked in decreasing order of oscillator strength modulation *(Inclusion thresholds:* $\Delta\tilde{\nu} \geq 500\ cm^{-1}, \Delta f_{rel} \geq 0.02\ or\ \Delta\mu_{rel} \geq 0.02\ , and\ \theta_{axis} \geq 0.25^{o}$*).*

| k | $\tilde{\nu}_k(cm^{-1})$ | $\max_Q\|\Delta\tilde{\nu}(Q)\|$ $(cm^{-1})$ | $\max_Q\|\Delta f_{rel}(Q)\|$ | $\max_Q\|\Delta\mu_{rel}(Q)\|$ | $\max_Q\theta_{axis}(Q)$ | Dominant Direction |
|---|---|---|---|---|---|---|
| 1 | 20.36 | 2148.66 | 0.1199 | 0.0083 | 5.5956 | oop |
| 4 | 31.47 | 764.61 | 0.0765 | 0.0495 | 0.3468 | oop |
| 2 | 25.82 | 1472.76 | 0.0581 | 0.0180 | 0.5652 | mixed |
| 172 | 1686.35 | 1046.10 | 0.0538 | 0.0538 | 0.2926 | short |
| 161 | 1476.53 | 529.09 | 0.0442 | 0.0095 | 0.0066 | below threshold |
| 166 | 1529.89 | 598.46 | 0.0360 | 0.0286 | 0.3623 | short |
| 163 | 1503.48 | 594.43 | 0.0167 | 0.0089 | 1.5816 | short |
| 164 | 1505.87 | 879.95 | 0.0158 | 0.0149 | 0.7913 | short |
| 148 | 1434.73 | 638.79 | 0.0120 | 0.0109 | 0.0084 | below threshold |
| 134 | 1352.28 | 796.87 | 0.0078 | 0.0228 | 0.0288 | below threshold |

**Table S34.** Torsional modes giving significant oscillator strength enhancement ($f(\varphi)/f_e$) and/or changes in transition dipole axis deviations ($\theta_{axis}$) with dominant directional responses. Here, modes are ranked in decreasing transition dipole axis deviations with respect to reference geometry axis *(Inclusion thresholds:* $\theta_{axis} \geq 1.0^{o}$, $\max_{\varphi\in D} f(\varphi)/\min_{\varphi\in D} f(\varphi) \geq 1.05$, *enhancement/suppression relative to 1.0* $(f(\varphi)/f_e)$*).*

| $\phi_D$ | $\Delta\phi_D(°)$ | $\frac{f(\varphi)}{f_e}$ | $\frac{\max_{\varphi\in D} f(\varphi)}{\min_{\varphi\in D} f(\varphi)}$ | $\max_Q\theta_{axis}$ | $Espan^D$ | $\max\varphi_i, \varphi_j$ | Dominant Direction |
|---|---|---|---|---|---|---|---|
| Bridge torsion | -15 to 15 | 0.9735 - 0.9993 | 1.0266 | 0.5513 | 404.08 | 1.0635 | oop |

## (VIII) Nile Red (neutral)

For Nile Red, the normal coordinate response map contains both coupled and comparatively selected channels, while the relaxed amino substituent torsion produces a predominantly orientational response within its sampled interval. These results distinguish deformation of the fused chromophore from rotation of the amino substituent relative to the core. Low frequency coordinates in the 25-40 $cm^{-1}$ region involve collective wagging of the molecular framework coupled to the amino and ethyl substituents, with different participation of the two ethyl groups and the opposing fused ring region. Over the retained normal-coordinate grids, one coordinate couples the transition-strength modulation to an approximately 0.99$^o$ short axis deviation, whereas the other actives gap, strength and axis responses, reaching approximately 1.57$^o$ out of plane deviation. A flapping coordinate near 85 $cm^{-1}$ with greater participation of the amino substituted side likewise couple's strength modulation to an out of plane axis response of approximately 0.90$^o$. Higher frequency coordinates reveals a different distribution of optical sensitivity. A breathing/stretching deformation near 1530 $cm^{-1}$ concentrated along the central heterocyclic framework and the amino substituted fused ring region predominantly couples electronic gap and transition strength modulation. Its maximum above absolute vertical gap change is approximately around 907 $cm^{-1}$ accompanied by oscillator strength and transition dipole magnitude changes of approximately 8.8 % and 6.6% respectively. Core stretching/compression near 1580 $cm^{-1}$ instead activates all three response channels, with a gap change of approximately 1434 $cm^{-1}$ and a short axis deviation of approximately 1.18$^o$. Thus, substantial gap sensitivity can coexist with different degrees of orientational response, depending upon on the structural distribution of the deformation. Additional core stretching/compression coordinates in the approximately 1660-1730 $cm^{-1}$ region differ in their participation of the carbonyl containing side and the amino substituted benzene region. These coordinates produce coupled gap, strength an predominantly short axis responses rather than a single uniform high frequency mechanism. A deformation near 1660 $cm^{-1}$ reaches an approximately 1.69$^o$ axis deviation, while a coordinate near 1705 $cm^{-1}$ produces the largest normal coordinate gap change, approximately 1926 cm+-1 with a 1.10$^o$ axis deviation. The neighbouring coordinate near 1725 $cm^{-1}$ yields particularly pronounced transition strength modulation, approximately 13.6% in oscillator strength but a smaller axis deviation of 0.75$^o$. The frequency regions organize qualitatively related core-deformation patterns. The relaxed torsional calculations probes rotation of the amino substituent relative to the conjugated core. Within this sampled domain, the coordinate produces and approximately 2.42$^o$ equilibrium references out of plane axis deviation, while the oscillator strength ratio remains 1.007 and the vertical gap span is less than 50 $cm^{-1}$. Therefore, amino-core twisting provides a comparatively selective transition axis channel, distinct from the stronger gap modulation associated with several core deformation coordinates. This rule to limited to spanned angular window, therefore extended sampling is required. Therefore, the Nile Red separated core associated spectral sensitivity from a comparatively selective amino-core orientational response. Therefore, environmental interactions can be examined for their effects on the identified core deformations and amino-core coordinate.

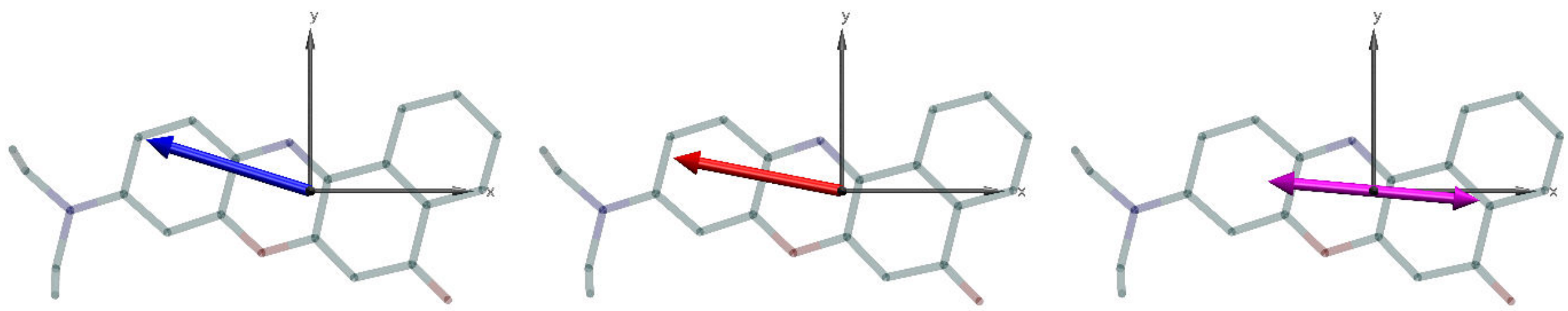

**Figure S39.** (Left and Centre) Dipole moment vectors for ground ($\boldsymbol{\mu}_{S_0}$) and first excited state ($\boldsymbol{\mu}_{S_1}$) of Nile Red neutral. (Right) Emission Transition Dipole Moment vector ($\boldsymbol{\mu}_{em}$).

**Table S35.** Modes with Huang-Rhys factors **($S_k$)** and reorganization energies **($\lambda_k$)** ranked in increasing order of frequency. We list only modes that collectively account for at least 95% of the total FC contributions *(Inclusion threshold: $C_S(T) \geq 0.95$ and $C_\lambda(T) \geq 0.95$).*

| **k** | $\tilde{\nu}_k(\mathbf{cm^{-1}})$ | $\mathbf{S_k}$ | $\boldsymbol{\lambda}_k(\mathbf{meV})$ |
|---|---|---|---|
| 1 | 26.26 | 0.0840 | 0.2736 |
| 2 | 37.68 | 0.0047 | 0.0221 |
| 3 | 49.53 | 0.0287 | 0.1761 |
| 4 | 84.77 | 0.0035 | 0.0365 |
| 5 | 95.21 | 0.0021 | 0.0248 |
| 6 | 121.00 | 0.0034 | 0.0509 |
| 9 | 169.28 | 0.0020 | 0.0428 |
| 10 | 184.87 | 0.0117 | 0.2689 |
| 11 | 191.32 | 0.0021 | 0.0489 |
| 12 | 216.82 | 0.0186 | 0.5005 |
| 13 | 232.87 | 0.0591 | 1.7073 |
| 14 | 245.91 | 0.0141 | 0.4292 |
| 15 | 273.75 | 0.0034 | 0.1165 |
| 17 | 317.80 | 0.0220 | 0.8688 |
| 19 | 356.53 | 0.0032 | 0.1425 |
| 20 | 383.08 | 0.1136 | 5.3968 |
| 22 | 441.77 | 0.0634 | 3.4703 |
| 25 | 477.35 | 0.0821 | 4.8597 |
| 26 | 502.07 | 0.0463 | 2.8808 |
| 27 | 505.51 | 0.0105 | 0.6557 |
| 28 | 532.29 | 0.0496 | 3.2713 |
| 29 | 537.25 | 0.0034 | 0.2271 |
| 30 | 568.07 | 0.0067 | 0.4722 |
| 31 | 599.20 | 0.1124 | 8.3491 |
| 32 | 602.50 | 0.0055 | 0.4139 |
| 34 | 657.70 | 0.0068 | 0.5515 |
| 35 | 681.33 | 0.0023 | 0.1982 |
| 38 | 727.94 | 0.0283 | 2.5499 |
| 40 | 765.99 | 0.0029 | 0.2789 |
| 42 | 800.43 | 0.0052 | 0.5178 |
| 45 | 825.35 | 0.0241 | 2.4705 |
| 50 | 900.81 | 0.0036 | 0.3981 |
| 56 | 1037.24 | 0.0029 | 0.3724 |
| 61 | 1106.23 | 0.0070 | 0.9659 |
| 62 | 1112.08 | 0.0019 | 0.2637 |
| 69 | 1220.99 | 0.0056 | 0.8501 |
| 70 | 1223.71 | 0.0099 | 1.5034 |
| 71 | 1255.87 | 0.0312 | 4.8641 |
| 73 | 1302.42 | 0.0021 | 0.3466 |
| 74 | 1317.43 | 0.0016 | 0.2602 |
| 76 | 1341.43 | 0.0034 | 0.5631 |
| 78 | 1375.23 | 0.0214 | 3.6423 |
| 82 | 1403.11 | 0.0069 | 1.1934 |
| 83 | 1404.40 | 0.0028 | 0.4892 |
| 84 | 1429.14 | 0.0341 | 6.0400 |
| 85 | 1446.69 | 0.0018 | 0.3166 |
| 86 | 1454.04 | 0.0145 | 2.6204 |
| 88 | 1457.70 | 0.0377 | 6.8139 |
| 89 | 1462.62 | 0.0894 | 16.2059 |
| 91 | 1490.96 | 0.0486 | 8.9832 |
| 92 | 1497.70 | 0.0263 | 4.8829 |
| 93 | 1517.89 | 0.0326 | 6.1364 |
| 94 | 1529.35 | 0.1284 | 24.3464 |
| 95 | 1556.62 | 0.0304 | 5.8675 |
| 96 | 1568.27 | 0.0165 | 3.2022 |
| 97 | 1578.38 | 0.0997 | 19.5012 |
| 98 | 1610.83 | 0.0333 | 6.6415 |
| 99 | 1637.58 | 0.0114 | 2.3219 |
| 100 | 1663.28 | 0.0340 | 7.0150 |
| 101 | 1705.73 | 0.1365 | 28.8590 |
| 102 | 1724.94 | 0.0795 | 16.9943 |

**Table S36.** Low and High frequency normal mode finite amplitude contributions to maximum changes in oscillator strength ($\mathbf{\Delta f_{rel}}$) and transition dipole magnitude ($\mathbf{\Delta \mu_{rel}}$). Here, modes are ranked in decreasing order of oscillator strength modulation *(Inclusion thresholds:* $\Delta f_{rel} \geq 0.02$ *or* $\Delta\mu_{rel} \geq 0.02$ , $\Delta\tilde{\nu} < 500\ cm^{-1}$ *and* $\theta_{axis} < 0.25^{o}$*).*

| k | $\tilde{\nu}_k(\mathrm{cm^{-1}})$ | $\max_Q\lvert\Delta f_{rel}(Q)\rvert$ | $\max_Q\lvert\Delta\mu_{rel}(Q)\rvert$ |
|---|---|---|---|
| 45 | 825.35 | 0.0350 | 0.0059 |
| 93 | 1517.89 | 0.0309 | 0.0056 |
| 38 | 727.94 | 0.0240 | 0.0013 |
| 50 | 900.81 | 0.0201 | 0.0063 |

**Table S37.** Low and High frequency normal mode finite amplitude contributions to maximum transition dipole axis deviations ($\mathbf{\theta_{axis}}$), along with dominant directional responses and oscillator strength changes ($\mathbf{\Delta f_{rel}}$). Here, modes are ranked in decreasing order of axis deviation *(Inclusion thresholds:* $\theta_{axis} \geq 0.25^{o}$ *and* $\Delta\tilde{\nu} < 500\ cm^{-1}$, *Mode 8 additionally meets strength modulation* $\Delta f_{rel}) \geq 0.02$ *or* $\Delta\mu_{rel} \geq 0.02$*).*

| k | $\tilde{\nu}_k(\mathrm{cm^{-1}})$ | $\max_Q\lvert\Delta f_{rel}(Q)\rvert$ | $\max_Q\theta_{axis}(Q)$ | Dominant Direction |
|---|---|---|---|---|
| 1 | 26.26 | 0.0253 | 0.9858 | short |
| 4 | 84.77 | 0.0260 | 0.8977 | oop |
| 88 | 1457.70 | 0.0345 | 0.4806 | short |
| 5 | 95.21 | 0.0235 | 0.4473 | short |
| 32 | 602.50 | 0.0113 | 0.4071 | short |
| 11 | 191.32 | 0.0207 | 0.3651 | oop |
| 34 | 657.70 | 0.0129 | 0.2882 | short |
| 9 | 169.28 | 0.0135 | 0.2690 | oop |

**Table S38.** Modes with large vertical electronic gap, along with their contributions to changes in oscillator strength ($\mathbf{\Delta f_{rel}}$), transition dipole moment ($\mathbf{\Delta \mu_{rel}}$) and transition dipole axis deviations ($\mathbf{\theta_{axis}}$), with dominant directional responses. Here, modes are ranked in decreasing order of oscillator strength modulation *(Inclusion thresholds:* $\Delta\tilde{\nu} \geq 500\ cm^{-1}$, $\Delta f_{rel} \geq 0.02$ *or* $\Delta\mu_{rel} \geq 0.02$ , *and* $\theta_{axis} \geq 0.25^{o}$*).*

| k | $\tilde{\nu}_k(\mathrm{cm^{-1}})$ | $\max_Q\lvert\Delta\tilde{\nu}(Q)\rvert$ $(\mathrm{cm^{-1}})$ | $\max_Q\lvert\Delta f_{rel}(Q)\rvert$ | $\max_Q\lvert\Delta\mu_{rel}(Q)\rvert$ | $\max_Q\theta_{axis}(Q)$ | Dominant Direction |
|---|---|---|---|---|---|---|
| 102 | 1724.94 | 870.2722 | 0.1364 | 0.0907 | 0.7455 | short |
| 94 | 1529.35 | 906.5671 | 0.0877 | 0.0663 | 0.1680 | Below threshold |
| 98 | 1610.83 | 683.1516 | 0.0779 | 0.0253 | 0.3093 | short |
| 95 | 1556.62 | 830.7510 | 0.0525 | 0.0110 | 1.0362 | short |
| 101 | 1705.73 | 1926.0519 | 0.0525 | 0.0713 | 1.0994 | short |
| 99 | 1637.58 | 775.0988 | 0.0479 | 0.0048 | 0.5395 | short |
| 91 | 1490.96 | 598.4634 | 0.0365 | 0.0300 | 1.0651 | short |
| 84 | 1429.14 | 690.4106 | 0.0351 | 0.0313 | 0.8930 | short |
| 2 | 37.68 | 775.9053 | 0.0344 | 0.0100 | 1.5685 | oop |
| 97 | 1578.38 | 1434.0537 | 0.0315 | 0.0333 | 1.1821 | short |
| 100 | 1663.28 | 700.0892 | 0.0295 | 0.0117 | 1.6938 | short |
| 89 | 1462.62 | 616.2076 | 0.0281 | 0.0073 | 0.2502 | short |

**Table S39.** Torsional modes giving significant oscillator strength enhancement ($f(\varphi)/f_e$) and/or changes in transition dipole axis deviations ($\theta_{axis}$) with dominant directional responses. Here, modes are ranked in decreasing transition dipole axis deviations with respect to reference geometry axis *(Inclusion thresholds:* $\theta_{axis} \geq 1.0^o$, $\max_{\varphi \in D} f(\varphi) / \min_{\varphi \in D} f(\varphi) \geq 1.05$*, enhancement/suppression relative to 1.0* $(f(\varphi)/f_e)$*).*

| $\boldsymbol{\phi_D}$ | $\boldsymbol{\Delta\phi_D}$ **(°)** | $\frac{\mathbf{f(\varphi)}}{\mathbf{f_e}}$ | $\frac{\max_{\varphi \in D} \mathbf{f(\varphi)}}{\min_{\varphi \in D} \mathbf{f(\varphi)}}$ | $\max_{Q} \boldsymbol{\theta_{axis,e}}$ | **Dominant Direction** |
|---|---|---|---|---|---|
| amino-ethyl | -30 to 30 | 0.9971–1.0016 | 1.0054 | 0.5535 | oop |

## Section S8.